\documentclass[aps,11pt,prd,groupedaddress,nofootinbib,notitlepage,eqsecnum,preprintnumbers]{revtex4-2}

\usepackage[utf8]{inputenc}
\usepackage{graphicx}
\usepackage{amsmath, amssymb}
\usepackage{mathtools}
\usepackage{physics}
\usepackage[english]{babel}
\usepackage[colorlinks=true, allcolors=blue]{hyperref}
\usepackage{url}
\usepackage{hyperref}
\usepackage{cleveref}
\usepackage{xcolor}
\usepackage[normalem]{ulem}

\graphicspath{{./figures/}}

\definecolor{elpurple}{HTML}{a300ff}

\begin{document}

\title{
From formation to evaporation: Induced gravitational wave probes of the primordial black hole reheating scenario
}

\author{\textsc{Guillem Dom\`enech$^{a,b}$}}
    \email{{guillem.domenech}@{itp.uni-hannover.de}}
\author{\textsc{Jan Tränkle$^{a}$}}
\email{{jan.traenkle}@{itp.uni-hannover.de}}

\affiliation{$^a$Institute for Theoretical Physics, Leibniz University Hannover, Appelstraße 2, 30167 Hannover, Germany}
\affiliation{$^b$ Max-Planck-Institut für Gravitationsphysik, Albert-Einstein-Institut, 30167 Hannover, Germany}

\begin{abstract}
We study the Primordial Black Hole (PBH) reheating scenario, where PBHs originate in a general cosmological background. In this scenario, ultralight PBHs with masses $M\lesssim 10^8$g temporarily dominate the Universe and reheat it via Hawking radiation before Big Bang Nucleosynthesis (BBN). We investigate whether the induced Gravitational Wave (GW) spectrum associated with PBH reheating contains information about the pre-PBH-dominated stage, namely the initial equation of state $w$ (after inflation). We first derive the transfer functions of curvature fluctuations for general $w$ with adiabatic and isocurvature initial conditions.
We find that, in general, a stiffer equation of state enhances the induced GW amplitude as it allows for a longer PBH dominated phase compared to the radiation dominated case. We also find that the spectral slope of GWs induced by primordial curvature fluctuations is sensitive to $w$, while the spectral slope of GWs induced by PBH number density fluctuations is not. Lastly, we derive constraints of the initial PBH abundance as a function of $w$, using BBN and Cosmic Microwave Background (CMB) observations. A stiffer equation of state leads to stricter constraints on the initial energy density fraction, as induced GWs are enhanced.
Interestingly, we find that such induced GW signals may enter the observational window of several future GW detectors, such as LISA and the Einstein Telescope. Our formulas, especially the curvature fluctuation transfer functions, are applicable to any early matter-dominated universe scenario.
\end{abstract}

\maketitle
\newpage

\section{Introduction}\label{sec:Introduction}
Observations of the Cosmic Microwave Background (CMB) \cite{Planck:2018jri} indicate that the Universe underwent a period of accelerated expansion, so-called cosmic inflation \cite{Brout:1977ix,Starobinsky:1979ty,Guth:1980zm,Sato:1981qmu}, which drove the Universe to almost perfect spatial flatness and homogeneity, while at the same time providing the seeds for inhomogeneous structures to grow \cite{Kodama:1984ziu,Mukhanov:1990me,Bassett:2005xm, Baumann:2009ds}.
On the other hand, the abundances of light elements like hydrogen and helium and the associated theory of Big Bang Nucleosynthesis (BBN), inform us that the Universe was thermalised and dominated by relativistic particles before it had cooled to a temperature of  4 MeV \cite{Kawasaki:1999na,Kawasaki:2000en,Hannestad:2004px,Hasegawa:2019jsa,Grohs:2023voo,Dodelson:2003ft}.

However, the physics governing the transition from the supercooled, empty Universe at the end of inflation to the hot, radiation dominated Universe at BBN is largely unconstrained. This period is referred to as reheating, and the details of the process strongly depend on the nature of the inflaton field, in particular on the shape of its potential and its coupling to standard model particles \cite{Allahverdi:2010xz, Amin:2014eta, Haque:2022kez}. This leaves a gap in our understanding of the primordial Universe, where different expansion histories are equally compatible with current data \cite{Allahverdi:2020bys}. It is, therefore, crucial to understand the implications of different models and test them against observations. A key observable that may be in reach of current and near-future detectors is the stochastic gravitational wave background (SGWB) originating from this epoch. Such a SGWB is predicted in a multitude of scenarios and is expected to posses unique features allowing to differentiate between the various models and probe the expansion history of the early Universe \cite{Giovannini:1998bp, Boyle:2007zx, Baumann:2007zm, Assadullahi:2009nf, Kuroyanagi:2011fy, Cui:2018rwi, Bernal:2019lpc,  Domenech:2019quo, Domenech:2020kqm, Gouttenoire:2021jhk, Soman:2024zor, Duval:2024jsg}. 

In this work, we consider one particularly exciting scenario to fill the Universe with radiation after inflation: the Primordial Black Hole (PBH) reheating scenario, considered by Carr \cite{Carr:1976zz} as early as 1976.
The key assumption in this scenario is the formation of a population of ultralight PBHs within a certain mass range and with a suitable initial abundance. After formation, the PBH population behaves like a non-relativistic matter fluid and, if the PBH lifetime is long enough, will become the dominant component of the Universe. Through the semi-classical Hawking process \cite{Hawking:1975vcx} the PBHs evaporate after some time, filling the Universe with relativistic particles and leaving it in the standard radiation dominated era preceding BBN \cite{RiajulHaque:2023cqe}.

PBH reheating is a plausible scenario, as it does not require the introduction of new fields or couplings and relies only on semi-classical results.\footnote{That is, except for PBH formation itself. In the case of PBHs formed through the collapse of large overdensities, an enhancement of the curvature perturbation at the PBH mass scale by several orders of magnitude compared to CMB scales is necessary to produce a significant number of PBHs \cite{Escriva:2022duf}.}
An appealing aspect of the scenario is that it can be potentially tested through the observation of gravitational waves (GWs) in the near future. In fact, there are already constraints from the overproduction of induced GWs \cite{Papanikolaou:2020qtd, Domenech:2020ssp, Domenech:2021wkk}. Furthermore, a number of recent papers \cite{Inomata:2020lmk, Papanikolaou:2020qtd, Domenech:2020ssp, Domenech:2021wkk, Kozaczuk:2021wcl, Papanikolaou:2022chm, Bhaumik:2022pil,Bhaumik:2022zdd,Domenech:2024cjn, Domenech:2024kmh, Balaji:2024hpu, Papanikolaou:2024kjb, He:2024luf, Bhaumik:2024qzd} showed that the GWs induced in the PBH reheating process by PBH number density fluctuations and primordial curvature perturbations could enter the frequency band of next-generation GW detectors.\footnote{See also \cite{Dolgov:2011cq, Zagorac:2019ekv, Hooper:2020evu, Ireland:2023avg} for discussions of the SGWB from evaporating PBHs due to graviton emission in Hawking radiation and binary PBH mergers.}

In most previous works \cite{Inomata:2020lmk, Papanikolaou:2020qtd, Domenech:2020ssp, Domenech:2021wkk, Kozaczuk:2021wcl, Papanikolaou:2022chm, Bhaumik:2022pil,Bhaumik:2022zdd, Domenech:2024cjn, Domenech:2024kmh, Balaji:2024hpu, Papanikolaou:2024kjb, He:2024luf} (except for Ref.~\cite{Bhaumik:2024qzd}) it is assumed, for simplicity, that the Universe transitions directly to radiation domination (RD) with an equation of state (EoS) parameter $w=1/3$ after inflation, and that the PBHs form during this RD period. However, in the PBH reheating scenario the inflaton need not decay into radiation at all. For example, inflation may end with coherent oscillations of the inflaton field, which for a quadratic potential leads to a dust-like EoS $w\approx0$, corresponding to a period of early matter domination (eMD) \cite{Assadullahi:2009nf, Erickcek:2011us, Amin:2011hj, Alabidi:2013lya, Lozanov:2022yoy}. Or, the inflaton may simply accelerate in a run-away potential, typical of quintessential inflation scenarios, where there is a period of kination with $w=1$ after inflation \cite{Spokoiny:1993kt, Joyce:1996cp, Peebles:1998qn, Brax:2005uf, Hossain:2014xha, WaliHossain:2014usl, Redmond:2018xty, Gouttenoire:2021jhk}.
For the inflaton $\varphi$ coherently oscillating in a polynomial potential $V(\varphi)\propto |\varphi|^{2n}$ with generic $n$, the value of $w$ is determined by the shape of the potential as $w=\frac{n-1}{n+1}$ \cite{Turner:1983he}, while for a scalar field with an exponential potential $V(\varphi)\propto e^{-\lambda \varphi}$ the relation is $3(1+w)=\lambda^2$ \cite{Lucchin:1984yf}. In both cases, $w$
can take any value between $-1$ and $1$ \cite{Dodelson:2003ft}. For example, a stiff EoS with $w=2/3$ can be realised by a potential $V(\varphi)\propto \varphi^{10}$ with $n=5$, or an exponential potential with $\lambda = \sqrt{5}$.
In general, $w$ may also vary over time during reheating and radiation domination \cite{Saha:2020bis, Lozanov:2016hid, Carr:2019kxo}.
See e.g.~Ref.~\cite{Allahverdi:2020bys} for a review of alternative expansion histories of the early Universe.

In the present paper we lift the assumption of early RD and instead treat the content of the primordial Universe as an adiabatic fluid with a generic equation of state. We discuss how different values of the EoS parameter $w$ impact PBH reheating and deform the allowed parameter space. In the process, we also compute the transfer functions for the curvature perturbation in the transitions from the primordial fluid dominated to the PBH dominated era, for both isocurvature and adiabatic initial conditions and for general constant $w$. It should be noted that recently Ref.~\cite{Bhaumik:2024qzd} studied a similar situation in the context of the memory burden effect \cite{Dvali:2020wft,Balaji:2024hpu, Barman:2024iht}, focusing on possible degeneracies between the value of $w$ and the memory burden parameters in the induced GW spectrum. Our work extends their analysis of the effect of a general EoS by providing an analytical computation of the full transfer function of the curvature perturbation and the induced GW spectra, which, in our opinion, provides more physical insight.

Using these results, we compute the GWs induced by PBH number density fluctuations, which are isocurvature in nature, and those induced by an adiabatic primordial curvature perturbation. We find that an EoS stiffer than radiation ($w> 1/3$) leads to an enhanced GW amplitude, which results in exciting observational prospects and opens the parameter space towards lower initial PBH abundances. Further, we show that a combined observation of the isocurvature and adiabatic induced GWs would allow us to determine the initial PBH mass and abundance, as well as the primordial EoS parameter and the parameters of the primordial curvature power spectrum.

This paper is organised as follows. In \cref{sec:PBH_reheating}, we introduce the PBH reheating scenario with a general cosmological background, focusing on the impact of the EoS parameter $w$ on the different timescales in the scenario.
The evolution of isocurvature-induced and adiabatic curvature perturbations is discussed in detail in \cref{sec:Curvature_perturbations}, treating the early $w$-dominated, the PBH dominated and the RD eras separately in \cref{sec:Fluctuations_wDE,sec:Fluctuations_eMD,sec:Fluctuations_RD}, respectively.
In \cref{sec:iGW_Spectrum} we compute the scalar induced GW spectra after PBH evaporation and discuss observational prospects.
We close with a summary and discussion in \cref{sec:Conclusion}. Technical details are deferred to several appendices. We work in units where $\hbar=c=k_B=1$ and we use the reduced Planck mass $M_{\rm Pl}=1/\sqrt{8\pi G_N}\approx 4.34 \times 10^{-6} \text{g}\approx 2.44\times 10^{18}\text{GeV}$.

\section{PBH reheating scenario}\label{sec:PBH_reheating}
Let us briefly review the PBH reheating scenario. We assume that the Universe after inflation is dominated by an adiabatic perfect fluid with a constant equation of state parameter ${w\coloneq P_w/\rho_w}$, with $\rho_w$ and $P_w$ denoting the energy density and pressure of the fluid, respectively.
In this work, we will allow $w$ to take any constant value within the range $0<w\leq 1$.
We exclude exact matter domination ($w=0$) because PBH formation during an eMD requires very long timescales to form an apparent horizon \cite{Escriva:2020tak}, and PBHs can not become dominant, which we assume below.
By keeping $w$ as a free parameter, we parameterise our ignorance of the content of the primordial Universe.

In order not to spoil the successful predictions of BBN, the Universe needs to transition to the radiation dominated era before cooling down to a temperature $T\sim 4\,\text{MeV}$ \cite{Kawasaki:1999na,Kawasaki:2000en,Hannestad:2004px,Hasegawa:2019jsa}. In the PBH reheating scenario, this is achieved by forming a population of microscopic black holes from large overdensities in the primordial fluid, which come to dominate the Universe for a transient period and decay via Hawking evaporation before BBN, filling the Universe with Hawking radiation.
The effects of evaporating PBHs in the early Universe have been first studied almost 50 years ago \cite{Carr:1976zz} and have attracted a lot of attention recently \cite{Inomata:2020lmk, Papanikolaou:2020qtd, Domenech:2020ssp, Domenech:2021wkk, Kozaczuk:2021wcl, Papanikolaou:2022chm, Bhaumik:2022pil, Bhaumik:2022zdd, RiajulHaque:2023cqe, Domenech:2024cjn, Domenech:2024kmh, Balaji:2024hpu, Papanikolaou:2024kjb, He:2024luf, Bhaumik:2024qzd, Dolgov:2011cq, Zagorac:2019ekv, Hooper:2020evu, Ireland:2023avg, Lennon:2017tqq, Hooper:2019gtx, Masina:2020xhk, Barman:2024slw}.
As we show in this paper, observation of the GWs induced in the PBH reheating scenario may allow us to constrain the value of $w$ and thus, the content of the post-inflationary Universe, in addition to possibly being the only direct probe of PBHs that have evaporated already.
In our derivations, we closely follow Ref.~\cite{Domenech:2020ssp} and use it to cross-check our results in the limit of $w=1/3$.

\subsection{PBH formation and evaporation}
In this work, we focus on PBHs formed from the gravitational collapse of large density fluctuations.
For reviews on PBH formation and PBHs in general, we refer the reader to \cite{Khlopov:2008qy, Escriva:2022duf}, and for current constraints, see \cite{Carr:2020gox, Carr:2023tpt, Gomez-Aguilar:2023bej}.
A PBH can emerge when a superhorizon fluctuation with an amplitude larger than a critical threshold value reenters the horizon after inflation. 
Then, some fraction $\gamma$ of the mass inside the Hubble horizon at that time will end up inside the PBH. Accordingly, the initial PBH mass is given by
\begin{equation} \label{eq:Mf}
    M_{\rm f} = \gamma \frac{4 \pi M_{\rm Pl}^2}{H_{\rm f}} \,,
\end{equation}
where $H$ is the Hubble parameter and the subscript "f" denotes evaluation at the time of PBH formation. $\gamma=\gamma(w,\delta_m)$ is an $\mathcal{O}(1)$ parameter quantifying the fraction of the mass inside the horizon that ends up in the PBH, and it depends on the amplitude of the fluctuation that leads to the collapse $\delta_m$, as well as the EoS parameter $w$ at formation \cite{Musco:2012au,Escriva:2020tak,Escriva:2021pmf, Escriva:2022duf}. For PBHs formed during radiation domination ($w=1/3$) and for the amplitude of the density contrast equal to the critical one, its value is approximately $\gamma\approx 0.2$, which we will adopt as a reference value.
For simplicity, we will assume that all PBHs form with the same mass $M_{\rm f}$. The effect of an extended mass function on PBH reheating has been studied in \cite{Papanikolaou:2022chm}.

On scales larger than their mean separation, the PBH population can be treated as a dust fluid with mean energy density $\rho_{\rm PBH}$. The initial energy density fraction in PBHs is then defined by
\begin{equation}
    \beta\coloneq \Omega_{\rm PBH,f} =  \frac{\rho_{\rm PBH,f}}{3 H_{\rm f}^2 M_{\rm Pl}^2} \,.
    \label{eq:DefinitionBeta}
\end{equation}
Note that the PBH reheating scenario is completely specified by the two parameters $(M_{\rm f},\beta)$.

After they have formed, PBHs start to evaporate due to Hawking radiation \cite{Hawking:1975vcx} and emit particles with an approximately thermal spectrum at the temperature ${T_{\rm PBH} = M_{\rm Pl}^2/M(t)}$, which is inversely proportional to the PBH mass $M(t)$. Due to the particle emission, a PBH loses mass with rate $\Gamma$, given by \cite{Page:1976df, Hooper:2019gtx}
\begin{equation}
    \Gamma\coloneq-\frac{d \ln M(t)}{dt} = A\frac{M_{\rm Pl}^4}{M(t)^3} \quad \text{with} \quad A=\frac{3.8 \pi}{480} g_H \,,
    \label{eq:EvaporationRate}
\end{equation}
where $g_H=g_H(T_{\rm PBH})$ are the spin-weighted degrees of freedom and $g_H\approx 108$ for $M\ll 10^{11}$g, assuming only the standard model of particles.
Integrating \cref{eq:EvaporationRate} reveals the time dependence of the PBH mass
\begin{equation}
    M(t)\approx M_{\rm f} \left(1-\frac{t}{t_{\rm eva}}\right)^{1/3} \,,
    \label{eq:MPBH(t)}
\end{equation}
where we assumed $t_{\rm eva}\gg t_{\rm f}$. The evaporation time is given by
\begin{equation}
    t_{\rm eva}=\frac{M_{\rm f}^3}{3 A M_{\rm Pl}^4} \approx 4.1 \times 10^{-28}\text{s}\left(\frac{M_{\rm f}}{1\text{g}}\right)^3 \,.
    \label{eq:teva}
\end{equation}
The subscript “eva” is used to denote evaluation at the time of PBH evaporation, i.e.~reheating of the Universe.
The solution \labelcref{eq:MPBH(t)} shows that the PBH mass initially decreases very slowly and then quickly drops to zero as $t\rightarrow t_{\rm eva}$, which is a consequence of the fact that the Hawking temperature $T_{\rm PBH}\propto 1/M$.
For the sake of simplicity, we will focus on non-rotating Schwarzschild PBHs in our analysis. This is reasonable since PBHs formed by the gravitational collapse of large fluctuations are generally expected to have very low initial spin if $M_{\rm f}$ is of the order of the horizon mass at formation \cite{Escriva:2022duf, Chiba:2017rvs}. Over the course of their lives, PBHs can potentially acquire spin by mergers and accretion. However, accounting for a spin would shorten the PBH lifetime only by a factor of order unity and would, therefore, slightly modify the allowed parameter space but leave our general conclusions unchanged \cite{Page:1976ki,Masina:2021zpu}.
Note that in this work we are assuming that the semi-classical approximation of Hawking's calculation holds until the black hole has evaporated completely. The impact of modifications to Hawking evaporation due to quantum effects like the memory burden \cite{Dvali:2020wft} on the PBH reheating scenario has been considered in \cite{Balaji:2024hpu, Barman:2024iht, Bhaumik:2024qzd}. 

Remarkably, we can put both upper and lower bounds on the allowed initial PBH mass $M_{\rm f}$ for the scenario to be compatible with current observations. 
On the one hand, for successful Big Bang Nucleosynthesis, the Universe should be radiation dominated and thermalised at the latest when it has reached a temperature $T_{\rm BBN} \approx 4 \rm{MeV}$ \cite{Kawasaki:1999na, Kawasaki:2000en, Hannestad:2004px, Hasegawa:2019jsa, Grohs:2023voo, deSalas:2015glj}.
We can compute the temperature of the Universe right after PBH evaporation by assuming that PBHs dominate right before evaporation and using the relation between the energy density and temperature of radiation \labelcref{eq:RelationT} to find \cite{Domenech:2020ssp}
\begin{equation} \label{eq:Teva}
    T_{\rm eva} \approx 2.76 \times 10^4 \text{GeV} \left(\frac{g_*(T_{\rm eva})}{106.75}\right)^{-1/4} \left(\frac{g_H}{108}\right)^{1/2} \left(\frac{M_{\rm f}}{10^4 \rm{g}}\right)^{-3/2} \,.
\end{equation}
Note that $T_{\rm eva}$ is determined solely by the initial PBH mass $M_{\rm f}$. $g_*(T)$ denotes the number of relativistic degrees of freedom and takes the value $g_*=106.75$ for high temperatures $T\gtrsim 100\text{GeV}$ and assuming the standard model of particle physics is valid up to those energies.

On the other hand, the 2018 Planck data imposes an upper bound on the Hubble parameter during inflation, ${H_{\rm inf} < 2.5 \times 10^{-5} M_{\rm Pl}}$ at 95\% CL \cite{Planck:2018vyg}, assuming the simplest models of inflation. As we are assuming that PBH formation takes place after the end of inflation, this corresponds to a lower bound on the PBH mass.
Thus, imposing $T_{\rm eva}>T_{\rm BBN}$ and $H_{\rm f}<H_{\rm inf}$ we find the allowed mass range
\begin{equation} \label{eq:Mf_bounds}
   0.44 \,\text{g} \, \left(\frac{\gamma}{0.2}\right) \lesssim \, M_{\rm f} \, \lesssim 3.6 \times 10^8 \text{g} \left(\frac{g_*(T_{\rm eva})}{106.75}\right)^{-1/6} \left(\frac{g_H}{108}\right)^{1/3} \, ,
\end{equation}
comprising roughly eight orders of magnitude.

\subsection{Evolution of background densities}
\begin{figure}
\centering
\includegraphics[width=\textwidth]{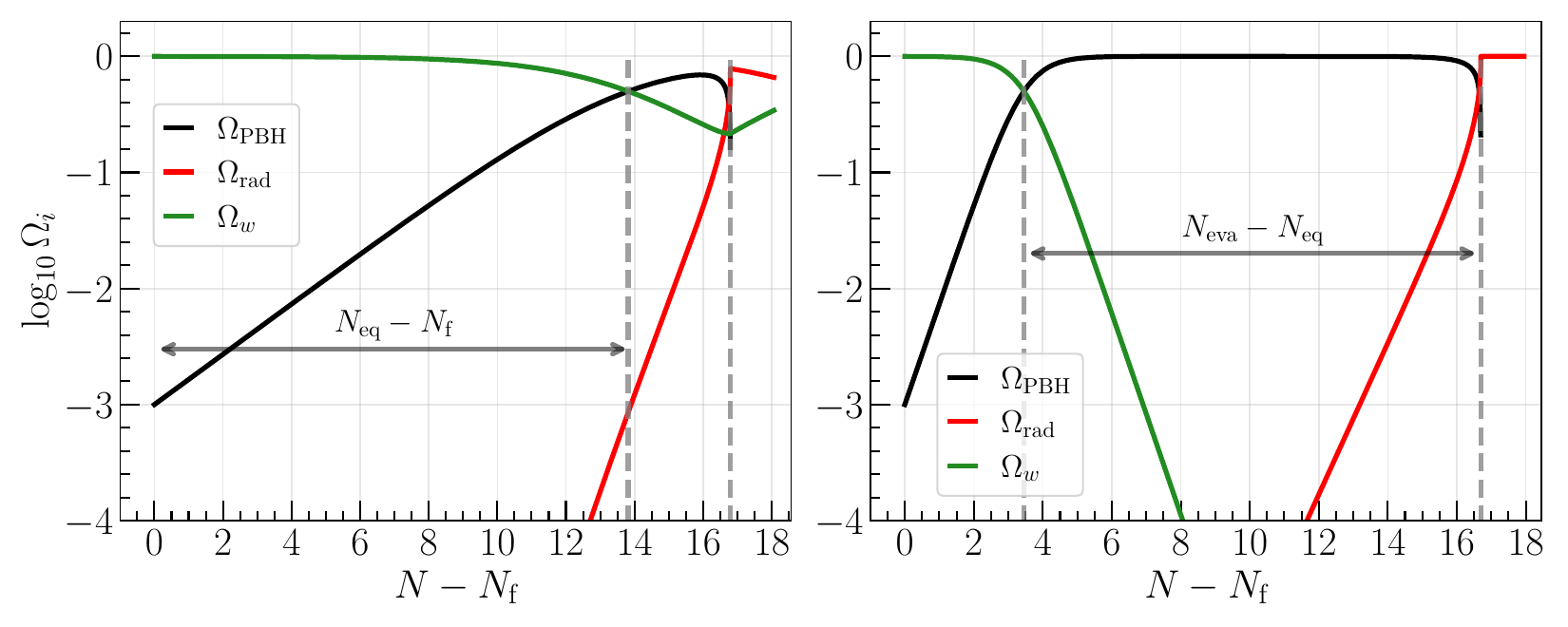}
\caption{Evolution of the energy density fractions of PBHs, radiation, and the primordial $w$-fluid in terms of the number of e-folds since PBH formation, $N-N_{\rm f}=\ln(a/a_{\rm f})$. In the left plot, we show the solution for $w=1/6$, the right one is for $w=2/3$, and we chose $M_{\rm f}=10$g and $\beta=10^{-3}$ in both cases for direct comparison. The dashed grey lines mark $N_{\rm eq}$ and $N_{\rm eva}$ as given in \cref{eq:Neq-Nf,eq:Neva-Nf}, respectively. Note that, although the chosen values are for illustration purposes only, for $w=1/6$ these values lie within the allowed parameter space, while for $w=2/3$ this particular combination violates the upper bound on $\beta$ set by BBN, see \labelcref{eq:beta_upper}. }
\label{fig:RhoPlot}
\end{figure}

Taking into account the energy transfer from the PBHs to radiation, the background energy densities are coupled and evolve according to
\begin{align}
    \dot{\rho}_{\rm PBH} + (3H + \Gamma)\rho_{\rm PBH} &= 0 \nonumber \\
    \dot{\rho}_{\rm rad} + 4H \rho_{\rm rad}- \Gamma\rho_{\rm PBH} &= 0 \nonumber \\
    \dot{\rho}_{w} + 3(1+w)H\rho_{w} &= 0 \,,
    \label{eq:EvolutionDensities}
\end{align}
where we assume that the products of Hawking evaporation only transfer energy to radiation and the overdot denotes a derivative with respect to $t$.
In order to gain some intuition for the background evolution, we numerically solve the system \labelcref{eq:EvolutionDensities} together with the Friedmann equation \labelcref{eq:FriedmannEq} in terms of the number of e-folds $N$ (defined by $dN=d \ln a$ in terms of the scale factor $a$). The resulting energy density fractions ${\Omega_i \coloneq \rho_i / (3H^2 M_{\rm Pl}^2)}$ are shown for two different values of $w$ in \cref{fig:RhoPlot}.
%
%
As one can observe, the time evolution can be split into three distinct epochs: an initial phase where the primordial $w$-fluid dominates, a second epoch where the PBH fluid comes to be the dominant component, and finally the radiation dominated era following PBH evaporation and preceding BBN.

This behaviour can be understood as follows.
As initially, the effect of Hawking radiation is small, $\rho_{\rm PBH}\propto a^{-3}$ and at some point, the PBH fluid starts to dominate over the primordial fluid, which is diluted faster due to the non-zero pressure. We denote the moment of $w$-PBH equality with the subscript “eq”, not to be confused with the much later, standard matter-radiation equality. After this equality, PBHs dominate the evolution of the Universe in a transient, early matter dominated era before evaporating in a final burst of radiation. Note that the final stage of the evaporation completes in less than one e-fold and can be approximately treated as almost instantaneous \cite{Inomata:2020lmk}.
See how, for smaller $w$, it takes significantly more e-folds to reach PBH domination. Note also that the number of e-folds until PBH evaporation $N_{\rm eva}$ is almost independent of $w$.
For $w<1/3$ we have to further assume that the primordial fluid decays to radiation and/or matter during the PBH dominated epoch, otherwise it starts dominating again after PBH evaporation. For our approximations to remain valid, we require the decay rate $\Gamma_w$ to lie within $H_{\rm eva}\lesssim \Gamma_w \lesssim H_{\rm eq}$ (assuming constant $\Gamma_w$). For values of $\Gamma_w$ outside this range, one can have additional phases of $w$- or radiation domination, and a more careful analysis would be required.

To better understand the situation, it is helpful to derive analytical estimates for the typical timescales involved.
For the number of e-folds from PBH formation to the early equality (i.e.~the beginning of PBH domination), we find
\begin{align}
\label{eq:Neq-Nf}
    N_{\rm eq}-N_{\rm f}  \simeq -\frac{1}{3w} \ln (\beta) \quad 
    \approx
        \begin{cases}
        23.03 -2 \ln \left(\frac{\beta}{10^{-5}}\right) & \quad (w=1/6)\\
        11.51 - \ln \left(\frac{\beta}{10^{-5}}\right) & \quad (w=1/3)\\
        5.76 - \frac{1}{2} \ln \left(\frac{\beta}{10^{-5}}\right)  & \quad (w=2/3)
    \end{cases} \,,
\end{align}
where we assumed $\rho_{w, \rm f}\gg \rho_{\rm PBH,f}$, or equivalently $\beta\ll 1$, and neglected the effect of Hawking radiation, i.e.~assumed that PBHs evaporate after equality. Note that \cref{eq:Neq-Nf} only depends on $\beta$, but not on $M_{\rm f}$. \Cref{eq:Neq-Nf} makes it clear that for a softer equation of state (i.e.~smaller $w$), it takes significantly longer until PBHs start dominating.

For the number of e-folds the Universe spends in PBH domination, we find
\begin{align}
\label{eq:Neva-Neq}
    N_{\rm eva}-N_{\rm eq} & \simeq \frac{2}{3} \ln \left((1+w)\frac{2\pi\gamma}{A} \left(\frac{M_{\rm f}}{M_{\rm Pl}}\right)^2\beta^{\frac{1+w}{2w}} \right) \nonumber \\
    & \approx \frac{4}{3}\ln\left(\frac{M_{\rm f}}{10^4 \text{g}}\right)+\frac{2}{3}\ln\left(\frac{\gamma}{0.2}\right) +
    \begin{cases}
        1.48 + \frac{7}{3} \ln\left(\frac{\beta}{10^{-5}}\right) & \quad (w=1/6)\\
        13.08 + \frac{4}{3} \ln\left(\frac{\beta}{10^{-5}}\right) & \quad (w=1/3)\\
        18.98+ \frac{5}{6} \ln\left(\frac{\beta}{10^{-5}}\right) & \quad (w=2/3)
    \end{cases}
    \,,
\end{align}
where we set $g_H=108$.
Note how for a stiffer primordial fluid (i.e.~larger $w$) the PBH dominated phase is significantly longer.
Later, we will see that this leads to an enhanced amplitude of the induced GWs because density fluctuations have more time to grow during the eMD era.

Demanding $N_{\rm eva}>N_{\rm eq}$, such that PBHs evaporate after equality, we have a lower bound on $\beta$
\begin{align} \label{eq:beta_lower}
    \beta > \beta_{\rm min}&\coloneq \left(\frac{A}{(1+w)2\pi\gamma}\left(\frac{M_{\rm Pl}}{M_{\rm f}}\right)^2\right)^{\frac{2w}{1+w}} \\
    & \approx \left(\left(\frac{g_H}{108}\right)\left(\frac{0.2}{\gamma}\right)\left(\frac{M_{\rm f}}{10^4 \text{g}}\right)^{-2}\right)^{\frac{2w}{1+w}} \times
    \begin{cases}
        5.31 \times 10^{-6} & \quad (w=1/6)\\
        5.48 \times 10^{-10} & \quad (w=1/3)\\
        1.27 \times 10^{-15} & \quad (w=2/3)
    \end{cases}  \,.
    \label{eq:beta_lower_vals}
\end{align}
\Cref{eq:beta_lower_vals} shows that a stiffer equation of state corresponds to a lower minimal value $\beta_{\rm min}$ required for reaching PBH domination.
It should be noted here that this lower bound on $\beta$ is necessary for the validity of the assumption of PBH domination used in the following calculations, but should not be understood as a physical constraint.

Combining \labelcref{eq:Neq-Nf} and \labelcref{eq:Neva-Neq} and demanding $N_{\rm eva}>N_{\rm eq}$, we find the number of e-folds for the whole PBH reheating process from formation to evaporation
\begin{align}
    \label{eq:Neva-Nf}
    N_{\rm eva}-N_{\rm f} & \simeq 24.4 + \frac{2}{3} \ln (1+w) + \frac{1}{3} \ln \left(\frac{\beta}{10^{-5}}\right) + \frac{4}{3} \ln \left(\frac{M_{\rm f}}{10^{4} \text{g}}\right) + \frac{2}{3}\ln\left(\frac{\gamma}{0.2}\right)
    \,,
\end{align}
which depends only very weakly (logarithmically) on $w$ and varies by less than half an e-fold between $w=0$ and $w=1$. This is due to the fact that, when PBH dominate the Universe, the evaporation temperature depends only on the mass of the PBH. In other words, it does not depend on the start of the PBH dominated phase. To summarize, $w$ determines when PBHs start dominating and the length of the PBH dominated phase, but the total number of e-folds until PBH evaporation is determined predominantly by the initial PBH mass $M_{\rm f}$ and abundance $\beta$.

To conclude this section, we derive some useful relations between the relevant scales in our system. A comoving scale $k_i$ is defined by the size of the comoving Hubble radius when that scale reenters the horizon, i.e.~$k_i = \mathcal{H}_i$, where $\mathcal{H}=aH$ is the comoving Hubble parameter.
The main relevant scale is the mean physical distance $d$ of the PBHs, which is determined by the details of PBH formation. It defines the comoving ultraviolet (UV) cutoff $k_{\rm uv}$ by \cite{Papanikolaou:2020qtd}
\begin{align} \label{eq:kUV}
    k_{\rm uv}\coloneq  \frac{a}{d} \quad
    \text{with} \quad d(t_{\rm f}) = \left(\frac{3}{4\pi}\frac{M_{\rm f}}{\rho_{\rm PBH,f}}\right)^{1/3} 
\,.
\end{align}
For scales larger than $k_{\rm uv}$, we can take a coarse-grained view and treat the population of PBHs as a non-relativistic matter fluid, also referred to as dust, but for smaller scales, the discrete nature leads to shot-noise effects and our effective treatment breaks down \cite{Papanikolaou:2020qtd}.
We find the following relations between the characteristic scales,
\begin{equation} \label{eq:Relationsk_kf}
    \frac{k_{\rm uv}}{k_{\rm f}}=\left(\frac{\beta}{\gamma}\right)^{1/3} \,,
    \quad \frac{k_{\rm eq}}{k_{\rm f}}\approx \sqrt{2} \ \beta^{\frac{1+3w}{6w}}
    \quad \text{and}\quad
    \frac{k_{\rm eva}}{k_{\rm f}}=\left(\frac{\beta A}{2\pi\gamma}\frac{M_{\rm Pl}^2}{M_{\rm f}^2}\right)^{1/3} \,,
\end{equation}
where we used \cref{eq:H(t),eq:af/aeq}. Then we also see that
\begin{equation}
    \frac{k_{\rm eq}}{k_{\rm uv}}=\sqrt{2}\gamma^{1/3}\beta^\frac{1+w}{6w}
     \quad \text{and}\quad
    \frac{k_{\rm eva}}{k_{\rm uv}}\approx
    4.32 \times 10^{-7}\left(\frac{g_H}{108}\right)^{1/3}\left(\frac{M_{\rm f}}{10^4\text{g}}\right)^{-2/3} \,,
    \label{eq:Relationsk_kUV}
\end{equation}
which allows us to express all relevant scales in terms of $k_{\rm uv}$. Note that the ratio $k_{\rm eva}/k_{\rm f}$ in \cref{eq:Relationsk_kf} is independent of $w$, under the assumption that PBHs dominate the Universe. This is because, if PBHs dominate the Universe, both $H_{\rm eva}$ and $H_{\rm f}$ depend only on $M_{\rm f}$ and the ratio ${a_{\rm eva}/a_{\rm f}= \left(n_{\rm f}/n_{\rm eva}\right)^{1/3}}$ (where $n$ is the PBH number density) depends on $\beta$ and $M_{\rm f}$, but not $w$. Thus, it follows that $k_{\rm eva}/k_{\rm f}=a_{\rm eva}H_{\rm eva}/(a_{\rm f}H_{\rm f})$ is independent of $w$, except for a possible $w$ dependence in $\gamma$. However, this does not apply to the $w\to 0$ limit as PBH would never dominate the Universe unless $\beta\sim 1$. Also note that we find overall agreement with Ref.~\cite{Bhaumik:2024qzd} except for the possible $w$-dependence in $k_{\rm f}$ and $k_{\rm eq}$ of Ref.~\cite{Bhaumik:2024qzd} through $\gamma$. Since there are uncertainties in the value of $\gamma$ we leave it as a free parameter not to obscure the discussion.

Finally, we can express the comoving scales in physical units using the temperature of the radiation bath \labelcref{eq:Teva}. We find e.g.~for the scale corresponding to PBH evaporation \cite{Domenech:2020ssp}
\begin{equation} \label{eq:kEva}
    k_{\rm eva}\approx 4.7 \times 10^{11} \text{Mpc}^{-1} \left(\frac{g_*(T_{\rm eva})}{106.75}\right)^{1/4}
    \left(\frac{g_{*,s}(T_{\rm eva})}{106.75}\right)^{-1/3}
    \left(\frac{g_H}{108}\right)^{1/2}
    \left(\frac{M_{\rm f}}{10^4 \rm{g}}\right)^{-3/2} \,,
\end{equation}
where we used entropy conservation, i.e.~$a T g_{*,s}^{1/3}\sim$const., and we introduced the number of entropic degrees of freedom $g_{*,s}$ \cite{Baumann:2022mni}.
\Cref{eq:kEva} makes it apparent that PBH reheating probes physics at scales many orders of magnitude smaller than the ones observed in CMB measurements, i.e.~at the order of the pivot scale $k_{\rm CMB}\approx 0.05 \text{Mpc}^{-1}$ \cite{Planck:2018vyg}.

\section{Curvature and isocurvature perturbations} \label{sec:Curvature_perturbations}
Let us now focus on the evolution of cosmological perturbations. We consider a perturbed Friedmann-Lema\^itre-Robertson-Walker (FLRW) metric in the Newton gauge,
\begin{equation} \label{eq:FLRW}
    ds^2 = a^2(\tau) \left[ - (1+2 \Psi)\text{d}\tau^2 + \left( \delta_{ij} + 2 \Phi \delta_{ij} + h_{ij}\right) \text{d}x^i \text{d}x^j \right] \,,
\end{equation}
where the two scalar potentials are related by $\Psi =- \Phi$ in the absence of anisotropic stress, $h_{ij}$ are the transverse-traceless tensor perturbations and the conformal time $\tau$ is defined by $\text{d}t=a\text{d}\tau$ in terms of the scale factor $a$ and the cosmic time $t$. For simplicity, we assume $\Psi =- \Phi$ from now on.
For details on cosmological perturbation theory, more generally, see Refs.~\cite{Dodelson:2003ft, Malik:2008im, Baumann:2022mni}. We provide details of the calculations in App.~\ref{app:einsteinequations}.

We begin by studying the evolution of the curvature perturbation $\Phi$ during the transition from the $w$- to the PBH-dominated era. In this early epoch where $a\ll a_{\rm eva}$ we can safely neglect the effect of Hawking radiation. At the background level, the Friedmann equation \labelcref{eq:FriedmannEq} yields
\begin{align}
k_{\rm eq}\tau=\frac{2 \sqrt{2} }{1+3
   w}\chi^{\frac{1+3w}{2}}\,
   _2F_1\left(\frac{1}{2},\frac{1+3w}{6w}
   ;\frac{1+9w}{6w};-\chi^{3\omega}\right)\quad{\rm with}\quad \chi\coloneq\frac{a}{a_{\rm eq}}\,,
   \label{eq:scalefactor}
\end{align}
where $_2F_1\left(a,b;c;d\right)$ is the hypergeometric function.

Turning our attention to the perturbations, we take the trace of the $ij$-component of the perturbed Einstein equations \labelcref{eq:ijEinsteinEq} and use \cref{eq:00EinsteinEq} to obtain the equation of motion for $\Phi$, namely
\begin{equation}
    \Phi'' +  3 \mathcal{H}(1+c_s^2)\Phi' + \left((1+3c_s^2)\mathcal{H}^2+2\mathcal{H}' +c_s^2 k^2\right)\Phi = \frac{1}{2}a^2c_s^2\rho_{\rm PBH} S \,,
    \label{eq:PhiEq}
\end{equation}
where $'\equiv d/d\tau$ and $k$ denotes the wavenumber of a given Fourier mode. The sound speed $c_s^2$ is defined as \cite{Malik:2008im}
\begin{equation}
    c_s^2 = \, \frac{c_{w}^2(1+w)\rho_w}{\rho_{\rm PBH}+(1+w)\rho_w} 
    \quad \text{where} \quad
    c_{w}^ 2=  \frac{\delta P_w}{\delta \rho_w} \,,
\end{equation}
with the pressure and density perturbations respectively denoted by $\delta P$ and $\delta \rho$. In \cref{eq:PhiEq} we also defined the isocurvature perturbation as \cite{Malik:2008im}
\begin{equation} \label{eq:S_Def}
    S\coloneq \frac{\delta\rho_{\rm PBH}}{\rho_{\rm PBH}}-\frac{\delta\rho_w}{(1+w)\rho_w} \,,
\end{equation}
which evolves according to
\begin{equation}
    S' = k^2 V_{\rm rel} + \frac{3\mathcal{H}}{(1+w)\rho_w}\delta P_{\rm nad} \qquad \text{with} \quad \delta P_{\rm nad}=(c_{w}^2-w)\delta\rho_w \,,
    \label{eq:SPrimeEq}
\end{equation}
where we introduced the relative velocity $V_{\rm rel}\coloneq V_{\rm PBH}-V_w$, and $\delta P_{\rm nad}$ is the non-adiabatic pressure \cite{Malik:2008im}. \Cref{eq:SPrimeEq} follows from \cref{eq:Energy_Cons_PBH,eq:Energy_Cons_w}.

In the case of an adiabatic perfect fluid, which we will assume hereon for simplicity, the sound speed is given by $c_{w}^2=w$ and $\delta P_{\rm nad}$ vanishes.
\Cref{eq:SPrimeEq} then shows how the relative velocity $V_{\rm rel}$ between the $w$ and PBH fluids is responsible for the time evolution of the isocurvature perturbation. It also shows that $S$ is constant on superhorizon scales.
Further, using \cref{eq:SPrimeEq} together with \cref{eq:Momentum_Cons_PBH,eq:Momentum_Cons_w}, we find a closed system of equations for $S$ and $\Phi$, namely
\begin{equation}
    S'' + (1+3(c_s^2-w))\mathcal{H}S' - k^2(c_s^2-w)S = \frac{2 c_s^2}{a^2(1+w)\rho_w}k^4 \Phi\,.
    \label{eq:SEq}
\end{equation}
\Cref{eq:PhiEq,eq:SEq} determine how $\Phi$ and $S$ evolve. In particular, it shows that an initial isocurvature perturbation generates a curvature perturbation as the Universe evolves. This has important implications, as curvature perturbations are the main source of induced GWs \cite{Domenech:2021ztg}.

At this point let us clarify the difference between adiabatic and isocurvature initial conditions. In the adiabatic case, there exists a slicing of spacetime where the energy density of all (matter) fields is homogeneous and fluctuations are carried only by the metric \cite{Domenech:2023jve}. This is e.g.~the case in the standard single-field inflationary scenario, where the matter fields all inherit the fluctuations of the same inflaton field. In this case, the Newtonian potential $\Phi$ has a non-zero initial value $\Phi_0$ of primordial origin.
In the opposite case of isocurvature initial conditions, the initial curvature perturbation vanishes. For this to be the case, the density fluctuations in the fields filling the Universe need to compensate each other, and there is thus no slicing in which each of their fluctuations vanish separately \cite{Domenech:2023jve}. Consequently, isocurvature is related to relative number density fluctuations and can only appear in a system of multiple fluids.

Interestingly, initial isocurvature fluctuations are inevitably generated after PBHs form. This is because during PBH formation, energy is transferred from the primordial fluid to the PBH fluid.
Simply put, when a PBH is formed from the collapse of a large density fluctuation in the primordial fluid, it leaves a “hole” in the background energy density, and thus the fluctuations in both fluids are equal and opposite, ${\delta_{\rm PBH,f}=-\delta_{w, \rm f}}$, leaving the total energy density unperturbed. Initially it is $\rho_w\gg\rho_{\rm PBH}$, and the initial isocurvature perturbation \labelcref{eq:S_Def} is determined by the PBH density contrast
\begin{equation}
    S_0 \approx \delta_{\rm PBH, f}\coloneq \frac{\delta \rho_{\rm PBH,f}}{\rho_{\rm PBH,f}} \,.
\end{equation}
Assuming that PBH formation is a rare event, the PBHs will be located randomly in space, and their spatial correlations are described by Poisson statistics \cite{Ali-Haimoud:2018dau,Carr:2018rid,Papanikolaou:2020qtd}.  PBH number density fluctuations, and therefore also isocurvature fluctuations, are then distributed according to the dimensionless initial power spectrum
\footnote{The dimensionless power spectrum $\mathcal{P}_{\delta}(k)$ is defined in terms of the two-point correlation function
\begin{equation}
  \langle \delta_{\rm PBH}(k) \delta_{\rm PBH}(k')\rangle = \frac{2 \pi^2}{k^3}\mathcal{P}_{\delta_{\rm PBH}}(k)\delta_D(k+k') \, .
\end{equation}}
\cite{Papanikolaou:2020qtd}
\begin{equation}
    \mathcal{P}_{S_0}(k)\approx\mathcal{P}_{\delta_{\rm PBH,f}}(k)=\frac{2}{3\pi}\left(\frac{k}{k_{\rm uv}}\right)^3 \Theta(k_{\rm uv}-k) \,,
    \label{eq:InitialIsocurvaturePS}
\end{equation}
which is valid on scales larger than the UV-cutoff $k_{\rm uv}$ defined in \labelcref{eq:kUV}.

We also consider that, in addition to isocurvature fluctuations there may be primordial adiabatic fluctuations, that is initial perturbations in the total energy density stemming from inflation.\footnote{Note that, in our scenario, large adiabatic fluctuations at the scale $k_{\rm f}$ are required for generating the PBHs in the first place. This also generates an induced GW background. However, we do not consider it here as it peaks at very high frequencies.} By extrapolation from CMB measurements, we parameterise the initial curvature power spectrum $\mathcal{P}_{\Phi_0}(k)$ with a power-law of the form
\begin{equation}
    \mathcal{P}_{\Phi_0}(k) = A_s \left(\frac{k}{k_{\rm *}}\right)^{n_s-1}  \Theta(k_{\rm cut}-k) \Theta(k-k_{\rm IR})\,,
    \label{eq:ScalarPS}
\end{equation}
where $k_{\rm *}$ is an arbitrary pivot scale. Note that the amplitude $A_s$ is arbitrary and, at the small scales we are considering, can in principle be significantly larger than the $\mathcal{O}(10^{-9})$ value measured at CMB scales $k_{\rm CMB}$ \cite{Planck:2018vyg}.  We also assume a UV cutoff of the power spectrum at some scale $k_{\rm cut}$.
The origin of such a cutoff is the fact that density fluctuations eventually become non-linear during a matter dominated phase. Thus, one often cuts the primordial spectrum at the scale that becomes non-linear at the end of the matter dominated phase, say at $k_{\rm NL}$ \cite{Assadullahi:2009nf,Inomata:2019zqy} (we give an estimate of the non-linear scale in \cref{sec:Suppresion_Seva}).\footnote{More conservatively, Refs.~\cite{Bhaumik:2022pil,Bhaumik:2022zdd,Bhaumik:2024qzd} set the cutoff at $k_{\rm eq}$.}  In this way, density fluctuations always remain in the linear regime. However, GW production will continue in the non-linear regime (see, e.g.~Ref.~\cite{Kawasaki:2023rfx} for early analytical approximations and Ref.~\cite{Fernandez:2023ddy} for a numerical simulation in a gradual transition). For this reason, we take $k_{\rm cut}$ as an arbitrary value between $k_{\rm NL}\leq k_{\rm cut}\leq k_{\rm uv}$ (beyond $k_{\rm uv}$ it is difficult to talk about fluctuations in the PBH dominated universe). For $k_{\rm cut}=k_{\rm NL}$ we may be underestimating the signal, while for $k_{\rm cut}=k_{\rm uv}$ we are probably overestimating it. Unfortunately, numerical simulations are needed to provide a more accurate estimate. In our plots we will set $k_*=k_{\rm cut}=k_{\rm uv}$ for direct comparison with the isocurvature case. Lastly, we impose an infrared (IR) cutoff at some scale $k_{\rm IR}>k_{\rm CMB}$ in order not to get into conflict with CMB observations.

We proceed to find approximation solutions to the equations of motion \labelcref{eq:PhiEq,eq:SEq} for the evolution of $\Phi$ and $S$ in the different epochs within the PBH reheating scenario, that is, the initial $w$-dominated phase, the following PBH-dominated era and finally the radiation era following PBH evaporation.
We will treat isocurvature ($\Phi(0)=0, \, S(0)=S_0$) and adiabatic ($\Phi(0)=\Phi_0, \, S(0)=0$) initial conditions separately, as they do not mix at linear order in perturbation theory. The general solution for $\Phi$ is given by the sum of the isocurvature and adiabatic solution branches. In the special case $w=1/3$, our solutions reduce to the ones collected in \cite{Domenech:2023jve}. We will first focus on the fluctuations that enter the horizon during the PBH dominated era (i.e.~$k\ll k_{\rm eq}$), and later on fluctuations that enter the horizon before the PBH dominated era ($k\gg k_{\rm eq}$).

\subsection{Superhorizon scales \texorpdfstring{$k\ll\mathcal{H}$}{k<<H}}  \label{sec:Fluctuations_Superhorizon}
Superhorizon scales are those scales $k$ that remain larger than the comoving Hubble radius $1/\mathcal{H}$ during the whole period under consideration, i.e.~$k\ll\mathcal{H}$.
Due to the rather complicated relation between the scale factor and conformal time, see \cref{eq:scalefactor}, it is convenient to introduce
\begin{equation}
    y\coloneq \left(\frac{a}{a_{\rm eq}}\right)^{3 w}=\chi^{3 w}
    \label{eq:yDefinition}
\end{equation}
as a new time variable. The full equations \labelcref{eq:PhiEq,eq:SEq} expressed in terms of $y$ are given in \cref{eq:PhiEq_y,eq:SEq_y}.
In the superhorizon regime we can neglect the scale-dependent $k^2$-term, which allows us to solve the equations analytically.

\subsubsection{Isocurvature fluctuations}
From \cref{eq:SPrimeEq} we know that $S$ is constant on superhorizon scales, such that $S=S_0$. For the curvature perturbation, on the other hand, we find
\begin{equation} \label{eq:Phi_Iso_Super}
    \Phi_{\rm iso}(y; k\ll \mathcal{H})=\frac{S_0}{3w} \left(\sqrt{y+1} \, _2F_1\left(\frac{1}{2},\frac{5}{6 w}+\frac{1}{2};\frac{5}{6 w}+\frac{3}{2};-y\right)-1\right) \,,
\end{equation}
where $_2F_1$ is the hypergeometric function.
At early times $y\ll1$, or equivalently $a\ll a_{\rm eq}$, the curvature perturbation \labelcref{eq:Phi_Iso_Super} grows linearly with $y$ as
\begin{equation}
   \Phi_{\rm iso}(y; k\ll \mathcal{H},a\ll a_{\rm eq})\approx \frac{S_0}{5+9w} y \,,
   \label{eq:Phi_Iso_Super_SmallChi}
\end{equation}
whereas after equality, it approaches a constant value,
\begin{equation}
    \Phi_{\rm iso}(k\ll \mathcal{H},a\gg a_{\rm eq})=\frac{S_0}{5} \,,
    \label{eq:PhiIsoSuperMD}
\end{equation}
independent of $w$. It should be noted that the independence of $w$ in \cref{eq:PhiIsoSuperMD} follows from the fact that the source of initial isocurvature becomes the source of curvature fluctuations. More precisely, as argued in Ref.~\cite{Papanikolaou:2020qtd}, the factor $1/5$ comes from relating the initial curvature fluctuation on uniform density slices at formation $\zeta_{\rm f}\sim S_0/3$ to the curvature fluctuation during the PBH dominated phase, namely $\Phi\approx 3/5\,\zeta=S_0/5$, where we used that $S$ is constant on superhorizon scales.
Also note that for $w\ll 1$, the limiting value \labelcref{eq:PhiIsoSuperMD} is approached only at very large values of $a\gg a_{\rm eq}$, while for larger $w$ it is reached earlier. This is a consequence of the fact that for small $w$, the transition to PBH domination is slower, cf.~\cref{fig:RhoPlot}.

\subsubsection{Adiabatic fluctuations}
For adiabatic fluctuations on superhorizon scales, the solution to \cref{eq:PhiEq} reads
\begin{equation}
    \Phi_{\rm ad}(y; k\ll \mathcal{H})=\frac{\Phi_0}{3 w +3} \left((3 w +5)-2 \sqrt{y+1} \, _2F_1\left(\frac{1}{2},\frac{5}{6 w }+\frac{1}{2};\frac{5}{6 w }+\frac{3}{2};-y\right)\right) \, .
    \label{eq:Phi_Ad_Super}
\end{equation}
At early times $\Phi_{\rm ad}(a\ll a_{\rm eq})=\Phi_0$ is constant.
After equality, $\Phi_{\rm ad}$ drops slightly but again goes to a constant value, given by
\begin{equation}
    \Phi_{\rm ad}(a\gg a_{\rm eq}; k\ll \mathcal{H})= \frac{3 w +5}{5w +5}\Phi_0 \, .
    \label{eq:PhiAdSuperMD}
\end{equation}
The prefactor can be understood from the relation of $\Phi$ to the gauge-invariant comoving curvature perturbation $\mathcal{R}(k\ll \mathcal{H})=\frac{3w +5}{3 w +3}\Phi$, which is conserved on superhorizon scales \cite{Lyth:2004gb,Baumann:2022mni}.

\subsection{Fluctuations during \texorpdfstring{$w$-domination ($a\ll a_{\rm eq}$)}{w-domination (a<<aEq)}} \label{sec:Fluctuations_wDE}
In this section, we will study how small-scale modes with $k\gg k_{\rm eq}$ evolve when they enter the horizon during the $w$-dominated era.
In order to study the evolution of these modes, we expand the equation of motion \labelcref{eq:PhiEq} for early times $a\ll a_{\rm eq}$, following the approach of \cite{Domenech:2021and}.
In this regime, the scale factor \labelcref{eq:scalefactor} is related to the conformal time by
\begin{align}
   \chi=\frac{a}{a_{\rm eq}} \approx \left(\sqrt{2}(1+b)\right)^{-1-b} \left(\frac{x}{\kappa}\right)^{1+b} \,,
   \label{eq:ScaleFactor_wDE}
\end{align}
where we introduced the useful parameter
\begin{equation}
    b\coloneq\frac{1-3w}{1+3w} \,\,
    \label{eq:bDefinition}
\end{equation}
for compactness of the equations. The value of $b$ ranges from $b=-1/2$ for $w=1$ over $b=0$ for radiation ($w=1/3$) to $b=1$ for $w=0$. In \cref{eq:ScaleFactor_wDE} we also defined
\begin{equation}
    x \coloneq k \tau \quad \text{and} \quad \kappa  \coloneq \frac{k}{k_{\rm eq}} \,.
\end{equation}
The parameter $\kappa$ determines whether a given mode $k$ enters before ($\kappa \gg 1$) or after ($\kappa\ll 1$) equality, and the time coordinate $x$ defines the superhorizon ($x\ll1$) and subhorizon ($x\gg1$) regimes. 
An expansion in $a\ll a_{\rm eq}$ is then equivalent to $x\ll \kappa$.
However, expanding directly in the variable $a/a_{\rm eq}$ is not well-defined for all values of $w$ (depending if $w>1/3$ or $w<1/3$).
Instead, we expand \cref{eq:PhiEq,eq:SEq} about $y\ll1$, where $y$ is the variable defined in \cref{eq:yDefinition}.
Transforming the resulting equations to the time variable $x$, one can separate the leading and subleading terms in $y$ as follows
\begin{align}
    \frac{d^2 \Phi}{d x^2}
    &+\frac{6 (1+w )}{(1+3 w)}\frac{1}{x}\frac{d \Phi}{d x}
    +w  \Phi \nonumber \\
    &+y\left(\frac{\left(12 w ^2+11 w +5\right)}{3 w ^2+4 w +1} \frac{1}{x}\frac{d\Phi}{d x}
    +\frac{w}{w +1}  \left(\frac{12 w }{(1+3w)^2}\frac{1}{x^2}-1\right) \Phi
    -\frac{6 w}{(1+3w)^2}\frac{1}{x^2}S \right) \approx 0 \, ,
    \label{eq:PhiEqX}
\end{align}
and similarly for the equation for $S$
\begin{align}
    \frac{d^2 S}{d x^2}
    &+\frac{2}{1+3 w}\frac{1}{x} \frac{dS}{d x}
    -\frac{w (1+3 w)^2}{6 (1+w)}x^2\Phi \nonumber \\
    &+y\left(\frac{\left(1+w+6 w ^2\right)}{1+4w+3w^2}\frac{1}{x} \frac{dS}{d x}+\frac{w}{1+w}S  +\frac{w (1+3w)^2}{6 (1+w)} x^2 \Phi\right)\approx 0 \,.
    \label{eq:SEqX}
\end{align}
The system \labelcref{eq:PhiEqX,eq:SEqX} can be solved with a perturbative ansatz ${\Phi=\Phi_0+\kappa^{-1+b}\Phi_1+...}$, and equivalently ${S=S_0+\kappa^{-1+b}S_1+...}$, where the power in $\kappa$ is determined by ${y\propto (x/\kappa)^{1-b}}$ from \labelcref{eq:yDefinition} and \labelcref{eq:ScaleFactor_wDE}.

\subsubsection{Isocurvature fluctuations}
\begin{figure}
  \centering
  \includegraphics[width=\textwidth]{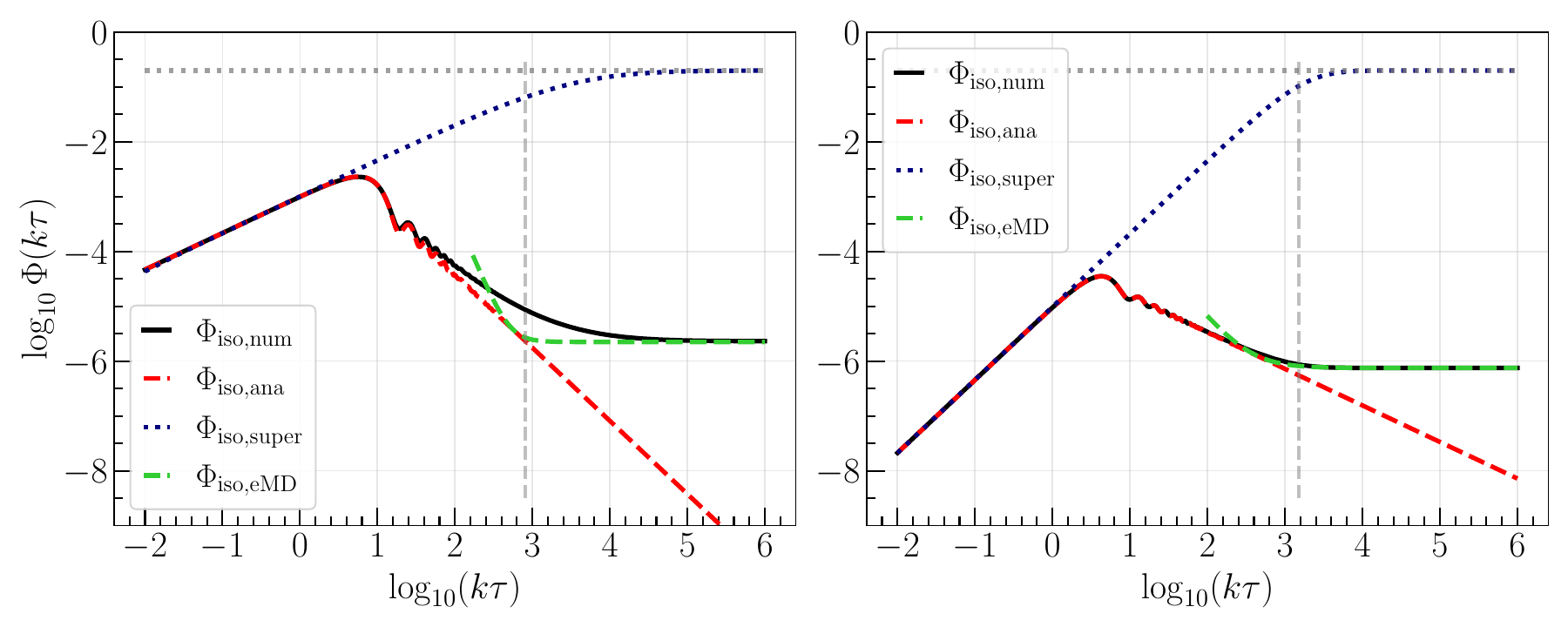}
  \caption{Evolution of the curvature perturbation $\Phi$ with isocurvature initial conditions from $w$- to PBH domination. We compare a numerical solution of \cref{eq:PhiEq,eq:SEq}, the superhorizon solution \labelcref{eq:Phi_Iso_Super} and the analytical result \labelcref{eq:Phi_Iso_wDE} for $w=1/6$ (left) and $w=2/3$ (right) with $\kappa = 10^3$ and $S_0=1$. The horizontal dotted line marks the limiting value $S_0/5$, and the vertical dashed line marks $\tau_{\rm eq}$. Additionally, we show the result of the matching for the plateau value \labelcref{eq:PhiIsoeMDInterpol}.}
  \label{fig:Phi_iso_Plot}
\end{figure}

We first consider isocurvature initial conditions with $\Phi_0=0$. Keeping only the leading order terms in $\Phi_1$ on the left-hand side of \cref{eq:PhiEqX} and treating $S_0$ as a constant source, we obtain
\begin{align}
    \Phi_{\rm iso}(x; a\ll a_{\rm eq})=& S_0 \, 2^{-\frac{b}{2}-4} \left(1+b\right)^{b} \left(3-3 b^2\right)^{\frac{1}{4} (-2 b-3)} \kappa ^{b-1} \Bigg(\frac{(1-b)^{b+\frac{5}{2}}}{b+4}x^{5/2} \Gamma \left(-b-\frac{3}{2}\right)  \nonumber \\ & \times 
    J_{-b-\frac{3}{2}}\left(c_w x\right) \, _1F_2\left(\frac{b}{2}+2;\frac{b}{2}+3,b+\frac{5}{2};-\frac{c_w^2 x^2}{4}\right)  \nonumber \\ &
    +12^{b+\frac{3}{2}} (b+1)^{b+\frac{3}{2}} x^{-2 b-\frac{1}{2}} \Gamma \left(b+\frac{3}{2}\right) \nonumber \\ & \times 
    J_{b+\frac{3}{2}}\left(c_w x\right) \, 
    _1F_2\left(\frac{1}{2}-\frac{b}{2};-b-\frac{1}{2},\frac{3}{2}-\frac{b}{2};-\frac{c_w^2 x^2}{4}\right)\Bigg)\,,
    \label{eq:Phi_Iso_wDE}
\end{align}
where $J_\nu(z)$ is the Bessel function of the first kind, $\Gamma(z)$ is the Euler gamma function, and $_1F_2$ is the generalised hypergeometric function.
In \cref{fig:Phi_iso_Plot}, the analytical solution \labelcref{eq:Phi_Iso_wDE} is shown in terms of $x$, together with a numerical solution of the full equations \labelcref{eq:PhiEq,eq:SEq} for comparison. As one can see, the analytical approximation agrees almost perfectly with the full numerical result, with deviations setting in only close to $\tau_{\rm eq}$. The approximation close to $\tau_{\rm eq}$ is better for larger $w$, where the transition between the two regimes is sharper.
We observe that the curvature perturbation grows while it is still outside the horizon and starts to oscillate and decay after it enters the horizon. Note how the scaling with $x$ and the amplitude of $\Phi$ vary with $w$.
We also show the superhorizon solution \labelcref{eq:Phi_Iso_Super} in \cref{fig:Phi_iso_Plot}, which agrees with \labelcref{eq:Phi_Iso_wDE} until $x\approx 1$, when the mode enters the horizon. The plateau of $\Phi$ after $a_{\rm eq}$ is discussed below in \cref{sec:Fluctuations_eMD}.
%
%

Expanding the Bessel and hypergeometric functions in \cref{eq:Phi_Iso_wDE} for small argument $x\ll1$ reproduces the superhorizon solution \labelcref{eq:Phi_Iso_Super_SmallChi}, as expected.
The behaviour of $\Phi$ after entering the horizon, i.e.~at $x\gg1$, is given by
\begin{align}
    \Phi_{\rm iso}(a\ll a_{\rm eq}; x\gg 1)\approx& 3 S_0\ 2^{\frac{b-3}{2}} (b+1)^{b+1}\kappa ^{b-1} x^{-b-1} \nonumber \\
    &-\frac{3 S_0 \pi ^{3/2} 2^{\frac{b-1}{2}} (b+1)^{b+2}   \csc (\pi  b) }{\sqrt{w} \Gamma \left(-\frac{b+2}{2}\right) \Gamma \left(\frac{b+3}{2}\right)} \kappa ^{b-1} x^{-b-2} \sin \left(c_w x\right) \,,
    \label{eq:Phi_Iso_wDE_LargeX}
\end{align}
which contains a decaying power-law term and a damped oscillating contribution.
Interestingly, when expressed in terms of the scale factor $\chi$, \cref{eq:Phi_Iso_wDE_LargeX} becomes independent of $w$ at leading order,
\begin{align}
    \Phi_{\rm iso}(a\ll a_{\rm eq}; x\gg 1)\approx  \frac{3 S_0}{4 \kappa ^2 \chi } -\frac{3\sqrt{2} \pi ^{3/2}  S_0  \csc (\pi  b)\chi ^{-\frac{b+2}{b+1}}}{4 \sqrt{w}  \Gamma \left(-\frac{b+2}{2}\right) \Gamma \left(\frac{b+3}{2}\right)\kappa ^3}\sin \left(\sqrt{\frac{2}{3}}\sqrt{1-b^2} \kappa  \chi ^{\frac{1}{b+1}}\right) \,,
    \label{eq:Phi_Iso_wDE_LargeX_Chi}
\end{align}
and is suppressed by $\kappa^{-2}$. This independence of $w$, and the factor $3/4$, in the leading order term of \eqref{eq:Phi_Iso_wDE_LargeX_Chi} can be derived by considering only the $k$-term in $\Phi$ in \cref{eq:PhiEq}. Then one sees that the leading order, constant term in $S$ is exactly compensated by the decaying leading order term of $\Phi$ in \eqref{eq:Phi_Iso_wDE_LargeX_Chi}. In simpler terms, $\Phi$ is given by the Poisson equation when PBH density fluctuations are dominant. Such behaviour is illustrated in \cref{fig:PhiChiPlot}.

The first order solution $S_1$ for the isocurvature perturbation can be given implicitly in terms of the first order solution \labelcref{eq:Phi_Iso_wDE} for $\Phi$. Keeping only the leading order terms in \cref{eq:SEqX} yields
\begin{align}
    S_{\rm iso}(x; a\ll a_{\rm eq})\approx S_0 
    & -\frac{2^{\frac{b-3}{2}} (1-b) (b+1)^{b-1} \kappa ^{b-1} }{3 (3-b) (b+2)} S_0 x^{3-b} \nonumber \\
    & - \frac{(1-b)}{3 (b+1)^2 (b+2)}\int_0^x dx_1 \, x_1^{-b-1} \left(\int_0^{x_1} dx_2 \, x_2^{b+3} \Phi_{\rm iso}(x_2)\right) \,.
    \label{eq:S_Iso_Implicit}
\end{align}
\Cref{eq:S_Iso_Implicit} is a good approximation for $x\lesssim 1$, where $S_1$ grows as $S_1\propto x^{3-b}$, see \cref{eq:S_Iso_Smallx}. Numerically we find that $S$ also grows at large $x$, whereas $\Phi$ decays according to \cref{eq:Phi_Iso_wDE_LargeX}. This, however, means that at $x\gg 1$, the terms in \cref{eq:SEqX} that are subleading in $y$ but involve $S_1$ will become relevant, and the perturbative ansatz breaks down. The evolution of $S$ is plotted in \cref{fig:Iso_plot}.

\subsubsection{Adiabatic fluctuations}
In the case of adiabatic initial conditions, the leading order terms of \cref{eq:PhiEqX,eq:SEqX} correspond to the homogeneous equation for $\Phi$ without any isocurvature contribution. The solution is the well-known result for an adiabatic fluctuation in a general cosmological background \cite{Baumann:2007zm} and is given by
\begin{equation}
   \Phi_{\rm ad}(x; a\ll a_{\rm eq}) = 2^{b+\frac{3}{2}} \Gamma \left(b+\frac{5}{2}\right) \left(c_w x \right)^{-b-\frac{3}{2}} J_{b+\frac{3}{2}}\left(c_w x \right) \Phi_0 \, .
   \label{eq:Phi_Adi_wDE}
\end{equation}
On superhorizon scales, i.e.~for small $x\ll1$, \labelcref{eq:Phi_Adi_wDE} goes to a constant, given by the initial value $\Phi_0$.
Expanding the Bessel function for large $x\gg1$ on the other hand, we find
\begin{equation}
    \Phi_{\rm ad}(a\ll a_{\rm eq}; x\gg 1) \approx \frac{2^{b+2}}{\sqrt{\pi }} \Gamma \left(b+\frac{5}{2}\right) (c_w x)^{-b-2} \cos \left(c_w x +\frac{\pi}{2}(1-b)\right) \Phi_0 \,,
    \label{eq:Phi_Adi_wDE_LargeX}
\end{equation}
which shows that after a mode has entered the horizon during $w$-domination it oscillates with frequency $c_w$ and decays away with a power-law whose spectral index depends on $w$ through $b$.
%
\begin{figure}
  \centering
  \includegraphics[width=\textwidth]{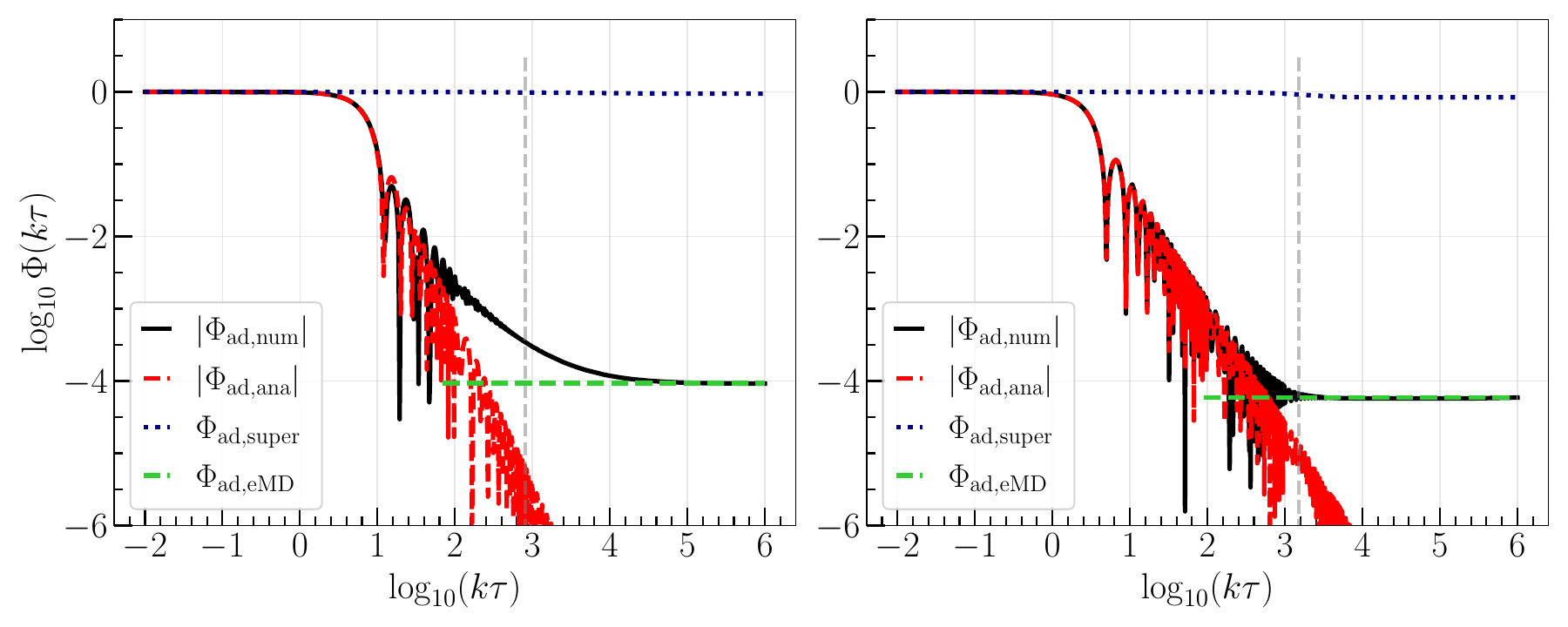}
  \caption{Evolution of the curvature perturbation $\Phi$ with adiabatic initial conditions from $w$- to PBH domination. We show a numerical solution of \cref{eq:PhiEq,eq:SEq}, the superhorizon solution \labelcref{eq:Phi_Ad_Super} and the analytical result \labelcref{eq:Phi_Adi_wDE} for $w=1/6$ (left) and $w=2/3$ (right) with $\kappa = 10^3$ and $\Phi_0=1$. The vertical dashed line marks $\tau_{\rm eq}$. Additionally, we plot the fit for the plateau value \labelcref{eq:PhiAdieMDInterpol}.}
  \label{fig:Phi_ad_Plot}
\end{figure}
%

In \cref{fig:Phi_ad_Plot}, we plot the result for $\Phi$ with adiabatic initial conditions. As one can see, the numerical and approximate analytical results agree excellently at early times, corresponding to $a\ll a_{\rm eq}$. The two solutions begin to differ around $\tau_{\rm eq}$, with deviations setting in earlier for smaller $w$ due to the more gradual transition. See how the gravitational potential starts decaying and oscillating strongly as soon as the mode enters the horizon.
The superhorizon solution \labelcref{eq:Phi_Ad_Super} agrees with the former two up to horizon crossing at $x=1$ and decreases only very slightly during the transition. One can also see how the plateau value at $a\gg a_{\rm eq}$ is well captured by the fit \labelcref{eq:PhiAdieMDInterpol} discussed below.

Feeding the homogeneous solution \labelcref{eq:Phi_Adi_wDE} for $\Phi$ back into equation \labelcref{eq:SEqX}, we obtain the first order solution for the isocurvature perturbation
\begin{align}
    S_{\rm ad}(x; a\ll a_{\rm eq}) =\frac{3 (c_w x)^4}{4 (1-b) (b+2) (b+4)}  \, _2F_3\left(\frac{b+4}{2},2;b+\frac{5}{2},\frac{b+6}{2},3;-\frac{(c_w x)^2}{4}\right) \Phi_0 \,,
   \label{eq:S_Adi_wDE}
\end{align}
where $_2F_3$ is the generalised hypergeometric function. For an illustration of the evolution of $S$, see \cref{fig:Iso_plot}.

\subsection{Fluctuations during PBH-domination \texorpdfstring{($a_{\rm eq}\ll a< a_{\rm eva}$)}{(aEq<<a<aEva)}} \label{sec:Fluctuations_eMD}
After equality and during PBH domination, the curvature perturbation becomes constant on all scales. This can be seen by expanding \cref{eq:PhiEq} in $a\gg a_{\rm eq}$ (or equivalently, expanding \cref{eq:PhiEq_y} for $y\gg1$) and solving the resulting equation, which yields
\begin{equation}
    \Phi(\chi; a\gg a_{\rm eq}) = C_1 + C_2 \, \chi ^{-5/2}
    \label{eq:Phi_eMD}
\end{equation}
in terms of the scale factor $\chi$. \Cref{eq:Phi_eMD} reproduces the well-known result for metric perturbations in a matter dominated Universe \cite{Baumann:2022mni}. The decaying branch $\propto C_2$ becomes unimportant quickly, while the constant $C_1$ determines the value of the plateau deep in PBH domination. What remains to be done is to determine the value of the constants, in particular that of $C_1$.

\subsubsection{Isocurvature fluctuations}
The most straightforward approach is to fix the coefficients $C_{1,2}$ by matching \labelcref{eq:Phi_eMD} to the leading order term of the analytical solution for $\Phi(a\ll a_{\rm eq}; x\gg 1)$ in \cref{eq:Phi_Iso_wDE_LargeX_Chi} at equality, i.e.~at $\chi=1$.
Matching only $C_1$ to the value at $\chi=1$ we recover the well-known result by Kodama and Sasaki \cite{Kodama:1986ud}, ${\Phi_{\rm iso}(a\gg a_{\rm eq})\approx 3 S_0 /(4 \kappa^{2})}$, which notably is independent of $w$.

However, numerically we find that the amplitude of the plateau does depend on $w$ and that the prefactor $3/4$ underestimates the numerical result. In order to improve upon the previous estimate, we match \cref{eq:Phi_eMD} and its first derivative to the first term of \cref{eq:Phi_Iso_wDE_LargeX_Chi}, keeping the matching point $\chi_{\rm m}$ as a free parameter. As \cref{eq:Phi_Iso_wDE_LargeX_Chi} was derived in an expansion in $y\ll1$, it is natural to introduce $y_{\rm m}=\chi_{\rm m}^{3w}$. In terms of $y_{\rm m}$ we find 
\begin{align}
    C_1&=\frac{9}{20}\frac{S_0}{\kappa ^2}y_m^{-\frac{1}{3w}} \quad \text{and}\quad C_2 =\frac{3}{10} \frac{S_0}{\kappa ^2}y_m^{\frac{1}{2w}} \,.
    \label{eq:PhieMDIsoMatch}
\end{align}
In order to determine the optimal value for $y_{\rm m}$, we numerically evolve the coupled equations \labelcref{eq:PhiEq,eq:SEq} for $\Phi$ and $S$ until deep in PBH domination and evaluate $\Phi(a\gg a_{\rm eq})$ there. By considering a range of values for $w$ and $\kappa$ we first confirm that $\kappa^{-2}$ is the correct $\kappa$-dependence independently of $w$. Next, by performing a fit to the numerical data, we find that $y_{\rm m}$ is well described by a linear function in the parameter $b$, defined in \labelcref{eq:bDefinition}, with a single free coefficient $\alpha_{\rm fit}$ as
\begin{equation}
    y_{\rm m} \simeq \alpha_{\rm fit} \ (3+b) \quad \text{with}\quad \alpha_{\rm fit} \approx 0.135 \,.
\end{equation}
This fit describes the numerical data to a precision of $\mathcal{O}(1\%)$ in the range $0.1 \leq w  \leq 1$, and the data was computed for $\kappa\in\left[10^3,10^4\right]$, which is large enough to capture the $\kappa\gg1$ behaviour at an acceptable numerical cost. We expect the fit to be a good description also at larger values of $\kappa$.
For smaller $w$, the numerics become increasingly costly because one has to evolve $\Phi$ very deep into PBH domination and needs large values of $\kappa$ to reach the plateau. Therefore, the fit should not be extrapolated to $w\rightarrow 0$.
With the value of the plateau at hand, we can take a simple interpolation between the super- and subhorizon solutions, \labelcref{eq:PhiIsoSuperMD} and \labelcref{eq:Phi_eMD}, to finally obtain
\begin{align}
    \Phi_{\rm iso, eMD}(k; a \gg a_{\rm eq}) & \simeq S_0 \left( 5 + \frac{1}{C(w)} \left(\frac{k}{k_{\rm eq}}\right)^2\right)^{-1} \label{eq:PhiIsoeMDInterpol} \\
    & \approx S_0 \times
    \begin{cases}
        1/5 & \quad k\ll k_{\rm eq}\\
       C(w) \, \kappa^{-2} & \quad k\gg k_{\rm eq}
    \end{cases} \,,
    \label{eq:PhiIsoeMDCases}
\end{align}
where we defined
\begin{align} \label{eq:Cw}
    C(w) &=\frac{9}{20} \alpha_{\rm fit}^{-\frac{1}{3w}}\left(3+\frac{1-3w}{1+3w}\right)^{-\frac{1}{3w}} \,.
\end{align}
The numerical value of $C(w)$ ranges from $C(1/6)\approx 2.23$ over $C(1/3)\approx 1.11$ to $C(2/3)\approx 0.75$.
The solution \labelcref{eq:PhiIsoeMDInterpol} is shown in \cref{fig:Phi_iso_Plot} above together with a numerical solution, demonstrating the excellent agreement.
As can be seen from \labelcref{eq:PhiIsoeMDCases}, while on superhorizon scales the curvature perturbation asymptotes to a constant independent of $k$, for small scales with $k\gg k_{\rm eq}$ the amplitude of $\Phi$ is suppressed by $\kappa^{-2}$, showing that modes which enter earlier during $w$-domination experience a stronger suppression. Note that for $k\gg k_{\rm eq}$ we find that $\Phi_{\rm iso}(a\gg a_{\rm eq}; k\gg k_{\rm eq})\propto \kappa^{-2}$, just like in the standard radiation case. This shows that the scale dependence in the transfer function is insensitive to the equation of state of the cosmological background preceding PBH domination. The reason for this result is that the $k^{-2}$ dependence is solely due to the Poisson equation, as PBH density fluctuations are both the source of initial isocurvature and the later curvature fluctuation.

Inspecting \labelcref{eq:Cw}, we observe that the suppression of the gravitational potential is smaller, i.e.~$C(w)$ is larger, the smaller $w$. This is due to the fact that for small $w$, the transition to PBH domination is more gradual than in the stiff case, meaning that the decay of $\Phi$ due to the non-zero pressure of the primordial fluid is weakened already before $a_{\rm eq}$.

\subsubsection{Adiabatic fluctuations}
To obtain an expression for the transfer function of the gravitational potential $\Phi$ with adiabatic initial conditions, we adopt a similar strategy as in the standard radiation to matter transition and follow the derivation in Ref.~\cite{Dodelson:2003ft}. The idea is to track the evolution of the matter density perturbations from horizon crossing through equality via a generalized Mészáros equation, and then relate them to the potential in the plateau regime, rather than solving for $\Phi$ directly. In this procedure, one takes advantage of the fact that PBH density fluctuations have vanishing propagation speed, which simplifies the calculations significantly by removing the $k$-dependence of the equation for the Fourier modes of $\delta_{\rm PBH}$. We provide the detailed calculation in \cref{sec:TransferFct_Adi}. The main result of \cref{sec:TransferFct_Adi}, though, is that for adiabatic initial conditions, the PBH density contrast evolves as $a^{-(1-3w)/2}$ both in super- and subhorizon regimes before PBH domination. For $w>1/3$, this turns into a super- and subhorizon growth of the PBH density contrast as $(k\tau)^{-b}$ (recall that $b<0$ for $w>1/3$). Thus, the transfer function for $k\gg k_{\rm eq}$ is enhanced by a factor $\kappa^{-b}$ for $w>1/3$ with respect to the isocurvature case. For $w<1/3$, we instead expect to recover the $\kappa^{-2}$ scaling stemming from the Poisson equation.

Let us here provide the final expression for the transfer function, which is given by
\begin{equation} \label{eq:Phi_Adi_eMD}
     \Phi_{\rm ad,eMD}(a\gg a_{\rm eq};k\gg k_{\rm eq}) = \frac{3}{2(1+w)}\Phi_0 c_3 \kappa^{-2} + \alpha_1 \kappa^{-2-b}\left( \frac{\alpha_2 - \kappa^b}{b}\right) \Phi_0 \,,
\end{equation}
where the constants $c_3, \, \alpha_1$ and $\alpha_2$ are given in \cref{eq:c34_Definition,eq:alphas}, respectively, and the parameter $b$ is defined in \cref{eq:bDefinition}. We refer the reader to \cref{sec:TransferFct_Adi} for the details.
We find that \cref{eq:Phi_Adi_eMD} is in excellent agreement with numerical results for $w \gtrsim 1/5$ for $\kappa \gtrsim 10^3$, while for $w\lesssim 1/5$ the large $k$-scaling is only reached for $\kappa \gtrsim 10^4$. In the very soft regime, the approximation breaks down, at the latest at $w = 1/9$ where the coefficient $\alpha_2$ diverges.
The first term $\propto \kappa^{-2}$ is always subdominant and can be neglected, except for very stiff EoS parameters $w\gtrsim 0.95$. For convenience we show the dominant terms of \cref{eq:Phi_Adi_eMD} in the stiff and soft limits, which read
\begin{equation}
    \label{eq:PhiAdLargeAFitCases}
    \Phi_{\rm ad,eMD}(a\gg a_{\rm eq};k\gg k_{\rm eq})\approx \Phi_0 \times
    \begin{cases}
        -\alpha_1 \kappa^{-2} & \quad w\ll1/3\\
        \frac{81}{8} \ln(\kappa)/\kappa^2 & \quad w=1/3\\
        \frac{\alpha_1 \alpha_2}{b} \kappa^{-2-b} & \quad w\gg1/3
    \end{cases} \, .
\end{equation}
In the radiation limit $b\rightarrow 0$, \cref{eq:Phi_Adi_eMD} reproduces the well-known $\ln(\kappa)/\kappa^2$ behaviour \cite{Dodelson:2003ft}, which results from the logarithmic growth of matter density fluctuations during radiation domination, while for stiff and soft $w$ the scale-dependence is given by a power-law. Note that this scaling is approached asymptotically for very large $\kappa$, especially in the soft case.

We match the small-scale transfer function \labelcref{eq:Phi_Adi_eMD} with the superhorizon solution \labelcref{eq:PhiAdSuperMD} in a simple interpolation, and provide an approximate transfer function for the primordially adiabatic mode as
\begin{align}
    \Phi_{\rm ad,eMD}(k; a \gg a_{\rm eq})  & \simeq \Phi_0 \left(\frac{5 b+10}{6 b+9} + \frac{1}{\alpha_1} \kappa^{2+b}\left(\frac{b}{\alpha_2 - \kappa^b}\right)\right)^{-1} \label{eq:PhiAdieMDInterpol} \\
    &\approx \Phi_0 \times
    \begin{cases}
        \frac{3}{5}\frac{3+2b}{2+b} & \quad k\ll k_{\rm eq}\\
       \frac{\alpha_1}{b} \kappa^{-2-b}\left(\alpha_2 - \kappa^b\right) & \quad k\gg k_{\rm eq}
    \end{cases} \,.
    \label{eq:PhieMDAdi}
\end{align}
Notice how in contrast to the isocurvature case, cf.~\cref{eq:PhiIsoeMDInterpol}, the $k$-dependent suppression has a dependence on the EoS parameter for $w>1/3$, implying that the adiabatic mode “remembers” the earlier $w$-dominated phase differently at different scales. Contrary to the isocurvature case, the primordial curvature fluctuation is sourced by fluctuations in the $w$-fluid. This is the source of the $w$ dependence in the exponent of $\kappa$.

However, in the computation of the induced GW spectrum, we will mainly be interested in scales ${\kappa <10^4}$. This is due to the cutoff at $k_{\rm uv}$, see \cref{eq:Relationsk_kUV}, and the possible overproduction of induced GWs. In this regime, i.e.~for ${\kappa <10^4}$, we find that \cref{eq:Phi_Adi_eMD} is not accurate for soft EoS parameters $w$. For this reason, we instead use a numerical fit in this regime, inspired by \cref{eq:Phi_Adi_eMD}, which is given by
\begin{equation}
    \Phi_{\rm ad,eMD,fit}(k_{\rm eq}<k\lesssim 10^4 k_{\rm eq})\simeq A(w) \kappa^{-2-b}\left(\frac{\kappa^{3 b/2}-1}{3 b/2}\right) \Phi_0
    \label{eq:PhiAdLargeAFit}\,,
\end{equation}
and provides a better fit to the numerical data. The additional factor $3/2$ in the exponent is found to reproduce the scaling of $\Phi$ in the soft regime more accurately than \cref{eq:Phi_Adi_eMD}.
The prefactor $A(w)$ is well described by a second order polynomial in the parameter $b$, given explicitly in \cref{eq:A(w)Fit}.
This fit captures the behaviour of $\Phi_{\rm ad}(a\gg a_{\rm eq})$ to $\mathcal{O}(1\%)$ precision in the range ${0.1 \leq w  \leq 1}$ and $\kappa\in\left[10^3,10^4\right]$, with largest deviations below 10\%. It further agrees excellently with the analytical result \labelcref{eq:Phi_Adi_eMD} for $w\geq 1/3$.

Let us note that the transfer function \eqref{eq:PhieMDAdi} is valid in any primordial universe scenario that consists of a period of $w$-domination after inflation followed by an early matter dominated phase, regardless of the composition of the dust fluid. For example, our formulas are valid for oscillations of heavy moduli field around the minimum of the potential \cite{Allahverdi:2020bys}, oscillons \cite{Amin:2011hj, Lozanov:2022yoy} and Q-balls \cite{Kawasaki:2023rfx}, which eventually dominate the early universe.

\subsection{Fluctuations after PBH evaporation \texorpdfstring{($a_{\rm eva}\ll a$)}{(aEva<<a)}} \label{sec:Fluctuations_RD}
As the PBHs evaporate due to Hawking radiation, the Universe rapidly transitions from matter to radiation domination. Assuming a monochromatic PBH mass function, all PBHs evaporate at the same time $\tau_{\rm eva}$,\footnote{To be precise, this is true only if we adopt the synchronous comoving gauge. In general, the evaporation is non-local if the PBH 3-velocity is non-zero \cite{Domenech:2020ssp}.} and as can be seen in \cref{fig:RhoPlot} the final evaporation happens within one e-fold. Thus, we begin by assuming an instantaneous transition to radiation domination.
The behaviour of $\Phi$ during the radiation period is given by a solution of the form \labelcref{eq:Phi_Adi_wDE} for $b\rightarrow 0$. By matching this solution to the plateau of the gravitational potential $\Phi_{\rm eMD}(k)$, given in \labelcref{eq:PhiIsoeMDInterpol} for the isocurvature case and in \labelcref{eq:PhieMDAdi} for the adiabatic one, we find \cite{Domenech:2020ssp}
\begin{equation}
    \Phi_{\rm RD}^{\rm instant}(k\tau) =  \frac{\Phi_{\rm eMD}}{c_s k \bar{\tau}}\left( C_1 j_1(c_s k \bar{\tau}) + C_2 y_1(c_s k \bar{\tau}) \right) \,,
    \label{eq:PhiRD}
\end{equation}
where we introduced the shifted conformal time $\bar{\tau}=\tau-\tau_{\rm eva}/2$ to maintain continuity of the background metric, and the constants $C_{1,2}$ are determined from requiring continuity of $\Phi$ and its first derivative to yield
\begin{align}
    C_1&= -\frac{1}{8} (c_s k \tau_{\rm eva})^3 y_2(c_s k \tau_{\rm eva}/2) \, ,\\
    C_2&= \frac{1}{8} (c_s k \tau_{\rm eva})^3 j_2(c_s k \tau_{\rm eva}/2) \, .
\end{align}
Here and above, $j_{n}$ and $y_{n}$ are the spherical Bessel functions of the first and second kind, respectively. More details on the matching can be found in Ref.~\cite{Domenech:2020ssp}.
Note that here and in the following sections, the sound speed takes the value $c_s=1/\sqrt{3}$ as usual during radiation domination.

In reality, the evaporation is not instantaneous, and our matching \eqref{eq:PhiRD} neglects the effect of the finite duration. This effect is small for long-wavelength modes, which vary on timescales larger than the evaporation rate $\Gamma$ defined in \cref{eq:EvaporationRate}, but is significant for the very short-wavelength modes with $k\gg \Gamma$ we are considering here. As a result, these modes decay already during the transition and get an additional suppression factor \cite{Inomata:2020lmk}
\begin{equation}
    {\cal S}_{\Phi,\rm eva}(k)=\frac{\Phi_{\rm RD}}{\Phi_{\rm RD}^{\rm instant}}\approx\left(\sqrt{\frac{2}{3}}\frac{k}{k_{\rm eva}}\right)^{-1/3} \,.
    \label{eq:SuppresionSeva}
\end{equation}
This analytic estimate has been shown to be in good agreement with numerical results \cite{Inomata:2020lmk}, especially for $k>k_{\rm eq}$.
More details on the derivation of \cref{eq:SuppresionSeva} can be found in \cite{Inomata:2019zqy, Inomata:2020lmk} and \cref{sec:Suppresion_Seva}.
The differences between a gradual and a sudden transition to the radiation era with respect to the induced GW signal have been discussed in \cite{Inomata:2019ivs, Inomata:2019zqy, Pearce:2023kxp}.

\section{Induced gravitational wave spectrum} \label{sec:iGW_Spectrum}
With the evolution of the curvature perturbation from PBH formation through PBH domination until inside the radiation era at hand, we can compute the gravitational waves induced in the process. GWs (subhorizon tensor modes) are sourced by scalar curvature perturbations at second order in perturbation theory. For more details on these scalar induced GWs we refer the reader to \cite{Ananda:2006af, Baumann:2007zm, Kohri:2018awv, Domenech:2019quo}  and the recent review \cite{Domenech:2021ztg}. Here, we will focus on the dominant contribution to the spectrum resulting from the fast transition to radiation domination after PBH evaporation \cite{Inomata:2019ivs}. The gravitational waves produced in the PBH reheating scenario with $w=1/3$ have been studied in \cite{Inomata:2020lmk, Papanikolaou:2020qtd, Domenech:2020ssp, Domenech:2021wkk}, and here again we follow closely the computation of \cite{Domenech:2020ssp}, recovering their results for $w=1/3$.

\subsection{GW spectral energy density}
The quantity of interest, which can later be compared with observations, is the GW spectral energy density $\Omega_{\rm GW}(k)$. It is defined in terms of the tensor power spectrum $\mathcal{P}_h$ as
\begin{equation} \label{eq:OmegaGWDefinition}
    \Omega_{\rm GW}(k) = \frac{k^2}{12 \mathcal{H}^2}\overline{\mathcal{P}_h(k,\tau)}\,,
\end{equation}
which for convenience is to be evaluated at a time $\tau$ during radiation domination when the GWs are deep inside the horizon and propagate as free waves, such that $h'\approx i k h$. After this time, the GW energy density redshifts like that of radiation, $\rho_{\rm GW}\propto a^{-4}$, and we can relate the GW spectral density in the radiation era $\Omega_{\rm GW, RD}(k)$ to the spectral energy density today as \cite{Ando:2018qdb}
\begin{equation}
    \Omega_{\rm GW,0}(k) h^2 \approx 0.387 \left(\frac{g_*(T_{\rm RD})}{106.75}\right)^{-1/3} \Omega_{r,0}h^2 \Omega_{\rm GW, RD}(k)\,,
\end{equation}
where we used entropy conservation and assumed $T_{\rm RD}\gtrsim0.5\rm{MeV}$ so that $g_{*,s}(T_{\rm RD})=g_{*}(T_{\rm RD})$. $\Omega_{r,0}h^2 \approx 4.18\times 10^{-5}$ is the energy density fraction of radiation today \cite{Planck:2018vyg}.

The overline in \cref{eq:OmegaGWDefinition} denotes an oscillation average, and the tensor power spectrum $\mathcal{P}_{h}$ is given by \cite{Kohri:2018awv}
\begin{align}
    \overline{\mathcal{P}_{h}}(k,\tau)=8 \int_{0}^{\infty} dv \int_{|1-v|}^{1+v} du \left(\frac{\left(1+v^2-u^2\right)^2-4 v^2}{4 u v}\right)^2
    {\cal P}_{\Phi}(u k)
    {\cal P}_{\Phi}(v k)
    \overline{I^2}(x,u,v)    \,,
    \label{eq:TensorPowerSpectrum}
\end{align}
where we already summed over the $(\cross)$ and $(+)$ polarisations.
The Kernel $I(x,u,v)$ carries the full-time dependence of $\overline{\mathcal{P}_{h}}(k,\tau)$ and can be split into contributions from the different epochs, namely the $w$-dominated, PBH-dominated and radiation-dominated eras.
As was shown in \cite{Inomata:2019ivs}, due to the (almost) sudden transition from matter to radiation domination caused by the fast final stages of the PBH evaporation, the contribution from right after PBH evaporation is the largest, and we will focus on this contribution here.
Thus, the curvature power spectrum ${\cal P}_{\Phi}$ should be understood as the power spectrum at the end of the eMD phase. It is related to \labelcref{eq:InitialIsocurvaturePS} in the case of isocurvature initial conditions and to \labelcref{eq:ScalarPS} for the adiabatic case via the transfer functions computed above.
The dominant contribution is therefore given by
\begin{align}
    \overline{\mathcal{P}_{h,\rm RD}}(k,\tau;\bar{x}\gg1) \approx & \frac{c_s^4}{2048} \frac{x_{\rm eva}^8}{\bar{x}^2}  \int_{0}^{\infty} dv \int_{|1-v|}^{1+v} du \left(4 v^2-\left(1+v^2-u^2\right)^2\right)^2 \overline{\mathcal{I}^2_{\rm osc}}(\bar{x},u,v) \nonumber \\
    & {\cal P}_{X}(u k) {\cal P}_{X}(v k) {\cal S}^2_{\Phi,\rm eva}(u k) {\cal S}^2_{\Phi,\rm eva}(v k)
    \left(T_{\Phi}^{\rm eMD}(u k)\right)^2 \left(T_{\Phi}^{\rm eMD}(v k)\right)^2 \,,
    \label{eq:TensorPowerSpectrumFull}
\end{align}
where $X=\{S_0,\Phi_0\}$ respectively for  isocurvature and adiabatic initial conditions, $\bar{x}=k\bar{\tau}$ is the shifted time coordinate in the radiation era and $x_{\rm eva}=k\tau_{\rm eva}=2 k / k_{\rm eva}$.
The derivation of \cref{eq:TensorPowerSpectrumFull} is discussed in more detail in \cref{sec:GW_kernel} and the oscillatory function $\mathcal{I}_{\rm osc}$ is defined explicitly in \cref{eq:OscIntegralRD}.
In \cref{eq:TensorPowerSpectrumFull}, we also introduced the notation  $T_{\Phi}^{\rm eMD}$ for the transfer function of the gravitational potential in the eMD era by dividing out the initial values $S_0$ and $\Phi_0$ from $\Phi_{\rm eMD}$ in the isocurvature and adiabatic cases \labelcref{eq:PhiIsoeMDInterpol,eq:PhiAdieMDInterpol}, respectively.
We now solve the remaining momentum integrals in $u$ and $v$ for the two cases separately in the following sections.

\subsection{Isocurvature induced GWs} \label{sec:iGW_Iso}
Considering the isocurvature mode, recall that the source to GWs is generated by the conversion of the isocurvature into a curvature perturbation, as discussed in \cref{sec:Curvature_perturbations}. As a consequence, the initial spectrum ${\cal P}_{\Phi}$ is given by ${\cal P}_{\delta_{\rm PBH,f}}$, defined in \labelcref{eq:InitialIsocurvaturePS}, and the transfer function $T_{\Phi}^{\rm eMD}$ is determined by $\Phi_{\rm iso, eMD}$ in \labelcref{eq:PhiIsoeMDInterpol}.
One can give an analytical estimate of the peak of the induced GW power spectrum by focusing on the contribution to the momentum integrals in \labelcref{eq:TensorPowerSpectrumFull} near the scales where ${u+v=c_s^{-1}}$. At this scale, where the sum of the frequencies of the two scalar modes equals the frequency of the tensor mode, the system has a resonance, and the GW production is enhanced at these frequencies \cite{Ananda:2006af, Domenech:2021ztg}. More details on the approximate evaluation of \cref{eq:TensorPowerSpectrumFull} near this scale may be found in \cref{sec:GW_integrals}, see in particular \labelcref{eq:OscIntegralRD2_res} for the relevant contribution to the Kernel $\overline{\mathcal{I}^2_{\rm osc}}$.
Using this approximation and assuming $k\gg k_{\rm eq}$ results in the following, compact expression for the GW spectral density near the resonant peak
\begin{equation}
    \label{eq:OmegaGWres}
    \Omega_{\rm GW, res}(k)\approx \Omega_{\rm GW, res}^{\rm peak}
    \left(\frac{k}{k_{\rm uv}}\right)^{11/3}
    \Theta_{\rm uv}(k)
\end{equation}
with the amplitude near the peak given by
\begin{align}
    \label{eq:OmegaGWresPeak}
   \Omega_{\rm GW, res}^{\rm peak} &= C^4(w)\frac{c_s^{7/3}(c_s^2-1)^2}{576 \times 6^{1/3}\pi}
    \left(\frac{k_{\rm eq}}{k_{\rm uv}}\right)^8
    \left(\frac{k_{\rm uv}}{k_{\rm eva}}\right)^{17/3} \\
    &\approx 9.58 \times 10^{30} C^4(w) \beta^{\frac{4(1+w)} {3w}}
    \left(\frac{g_H}{108}\right)^{-17/9}
    \left(\frac{\gamma}{0.2}\right)^{8/3}
    \left(\frac{M_{\rm f}}{10^4\text{g}}\right)^{34/9} \,. \label{eq:OmegaGWresPeak_Num}
\end{align}
The function $\Theta_{\rm uv}(k)$ is defined explicitly in \cref{eq:ThetaUV} and acts as a smooth cutoff of the GW spectrum for large frequencies. For practical purposes, it roughly behaves like a Heaviside step-function, namely
\begin{align} \label{eq:ThetaUV_Heaviside}
    \Theta_{\rm uv}(k)\approx 1.167 \times \Theta(1-k/k_{\rm uv}) \,.
\end{align}
However, \cref{eq:ThetaUV_Heaviside} results in a sharp cutoff at $k_{\rm uv}$, whereas \labelcref{eq:ThetaUV} smoothly goes to zero between ${\frac{2 c_s}{1+c_s}\leq k/k_{\rm uv} \leq 2 c_s}$ before vanishing at and above $k = 2 c_s k_{\rm uv}$.

Consequently, the spectrum grows as $k^{11/3}$ until it peaks near $k_{\rm uv}$, and then sharply drops above $k_{\rm uv}$ before going to zero at $2c_s k_{\rm uv}$.
The comoving scale $k_{\rm uv}$ corresponds to a frequency $f_{\rm uv}=k_{\rm uv}/(2\pi)$ today, which evaluates to \cite{Domenech:2020ssp}
\begin{equation} \label{eq:fUV}
    f_{\rm uv}\approx 1.7 \,\text{kHz} \left(\frac{g_{*,s}(T_{\rm eva})}{106.75}\right)^{-1/3}\left(\frac{g_*(T_{\rm eva})}{106.75}\right)^{1/4} \left(\frac{g_H}{108}\right)^{1/6} \left(\frac{M_{\rm f}}{10^4 \rm{g}}\right)^{-5/6} \,,
\end{equation}
where we again used entropy conservation. Note that the peak frequency is determined solely by the PBH mass $M_{\rm f}$.
Considering the allowed mass range $M_{\rm f}=\mathcal{O}(1-10^8)\rm{g}$ given by \labelcref{eq:Mf_bounds}, we note that the peak frequency falls within the range $f_{\rm uv}\in (0.3, 7 \times 10^6)$Hz and therefore may enter the observational window of several current and future GW detectors \cite{Domenech:2020ssp}, as shown in \cref{fig:OmegaGW_Forecast}.

As \cref{eq:OmegaGWresPeak_Num} shows, the equation of state of the primordial fluid enters the GW amplitude through the prefactors, $C^4(w)$ and $\beta^{\frac{4(1+w)}{3w}}$. On the one hand, the prefactor $C^4(w)$ due to the transfer function yields an enhancement of the amplitude of the GW power spectrum for small values of $w$, because in this case, the transition to PBH domination is more gradual and the gravitational potential $\Phi$ decays less, see \labelcref{eq:Cw}.
On the other hand, the factor $\beta^{\frac{4(1+w)}{3w}}$ has the opposite effect of relatively enhancing the amplitude for stiffer $w$ and suppressing it for softer $w$.
It stems from the ratio $\left(\frac{k_{\rm eq}}{k_{\rm uv}}\right)^8$ in \labelcref{eq:OmegaGWresPeak} due to the relation \labelcref{eq:Relationsk_kUV}, and leads to a strong $w$-dependence of the amplitude as generally $\beta\ll1$. This can be understood as a result of the longer (shorter) PBH dominated phase for a stiffer (softer) equation of state, and it is, in fact, the dominant effect, so that overall a stiffer equation of state ($w>1/3$) leads to an enhanced GW amplitude, while a softer one ($w<1/3$) results in suppression.

Note that the scaling of the spectrum near the peak $\Omega_{\rm GW, res}\propto k^{11/3}$ is independent of $w$, because the transfer function $\Phi_{\rm eMD, iso}$ \labelcref{eq:PhiIsoeMDInterpol} has the same scale-dependence (i.e.~$\kappa^{-2}$ for $k\gg k_{\rm eq}$) for all $w$.
Additionally, the peak frequency \labelcref{eq:fUV} depends only on $M_{\rm f}$.
This, unfortunately, implies that an observation of the isocurvature-induced peak \labelcref{eq:OmegaGWres} alone is not sufficient to determine $w$ because the amplitude is degenerate with respect to $\beta$ and $w$.

Next, we consider the IR regime corresponding to low frequencies or large scales.
In this regime, $k\ll k_{\rm uv}$, the momentum integrals in \cref{eq:TensorPowerSpectrumFull} are dominated by the large momentum tails, i.e.~$u\approx v\gg 1$. In this regime, we keep only the relevant terms given in \labelcref{eq:OscIntegralRD2_LV} for the kernel $\overline{\mathcal{I}^2_{\rm osc}}$, and one can again solve the integrals analytically. The dominant contribution to the GW spectrum at low frequencies is given by
\begin{align}
    \label{eq:OmegaGWIR}
    \Omega_{\rm GW, IR}(k) &= C^4(w) \frac{c_s^4 }{120 \pi ^2}\left(\frac{2}{3}\right)^{1/3} 
    \left(\frac{k_{\rm eq}}{k_{\rm uv}}\right)^8
    \left(\frac{k_{\rm uv}}{k_{\rm eva}}\right)^{14/3}
    \left(\frac{k}{k_{\rm uv}}\right) \\
    &\approx 9.03 \times 10^{24} C^4(w) \beta^{\frac{4(1+w)} {3w}}
    \left(\frac{g_H}{108}\right)^{-14/9}
    \left(\frac{\gamma}{0.2}\right)^{8/3}
    \left(\frac{M_{\rm f}}{10^4\text{g}}\right)^{28/9}\left(\frac{k}{k_{\rm uv}}\right) \,.
    \label{eq:OmegaGWIR_Num}
\end{align}
Comparing \labelcref{eq:OmegaGWIR,eq:OmegaGWIR_Num} to \labelcref{eq:OmegaGWresPeak,eq:OmegaGWresPeak_Num} one sees that the amplitude of the IR tail is significantly suppressed compared to the resonant peak, namely by a factor of the order $(k_{\rm eva}/k_{\rm uv})$.
In the low-frequency regime, the EoS parameter $w$ enters only through the prefactors and, thus, does not break the degeneracy (as for the resonant peak).
Also an observation of the transition frequency $k_{\rm tr}$ at the knee of the spectrum, where $\Omega_{\rm GW, IR}(k_{\rm tr})=\Omega_{\rm GW, res}(k_{\rm tr})$, would not provide additional information, as the $w$- and $\beta$-dependent prefactors in \cref{eq:OmegaGWresPeak,eq:OmegaGWIR} are the same, leaving $k_{\rm tr}$ as a function of $M_{\rm f}$ only \cite{Domenech:2020ssp}.
%
\begin{figure}
\centering
\includegraphics[width=0.75\textwidth]{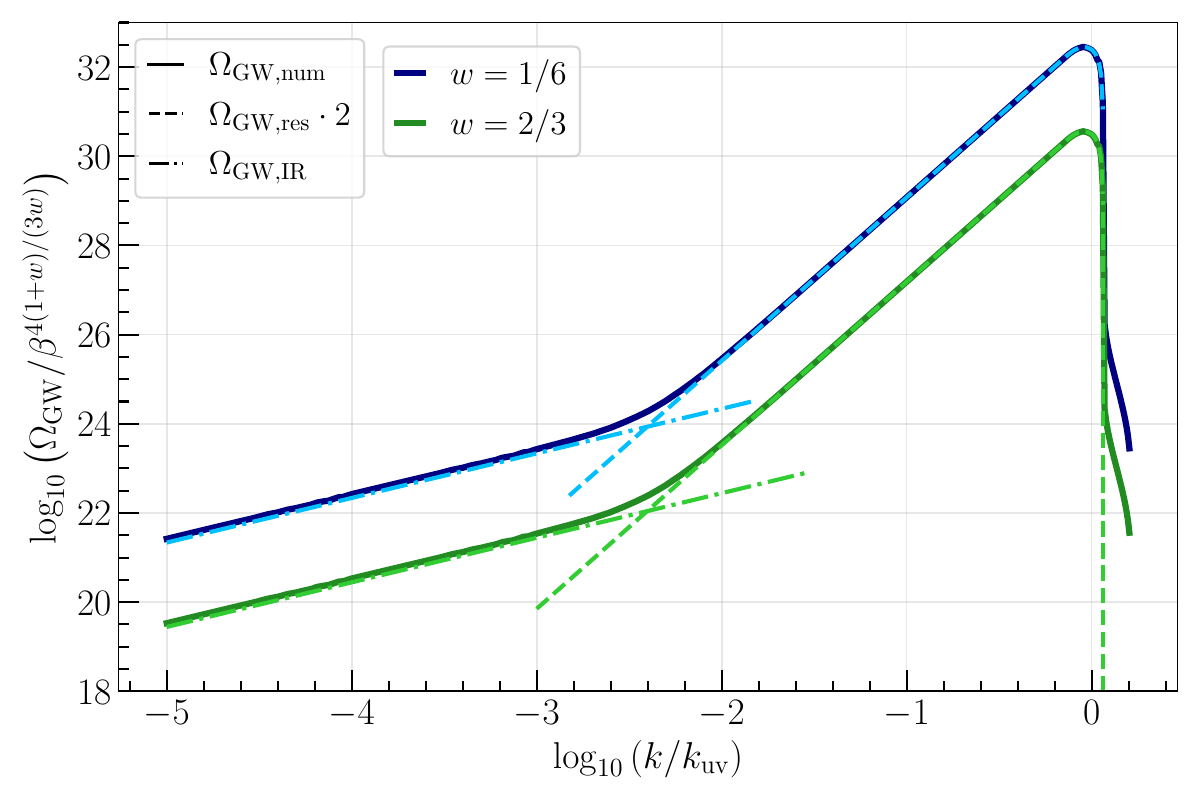}
\caption{The isocurvature induced GW spectrum as a function of $k/k_{\rm uv}$ for $M_{\rm f}=10^4 g$. The spectrum is normalised by $\beta^{\frac{4(1+w)}{3w}}$, leaving the amplitude independent of $\beta$. We compare a numerical solution of the full momentum integrals \labelcref{eq:TensorPowerSpectrumFull} with the analytical approximations for the resonant peak \labelcref{eq:OmegaGWres} and the IR tail \labelcref{eq:OmegaGWIR} for $w=1/6$ (blue curve) and $w=2/3$ (green).
For this plot we assume that $k> k_{\rm eq}$ in the entire range shown. }
\label{fig:Omega_GW}
\end{figure}
%

In figure \labelcref{fig:Omega_GW}, we show the resulting spectral energy density of the isocurvature-induced GWs. We plot the spectrum obtained by numerically solving the full momentum integrals in \cref{eq:TensorPowerSpectrumFull}, together with the analytical approximations for the resonant peak \labelcref{eq:OmegaGWres} and the IR-tail \labelcref{eq:OmegaGWIR}. As the plot illustrates, we find excellent agreement between the spectral shapes in the respective regimes. See how the spectrum grows linearly $\sim k$ in the IR, then rises as $k^{11/3}$, reaching its peak value near $k_{\rm uv}$, and quickly drops at ${k/k_{\rm uv}=2 c_s}$. Note that because we normalised the spectra by $\beta^{\frac{4(1+w)}{3w}}$, the amplitude for $w=1/6$ (blue curve) is larger than for $w=2/3$ (green) due to the value of $C(w)$ \labelcref{eq:Cw}. Including the dependence on $\beta$ reverses the effect by multiple orders of magnitude.

In passing, we find that the resonant approximation underestimates the peak amplitude by a factor of two, compared to the full numerical solution. The origin of this minor discrepancy is that the sum of all the finite terms contributing to the Kernel $\overline{\mathcal{I}^2_{\rm osc}}$ resulting from \cref{eq:OscIntegralRD} results in an equal contribution to the GW spectrum as the resonant term with the divergent cosine integral \labelcref{eq:OscIntegralRD2_res} alone. We confirmed this by numerically computing the GW spectrum stemming from the different contributions to the Kernel separately. Thus, the missing factor of 2 can easily be accounted for and has been already included in \cref{fig:Omega_GW}, leaving our analytical approximation in perfect agreement with the numerical result.

\begin{figure}
\centering
\includegraphics[width=\textwidth]{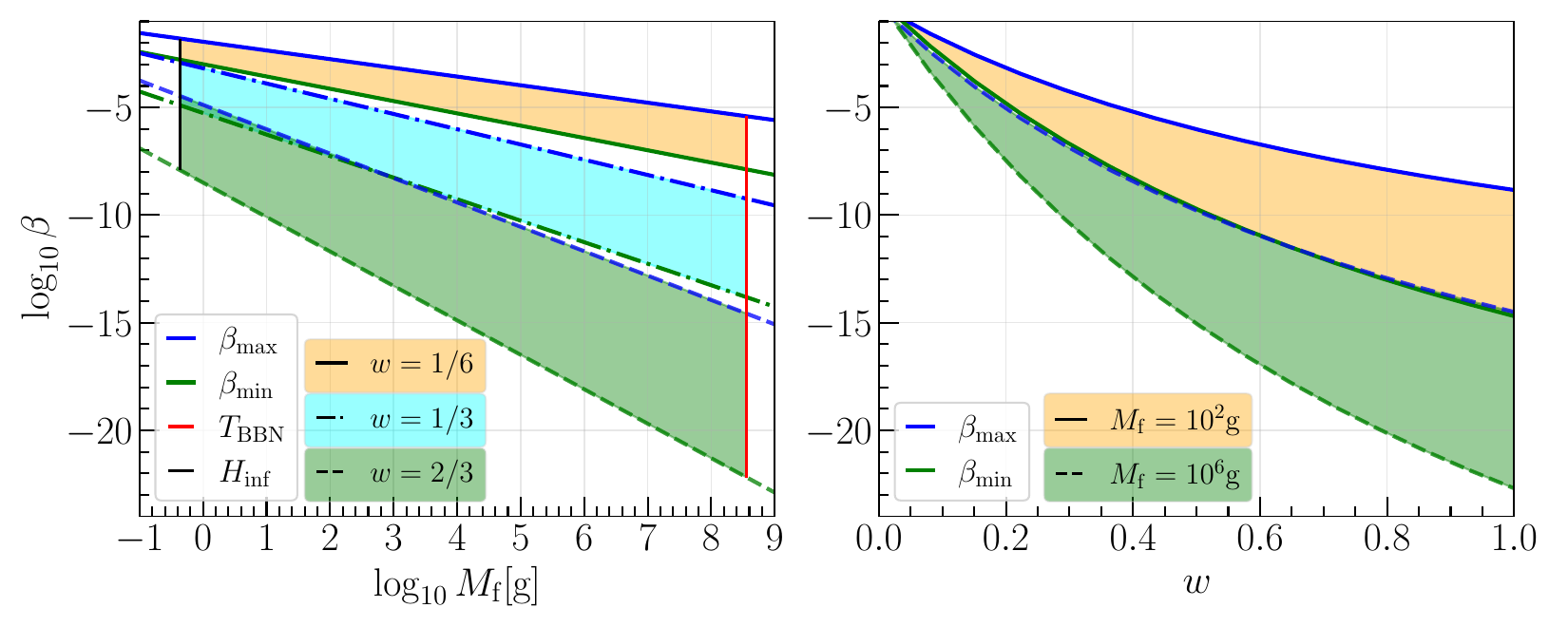}
\caption{The allowed parameter space in the $(M_{\rm f},\beta)$-plane (left) for 3 values of $w$, and in the $(w,\beta)$-plane (right) for 2 values of $M_{\rm f}$. The lower and upper bounds for $\beta$ are set by \labelcref{eq:beta_lower} and \labelcref{eq:beta_upper}, respectively. The bounds on the PBH mass are given by \cref{eq:Mf_bounds}. Note that for the plot we fixed $\gamma=0.2$. The potential dependence of $\gamma$ on $w$ only introduces $\mathcal{O}(1)$ factors.}
\label{fig:parameter_space}
\end{figure}

Once the tensor modes are deep within the horizon, they behave as radiation, and the GW energy density $\Omega_{\rm GW}$ contributes to the total energy content of the Universe like an additional relativistic species. The effective number of extra relativistic degrees of freedom is, however, constrained by CMB observations to $\Delta N_{\rm eff} < 0.30$ at 95\% \cite{Planck:2018vyg}. This in turn, implies that the GW energy density in the radiation dominated epoch at BBN cannot exceed \cite{Caprini:2018mtu}
\begin{equation} \label{eq:Neff_bound}
    \Omega_{\rm GW, BBN} \lesssim \frac{7}{8} \left(\frac{4}{11}\right)^{4/3} \Delta N_{\rm eff}\approx 0.068 \,.
\end{equation}
Note that the bound \cref{eq:Neff_bound} is an integrated constraint on the total GW energy density given by ${\Omega_{\rm GW}=\int d \ln k \, \Omega_{\rm GW}(k)}$. As the GW spectrum is very peaked, we estimate the total GW energy at BBN from our analytical expression for the resonant peak \labelcref{eq:OmegaGWres}, yielding
\begin{equation} \label{eq:OmegaGW_BBN}
    \Omega_{\rm GW, BBN} \approx \int_0^{\infty} d \ln k \, \Omega_{\rm GW, res}(k) \approx 0.31 \, \Omega_{\rm GW, res}^{\rm peak} \,.
\end{equation}
Combining \cref{eq:Neff_bound,eq:OmegaGW_BBN} and using \labelcref{eq:OmegaGWresPeak_Num}, we can derive an upper bound on the initial PBH abundance $\beta$. It is given by
\begin{align} \label{eq:beta_upper}
    \beta < \beta_{\rm max}&\coloneq (2.3 \times 10^{-32})^{\frac{3w}{4(1+w)}}C(w)^{-\frac{3w}{1+w}}
    \left(\frac{g_H}{108}\right)^{\frac{17w}{12(1+w)}}
    \left(\frac{\gamma}{0.2}\right)^{-\frac{2w}{1+w}}
    \left(\frac{M_{\rm f}}{10^4\text{g}}\right)^{-\frac{17w}{6(1+w)}} \\
    & \approx
    \begin{cases}
        2.9 \times 10^{-4} 
        \left(\frac{g_H}{108}\right)^{17/84}
    \left(\frac{\gamma}{0.2}\right)^{-2/7}
    \left(\frac{M_{\rm f}}{10^4\text{g}}\right)^{-17/42} & \quad (w=1/6)\\
        1.1 \times 10^{-6}
        \left(\frac{g_H}{108}\right)^{17/48}
    \left(\frac{\gamma}{0.2}\right)^{-1/2}
    \left(\frac{M_{\rm f}}{10^4\text{g}}\right)^{-17/24} & \quad (w=1/3)\\
        4.5 \times 10^{-10}
        \left(\frac{g_H}{108}\right)^{17/30}
    \left(\frac{\gamma}{0.2}\right)^{-4/5}
    \left(\frac{M_{\rm f}}{10^4\text{g}}\right)^{-17/15} & \quad (w=2/3)
    \end{cases}\, ,   
\end{align}
showing that a stiffer $w$ leads to a significantly stronger upper bound on $\beta$, because the GW amplitude is enhanced by the longer PBH dominated phase, whereas a softer $w$ relaxes the bound.
The resulting allowed parameter space, taking into account the lower and upper bounds on the initial PBH mass $M_{\rm f}$ \labelcref{eq:Mf_bounds} and abundance \labelcref{eq:beta_lower,eq:beta_upper}, respectively, is plotted in \cref{fig:parameter_space}.
See how, for a stiffer equation of state parameter $w$, the allowed range for $\beta$ shifts to lower values and how the open parameter space narrows as $w\rightarrow 0$. Note also how larger initial PBH masses $M_{\rm f}$ require lower values of $\beta$.
%
%

\subsection{Adiabatic induced GWs} \label{sec:iGW_Adi}
In addition to the GWs induced by the isocurvature perturbation, we now discuss the effect of the primordially adiabatic mode. Recall that we have parameterised its initial power spectrum $\mathcal{P}_{\Phi}(k)$ by the power-law ansatz \labelcref{eq:ScalarPS}.
Analogously to the isocurvature case, we compute the GW spectrum by inserting the scalar power spectrum $\mathcal{P}_{\Phi}$ and the transfer function \labelcref{eq:PhiAdieMDInterpol} of the adiabatic mode into the tensor power spectrum \labelcref{eq:TensorPowerSpectrumFull}. In contrast to the isocurvature case, the power-law in the momenta now depends on the scalar spectral index $n_s$, as well as on the EoS parameter $w$ through the transfer function \labelcref{eq:PhiAdieMDInterpol}.
Inspecting the large-$k$ branch of the transfer function \labelcref{eq:Phi_Adi_eMD} one notices that near $w=1/3$ both powers of $\kappa$ are equally relevant and in the special case of radiation, the logarithmic correction to the power-law arises. For the computation of the induced GWs, it is, therefore, convenient to parameterise the transfer function for our scales of interest with an effective power law
\begin{equation}
    \Phi_{\rm ad,eMD,eff}(k\gg k_{\rm eq}) = A_\Phi(w) \kappa^{n(w)} \, ,
\end{equation}
with the power-law exponent given by
\begin{equation} \label{eq:PhiAd_Eff_PL}
    n(w)\approx -
        \begin{cases}
        2-b/2 & \quad w\lesssim 1/5\\
        1.83 + 0.285 b -0.790 b^2 & \quad 1/5<w<2/3\\
        2+b & \quad w\gtrsim2/3
    \end{cases} \,,
\end{equation}
and $n(w;k\ll k_{\rm eq})=0$.
The scaling in the soft and stiff regimes results from \cref{eq:PhiAdLargeAFit} and a numerical fit for the intermediate regime, respectively.
The amplitude $A_\Phi$ is given by $A_\Phi=\pm 2 A(w)/3b$ as in \cref{eq:PhiAdLargeAFit} for the stiff (-) and soft (+) regimes, and by the fit given in \cref{eq:A(w)Fit} around $w=1/3$. 
Schematically, this parametrisation results in the following momentum integrals
\begin{align} \label{eq:TensorPS_Adi_Gen}
    \overline{\mathcal{P}_{h,\rm RD}}(k,\tau,\bar{x}\gg1) \propto & \int_{0}^{\infty} dv \int_{|1-v|}^{1+v} du \left(4 v^2-\left(1+v^2-u^2\right)^2\right)^2 (u v)^{n_{\rm eff}} \, \overline{\mathcal{I}^2_{\rm osc}}(\bar{x},u,v)
\end{align}
with the effective spectral index
\begin{align}
    n_{\rm eff}&=-\frac{5}{3} + n_s + 2 n(w) \label{eq:neff}
\end{align}
collecting the $k$-dependencies arising from the suppression factor \labelcref{eq:SuppresionSeva}, the scalar power spectrum \labelcref{eq:ScalarPS} and the transfer function \labelcref{eq:PhiAd_Eff_PL}. Note that for $k>k_{\rm eq}$, $n_{\rm eff}$ is negative unless $n_s>14/3$.
For scales $k\gg k_{\rm eq}$ we first obtain the resonant contribution
\begin{align}
    \label{eq:OmegaGWresAdi}
    \Omega_{\rm GW, res}^{\rm ad}(k) 
    =& \Omega_{\rm GW, res}^{\rm ad, peak}
    \left(\frac{k}{k_{\rm uv}}\right)^{2 n_{\rm eff}+7} \tilde{\Theta}_{\rm uv}(k) \,,
\end{align}
where the amplitude at $k=k_{\rm uv}$ is given by
\begin{align}
    \label{eq:OmegaGWresPeakAdi}
   \Omega_{\rm GW, res}^{\rm ad, peak} &= A_s^2 A_\Phi(w)^4\frac{\pi}{2^{11}}\left(\frac{2}{3}\right)^{1/3}\frac{\left(c_s^2-1\right)^2 }{(2 c_s)^{1+2 n_{\rm eff}}}
    \left(\frac{k_{\rm uv}}{k_{\rm eq}}\right)^{4 n(w)}
    \left(\frac{k_{\rm uv}}{k_{\rm eva}}\right)^{17/3} \\
    &\approx 9.72\times 10^{32} \frac{ 3^{2 n(w)+n_s}}{4^{3 n(w)+n_s}} A_s^2 A_\Phi(w)^4 \gamma ^{-4 n(w)/3}  \beta ^{-\frac{2 n(w) (w +1)}{3 w }}\left(\frac{g_H}{108}\right)^{-17/9} \left(\frac{M_{\rm f}}{10^4\text{g}}\right)^{34/9}\,.
\end{align}
Like $\Theta_{\rm uv}$ in the isocurvature case, see \cref{eq:ThetaUV,eq:ThetaUV_Heaviside}, $\tilde{\Theta}_{\rm uv}$ acts as a smooth cutoff and interpolates between zero and an $\mathcal{O}(1)$-value that depends on $n_{\rm eff}$. The explicit expression for $\tilde{\Theta}_{\rm uv}(k)$ is given in \labelcref{eq:ThetaUV_ad}.
An important observation is that the spectral slope in \labelcref{eq:OmegaGWresAdi} depends on $w$ through the effective spectral index $n_{\rm eff}$ \labelcref{eq:neff}.
Let us mention here that the isocurvature result \labelcref{eq:OmegaGWres} is recovered from \cref{eq:OmegaGWresAdi} with the replacements $n_s\rightarrow 4$, $n(w)\rightarrow -2$, $A_s\rightarrow 2/(3\pi)$ and $A(w)\rightarrow C(w)$.

Interestingly, the slope of the resonant part of the spectrum, $2n_{\rm eff}+7$, is negative for $n_s\lesssim 1$, which entails that the peak in this case is not at the UV scale $k_{\rm uv}$. Numerically we find the peak close to $k_{\rm eq}$, corresponding to a frequency today
\begin{equation} \label{eq:fEq}
    f_{\rm eq}=\sqrt{2}\gamma^{1/3}\beta^\frac{1+w}{6w} f_{\rm uv} \,,
\end{equation}
due to the relation \labelcref{eq:Relationsk_kUV} and with $f_{\rm uv}$ as given in \cref{eq:fUV}.
In order to get an analytic estimate for the peak region near $k_{\rm eq}$ we use the large-$v$ approximation  \labelcref{eq:OscIntegralRD2_LV} for the Kernel $\overline{\mathcal{I}^2_{\rm osc}}$ and split the remaining integral into two parts, where we consider the large and small $k$-branches of the transfer function \labelcref{eq:PhieMDAdi} separately. The relevant contribution stems from the lower boundary of the $v>v_{\rm eq}$ part of the integral. We evaluate the integral at $v_{\rm min}=\xi_1 k_{\rm eq}/k$, where we introduced the $\mathcal{O}(1)$ factor $\xi_1(\beta,w)$ to account for the fact that our split of the transfer function neglects the transition near $k_{\rm eq}$. This yields the spectrum in the intermediate, near-peak regime
\begin{equation}    \label{eq:OmegaGWmidAdi}
    \Omega_{\rm GW, mid}^{\rm ad}(k)= -\frac{ A_s^2 A_\Phi(w)^4 c_s^4}{768 (1+2n_{\rm eff})}\left(\frac{3}{2}\right)^{2/3} \xi_1^{1+2n_{\rm eff}}
    \left(\frac{k_{\rm eq}}{k_{\rm uv}}\right)^{2 n_s-7/3}
    \left(\frac{k_{\rm uv}}{k_{\rm eva}}\right)^{14/3}
    \left(\frac{k}{k_{\rm uv}}\right)^{5} \,.
\end{equation}
We find that the $k^5$ scaling agrees well with the scaling found numerically, but the amplitude is quite sensitive to the value of $\xi_1$ due to the $w$-dependent exponent. Therefore, we need to rely on the numerical solutions with the full transfer function to determine the correct amplitude.

In the deep IR, i.e.~for low frequencies with $k\ll k_{\rm eq}$, we again use the large-$v$ approximation \labelcref{eq:OscIntegralRD2_LV}. In this case, we find the dominant contribution to the integral \cref{eq:TensorPS_Adi_Gen} to be given by the growing branch of the $v<v_{\rm eq}$ part of the integral, resulting in
\begin{equation} \label{eq:OmegaGWIRAdi}
    \Omega_{\rm GW, IR}^{\rm ad}(k)=\frac{ A_s^2 c_s^4 (3 w +5)^4 }{10^4  (6 n_s+5) (w +1)^4}
    \left(\frac{3}{2}\right)^{2/3}
    \left(\frac{\xi_2 k_{\rm eq}}{k_{\rm uv}}\right)^{5/3 + 2n_s}
    \left(\frac{k_{\rm uv}}{k_{\rm eva}}\right)^{14/3}
    \left(\frac{k}{k_{\rm uv}}\right) \,,
\end{equation}
where we evaluated the integral at the upper boundary $v_{\rm max}=\xi_2 k_{\rm eq}/k$. The $\mathcal{O}(1)$ factor $\xi_2=\xi_2(w)$ in this case can be estimated analytically from the crossing of the two branches of \labelcref{eq:PhieMDAdi} and is given in \cref{eq:xi_matching}.
This approximation agrees well with the numerical result, up to a factor of 1/2.
%
\begin{figure}
\centering
\includegraphics[width=0.75\textwidth]{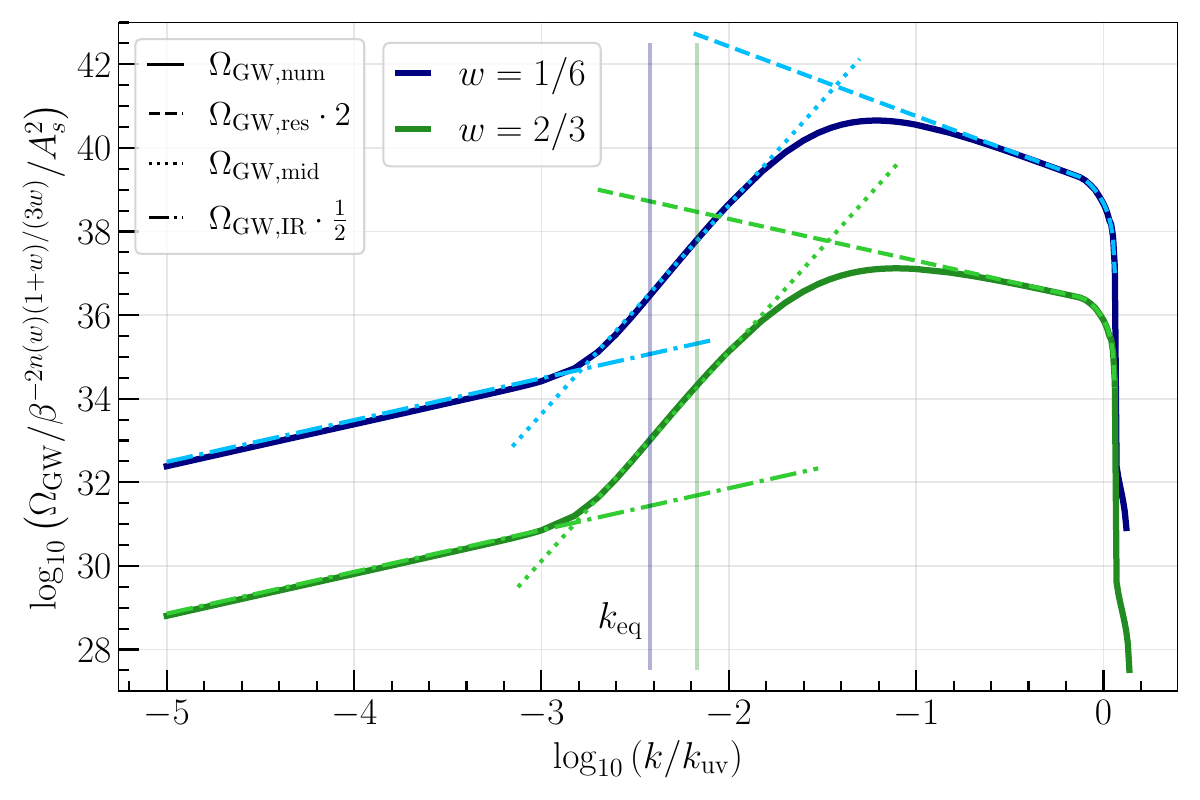}
\caption{The GW spectrum induced after PBH evaporation by a scale-invariant ($n_s=1$) primordial power spectrum as a function of $k/k_{\rm uv}$ for $M_{\rm f}=10^4 g$. We fixed $\beta=10^{-2}$ and $\beta=10^{-5}$ in the cases of $w=1/6$ (blue lines) and $w=2/3$ (green), respectively, for better comparison of the spectral shapes, and the thick vertical lines mark the corresponding $k_{\rm eq}$. We set $\xi_1=1.1$ and $\xi_1=0.48$, respectively, yielding good agreement of the amplitude in the intermediate regime.
The spectra are normalised by $\beta^{\frac{-2n(w)(1+w)}{3w}}$, such that the resonant contribution becomes independent of $\beta$. See how the analytical approximations \labelcref{eq:OmegaGWresAdi,eq:OmegaGWmidAdi,eq:OmegaGWIRAdi} agree well with the numerical result in the respective regimes. }
\label{fig:Omega_GW_Adi}
\end{figure}
%
In \cref{fig:Omega_GW_Adi}, we show the induced GW spectrum for a scale-invariant primordial power spectrum with $n_s=1$ for two values of $w$. As the plot shows, the analytical expressions for the resonance and the IR tail \labelcref{eq:OmegaGWresAdi,eq:OmegaGWIRAdi} agree well with the numerical result, and for the near-peak regime \labelcref{eq:OmegaGWmidAdi} we get an accurate result by fixing $\xi_1$ by hand.
Note that for $n_s\lesssim 1$, the peak is rather broad and close to $k_{\rm eq}$, as opposed to the isocurvature case where the spectrum is sharply peaked at $k_{\rm uv}$.

Importantly, the scale $k_{\rm eq}$, whose value compared to $k_{\rm uv}$ is given by the initial PBH abundance $\beta$ and $w$ from \cref{eq:Relationsk_kUV}, determines where the spectrum tips over from the resonant slope to the IR tail. In principle, this gives another observable complementary to the peak of the isocurvature-induced GWs and, in combination, could break the degeneracy between $\beta$ and $w$. Note that in the presence of adiabatic primordial fluctuations\footnote{Note that the PBH isocurvature is an inevitable consequence of the presence of PBHs, but the presence of primordial adiabatic fluctuations assumes an extrapolation of CMB measurements. However, we do not know the spectrum of primordial fluctuations at small scales. Therefore, whether there is an adiabatic-induced GW signal in the PBH reheating scenario depends on whether there are sizeable primordial fluctuations on such small scales.} the total induced GW spectrum in the PBH reheating scenario will have a double peak structure. This was first pointed out in Ref.~\cite{Bhaumik:2022pil}.
If both signals were observed, one could first determine the initial PBH mass $M_{\rm f}$ from the peak frequency $f_{\rm uv}$ \labelcref{eq:fUV}. Then, the amplitude of the resonant peak \labelcref{eq:OmegaGWresPeak} determines a line of possible $\beta(w)$. The peak frequency of the adiabatic signal could then be used to determine $f_{\rm eq}$ by comparing the analytical expressions \cref{eq:OmegaGWresAdi,eq:OmegaGWmidAdi,eq:OmegaGWIRAdi} or a numerical template for the adiabatic spectrum with the observational data. This finally singles out the true combination $(\beta,w)$ by \labelcref{eq:fEq}. One can determine the remaining amplitude and tilt of the primordial curvature power spectrum, $A_s$ and $n_s$, from the amplitude and slope of the resonant part of the adiabatic GW spectrum \labelcref{eq:OmegaGWresAdi}.

%
\begin{figure}
\centering
\includegraphics[width=0.75\textwidth]{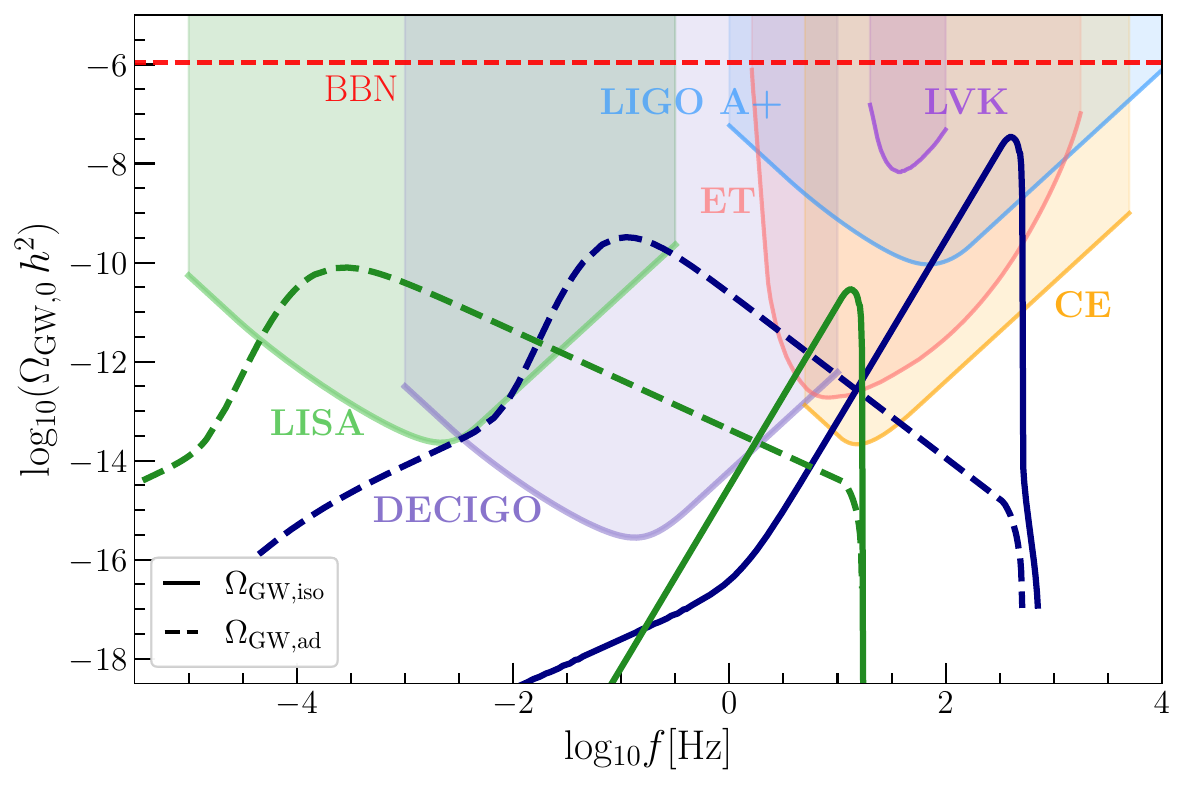}
\caption{The induced GW spectra for 2 sets of example values. We compare the spectra for $w=1/6$, $M_{\rm f}=5\times 10^4 \rm{g}$, $\beta=9\times 10^{-5}$ (blue curves) and $w=2/3$, $M_{\rm f}=3\times 10^6 \rm{g}$, $\beta=2\times 10^{-14}$ (green). In both cases we set $n_s=1$ and $A_s=2.1\times 10^{-9}$.  The solid lines represent the isocurvature induced GWs, while the dashed lines show the adiabatic induced GWs.
We confront the spectra with the power-law integrated sensitivity curves of several next-generation GW detectors \cite{Thrane:2013oya, Schmitz:2020syl, SensitivityCurves, CE, A+}. We also show the current upper bound on an isotropic GW background from the LIGO-Virgo-Kagra (LVK) collaboration \cite{KAGRA:2021kbb} and the integrated BBN bound \labelcref{eq:Neff_bound} evaluated today.
}
\label{fig:OmegaGW_Forecast}
\end{figure}
%
In \cref{fig:OmegaGW_Forecast}, we show the induced GW spectra evaluated today for two values of $w$ and different initial PBH masses $M_{\rm f}$ and abundances $\beta$ within the allowed parameter space. As can be seen, both the adiabatic and isocurvature induced GW signals can enter the observational windows of future detectors. While the isocurvature induced GWs are a target for high frequency detectors such as the Einstein Telescope (ET) \cite{Branchesi:2023mws} and Cosmic Explorer (CE) \cite{Evans:2021gyd}, the adiabatic GWs may be observable by LISA \cite{LISACosmologyWorkingGroup:2022jok} and DECIGO \cite{Kawamura:2020pcg}. Interestingly, for the values $n_s=1$ and $A_s=2.1\times 10^{-9}$ which are close to the values measured at CMB scales \cite{Planck:2018vyg}, the adiabatic induced GWs are of comparable amplitude as the isocurvature ones and remain below the BBN bound.

Remarkably, the above considerations imply that by measuring both the adiabatic and isocurvature-induced GW signals, one could determine the equation of state of the post-inflationary Universe, along with the parameters of the primordial curvature power spectrum at very small scales. Additionally, one could pin down the initial PBH mass and abundance, establishing whether evaporating primordial black holes were responsible for reheating the early Universe.

\section{Summary and conclusion} \label{sec:Conclusion}
In this paper, we have investigated a generalised PBH reheating scenario, where PBHs initially form in an epoch with a general constant equation of state after inflation. We found that a stiffer EoS $(w>1/3)$ leads to a significantly longer PBH dominated phase, see \cref{eq:Neva-Neq}, because the primordial fluid is diluted faster due to the larger pressure. As \cref{eq:beta_lower} shows, this, in turn, leads to a much lower minimal initial abundance $\beta$ required to reach PBH domination.

We then investigated in detail the evolution of isocurvature-induced and primordial curvature perturbations and the resulting induced GWs. We derived the general transfer functions for the curvature perturbation for both isocurvature and adiabatic initial conditions, which can be found in \cref{eq:PhiIsoeMDInterpol,eq:PhiAdieMDInterpol}, respectively. Our main result is that the amplitude of the curvature perturbation deep inside the PBH dominated stage for $k\gg k_{\rm eq}$ is proportional to $(i)$ $\kappa^{-2}$ for isocurvature initial conditions, $(ii)$ $\kappa^{-2-b}$ for adiabatic initial conditions and $w>1/3$ ($b<0$) and $(iii)$ $\kappa^{-2}$ for adiabatic initial conditions and $0<w<1/3$ ($1>b>0$). In all cases, the $\kappa^{-2}$ scaling comes from the Poisson equation relating curvature fluctuations with PBH density fluctuations. In the isocurvature case,  the PBH density contrast is dominant in the beginning, which directly carries over to the curvature perturbation in the PBH dominated phase. This is why there is no $w$-dependence in the exponent of $\kappa$. In the adiabatic case, PBH density fluctuations are sourced by initial curvature fluctuations and keep a “memory” of the equation of state $w$. The additional power $\kappa^{-b}$ for $w>1/3$ is due to a subhorizon growth of the PBH density contrast. For $w<1/3$ the situation is more subtle, as the scaling $\kappa^{-2}$ is achieved only for very small scales. In intermediate regimes there is in fact a $\kappa^{-b}$ correction to the $\kappa^{-2}$ scaling (see the discussion in sec.~\ref{sec:Fluctuations_eMD}). Lastly, note that in all cases, there is a $w$-dependent prefactor in the transfer function.

In \cref{sec:iGW_Spectrum} we presented our results on the induced GWs after PBH evaporation. We found that, in general, the induced GW spectrum is enhanced for $w>1/3$ due to the longer PBH dominated phase. For the PBH isocurvature induced GWs, we find that it is sharply peaked at the UV-cutoff scale $k_{\rm uv}$ which is determined by the initial PBH mass $M_{\rm f}$ and corresponds to a peak frequency today of about $f_{\rm uv}=\mathcal{O}(1-10^6)$Hz. The peak amplitude \labelcref{eq:OmegaGWresPeak_Num} is determined by the PBH parameters $M_{\rm f}$ and $\beta$, as well as the EoS parameter $w$. However, we find that the slope of the GW spectrum is insensitive to $w$, since the transfer function for the curvature perturbation only depends on $\kappa^{-2}$. 
Then, by demanding that the amount of induced GWs does not violate the bound on the effective number of relativistic species $\Delta N_{\rm eff}$ at BBN, we placed an upper bound \labelcref{eq:beta_upper} on the initial PBH abundance. The resulting parameter space is plotted in \cref{fig:parameter_space}, showing how a stiffer EoS broadens the allowed range for $\beta$ and shifts it to lower values.

For adiabatic induced GWs, we found that, contrary to the PBH isocurvature induced GWs, the spectral features depend on $w$, in addition to the standard dependence on the amplitude and tilt of the primordial curvature power spectrum and the PBH mass and abundance. Furthermore, the peak of the GW spectrum appears closer to $k_{\rm eq}$, providing complementary information to the isocurvature case. Interestingly, a combined observation of the isocurvature and adiabatic induced GW spectra would be able to break the degeneracy between model parameters. In particular, by measuring the peak and amplitude of the isocurvature and adiabatic induced GW spectra, one is able to determine the initial energy density fraction $\beta$ and the initial equation of state $w$. These GW signals can be observed by future detectors such as LISA, DECIGO, the Einstein Telescope and Cosmic Explorer, as illustrated in \cref{fig:OmegaGW_Forecast}.

Our analysis could be extended in several aspects. Firstly, we conducted our analysis using linear cosmological perturbation theory. However, during the eMD period, density fluctuations grow proportionally to the scale factor, and the smallest scales will reach the non-linear regime before PBH evaporation. We give an estimate for the scale $k_{\rm NL}$ at which this happens in \cref{eq:kNL}. Our results should, therefore, be understood as optimistic estimates, and the derived upper bound on $\beta$ can be viewed as rather conservative.
Let us note, however, that on the small scales considered here, $\Phi$ can remain linear even if the density fluctuations become larger than unity, which could be argued to justify using linear theory as an estimate. To improve upon our computations, one could model the effect of non-linear density fluctuations by employing empirical fits to the results of $N$-body simulations of structure formation, as done in \cite{Kawasaki:2023rfx}. Alternatively, one could employ analytical approaches such as kinetic field theory (KFT) \cite{Bartelmann:2019unp, Konrad:2022tdu} to compute non-linear density fluctuation power spectra. Otherwise, sophisticated numerical simulations as in \cite{Fernandez:2023ddy} are needed.

Secondly, in our derivation, we assumed a monochromatic PBH mass function. This led to the simultaneous evaporation of all PBHs and the sudden transition to the RD era. However, depending on the details of the PBH formation, the mass function may be rather broad, and the transition, therefore, more gradual. The effects of an extended mass function and a gradual transition from an eMD to the RD have been studied in \cite{Papanikolaou:2022chm} and \cite{Inomata:2019zqy, Pearce:2023kxp}, respectively.
It would be interesting to apply similar analyses also in our more general setup, although we expect the effects to be comparable.

Finally, we treated the isocurvature and adiabatic modes separately in our computations, which is justified at the linear level. However, at the second order in the GW source term, one gets an additional cross-term with contributions from both modes. In order to compute the resulting tensor power spectrum induced by this term, one would need to know the mixed two-point correlator $\langle \Phi_0 S_0 \rangle$. Assuming PBH formation is a rare and random event, the correlation would be vanishing, but in the case of sizeable primordial non-Gaussianity\footnote{The effect of local-type non-Gaussianities on the PBH isocurvature induced GWs have recently been studied in \cite{Papanikolaou:2024kjb, He:2024luf}. There, it was found that the presence of non-Gaussianities modifies the initial Poisson distribution of PBH density fluctuations \labelcref{eq:InitialIsocurvaturePS} and leads to a second peak in the induced GW spectrum.}
or PBH clustering, it could carry information on the common origin of the two modes, considering that the isocurvature perturbation (i.e.~the PBH density fluctuations) is correlated with the curvature fluctuation that leads to PBH formation in the first place.
It could be interesting to compute the GW spectrum induced by this mixed term to look for any interesting interference effects and additional peaks and use the (non-)observation of the signal as a consistency check.

We plan to study the GWs induced by the isocurvature mode itself (as in Ref.~\cite{Domenech:2021and}) during the $w$-dominated phase in a subsequent work, which will have applications to the universal GW signatures of cosmological solitons studied in Refs.~\cite{Lozanov:2023knf,Lozanov:2023aez}.
To conclude, let us remark that, as the GW amplitude is strongly dependent on $\beta$, a slight increase in the value of $\beta$ could easily compensate for the suppression due to non-linear effects or an extended mass function. Therefore, our main conclusions should remain unchanged by the above considerations, and we may still hope to observe the GW signals with future detectors and uncover the physics at play during the reheating of the Universe.

\begin{acknowledgments}
  We would like to thank A. Ganz for helpful discussions. We also thank D. Paul for pointing out a typo in a previous version.
  This research is supported by the DFG under the Emmy-Noether program grant no.~DO 2574/1-1, project number 496592360, and by the JSPS KAKENHI grant No.~JP24K00624.
  In the derivation of perturbation equations, we used the \texttt{xPand (xAct)} package \cite{Pitrou:2013hga} for \texttt{Mathematica}, and unit conversions were done using the \texttt{NaturalUnits} package \cite{Tomberg:2021ajh}.
\end{acknowledgments}

\appendix
\section{Background and perturbation equations \label{app:einsteinequations}}
The expansion of the background spacetime is governed by the Hubble parameter ${H=\dot{a}/a}$, where $a$ is the scale factor of the FLRW metric \labelcref{eq:FLRW}. $H$ satisfies the Friedmann equation
\begin{equation}
    3 H^2 M_{\rm Pl}^2 = \rho_{\rm PBH}+\rho_{\rm rad}+\rho_{w} \,,
    \label{eq:FriedmannEq}
\end{equation}
with the energy densities $\rho_n$, $n\in\lbrace \text{PBH},\, \text{rad},\, w \rbrace$, of the PBH, radiation and primordial fluids, respectively.
The Hubble parameter is related to cosmic time $t$ as
\begin{equation} \label{eq:H(t)}
    H=\frac{2}{3(1+w)} \frac{1}{t}
\end{equation}
when the Universe is dominated by a fluid with EoS parameter $w$. This relation is useful e.g.~for expressing the time of PBH formation $t_{\rm f}$ in terms of the initial PBH mass \labelcref{eq:Mf}, or when evaluating the Hubble parameter at the evaporation time $t_{\rm eva}$ \labelcref{eq:teva}.

Neglecting the radiation component $\rho_{\rm rad}$ and the energy transfer between components, it is convenient to write the total energy density at early times $a\ll a_{\rm eva}$ as
\begin{equation}
    \rho = \frac{\rho_{\rm eq}}{2} \left( \left(\frac{a_{\rm eq}}{a}\right)^3 + \left(\frac{a_{\rm eq}}{a}\right)^{3(1+w)} \right) \,,
    \label{eq:BackgroundRho}
\end{equation}
with the total density at equality $\rho_{\rm eq}$. From the different dependencies of the primordial background and PBH fluid densities on the scale factor $a$, one can read off the useful relation
\begin{equation}
    \left(\frac{a_{\rm f}}{a_{\rm eq}}\right)^{3 w}\approx \beta
    \label{eq:af/aeq}
\end{equation}
for the scale factor at PBH formation and equality.
Solving \labelcref{eq:FriedmannEq} with the total energy density \labelcref{eq:BackgroundRho} one obtains the scale factor \labelcref{eq:scalefactor} quoted in the main text. Equivalently, we may also write 
\begin{equation}
    \frac{a}{a_{\rm eq}} = \mathcal{F}^{-1}\left[\frac{k_{\rm eq} \, \tau}{\sqrt{2}} + c_1\right] \quad \text {with} \quad \mathcal{F}(x) = 2 \, _2F_1\left(\frac{1}{2},-\frac{1}{6w};1-\frac{1}{6w};-\frac{1}{x^{3w}}  \right)\sqrt{x}\,,
    \label{eq:scalefactor2}
\end{equation}
where the constant $c_1$ is set by requiring $a(0)=0$ to yield
\begin{equation} \label{eq:c1}
    c_1=\frac{2}{\sqrt{\pi }}\Gamma \left(1-\frac{1}{6 w }\right) \Gamma \left(\frac{1}{2} +\frac{1}{6 w }\right) \,.
\end{equation}
This form for $a/a_{\rm eq}$ is more suitable in the $a\gg a_{\rm eq}$ limit as $_2F_1\left(a,b;c;-{\chi^{-3w}}  \right)\sim 1$ for $\chi\gg 1$. It is easy to see that in this limit $a\propto \tau^2$. Lastly, when the Universe is dominated by radiation after PBH evaporation, the radiation temperature is related to the Hubble parameter as
\begin{equation} \label{eq:RelationT}
    \rho_{\rm rad}(t_{\rm eva})=\frac{\pi^2}{30}g_*(T_{\rm eva}) T_{\rm eva}^4 \approx 3 M_{\rm Pl}^2 H_{\rm eva}^2\,.
\end{equation}

Perturbing the metric around the FLRW background as in \cref{eq:FLRW} and the fluid 4-velocities as $u_n^\mu=\bar{u}_n^\mu + \delta u _n^\mu$ results in the perturbed Einstein equations, whose $00$, $0i$ and $ij$ trace components at linear order in perturbations read
\begin{align}
    6 \mathcal{H} \Phi' + 6 \mathcal{H}^2 \Phi - 2 \Delta \Phi =& a^2 \left( \delta \rho_{\rm PBH} + \delta \rho_{w} \right) \label{eq:00EinsteinEq} \\
    \Phi' + \mathcal{H} \Phi =& \frac{1}{2} a^2 \left( V_{\rm PBH}\rho_{\rm PBH} + (1+w)V_{w} \rho_w \right) \\
    \Phi'' + 3\mathcal{H} \Phi' + (\mathcal{H}^2 + 2 \mathcal{H}')\Phi =& -\frac{1}{2} a^2 c_{w}^2 \delta \rho_w \,,
     \label{eq:ijEinsteinEq}
\end{align}
where again we are considering $a\ll a_{\rm eva}$ such that the radiation component is negligible and we set $M_{\rm Pl}=1$ for convenience. While at the background level the fluid velocities are simply given by $(u_n^\mu)=(1/a,\Vec{0})$, at first order in perturbations we have $(\delta u_n^\mu)=(\Phi/a,\partial^i V_n / a)$  \cite{Malik:2008im} where we have kept only the scalar part of the spatial velocity perturbations.
The covariant conservation of the energy-momentum tensors $\nabla_\mu T^{\mu\nu}_n=0$ of the PBH and $w$-fluids further leads to the energy conservation equations, resulting from the $\nu=0$ component and given by
\begin{align}
    \delta \rho_{\rm PBH}' + 3 \mathcal{H} \delta \rho_{\rm PBH} + \rho_{\rm PBH} ( 3 \Phi' + \Delta V_{\rm PBH} ) =& 0 \label{eq:Energy_Cons_PBH}  \\
    \delta \rho_w ' + 3 (1+c_w^2)\mathcal{H} \delta\rho_w + (1+w) \rho_w (3\Phi' + \Delta V_w) =& 0 \,. \label{eq:Energy_Cons_w}
\end{align}
The spatial components $\nu=i$ lead to the momentum conservation equations
\begin{align}
    V_{\rm PBH}' + \mathcal{H} V_{\rm PBH} - \Phi =& 0 \label{eq:Momentum_Cons_PBH} \\
    V_w' + (1-3w)\mathcal{H} V_w + \frac{c_w^2}{1+w}\frac{\delta \rho_w}{\rho_w} - \Phi =& 0 \,. \label{eq:Momentum_Cons_w}
\end{align}

%
\begin{figure}
  \centering
  \includegraphics[width=0.75\textwidth]{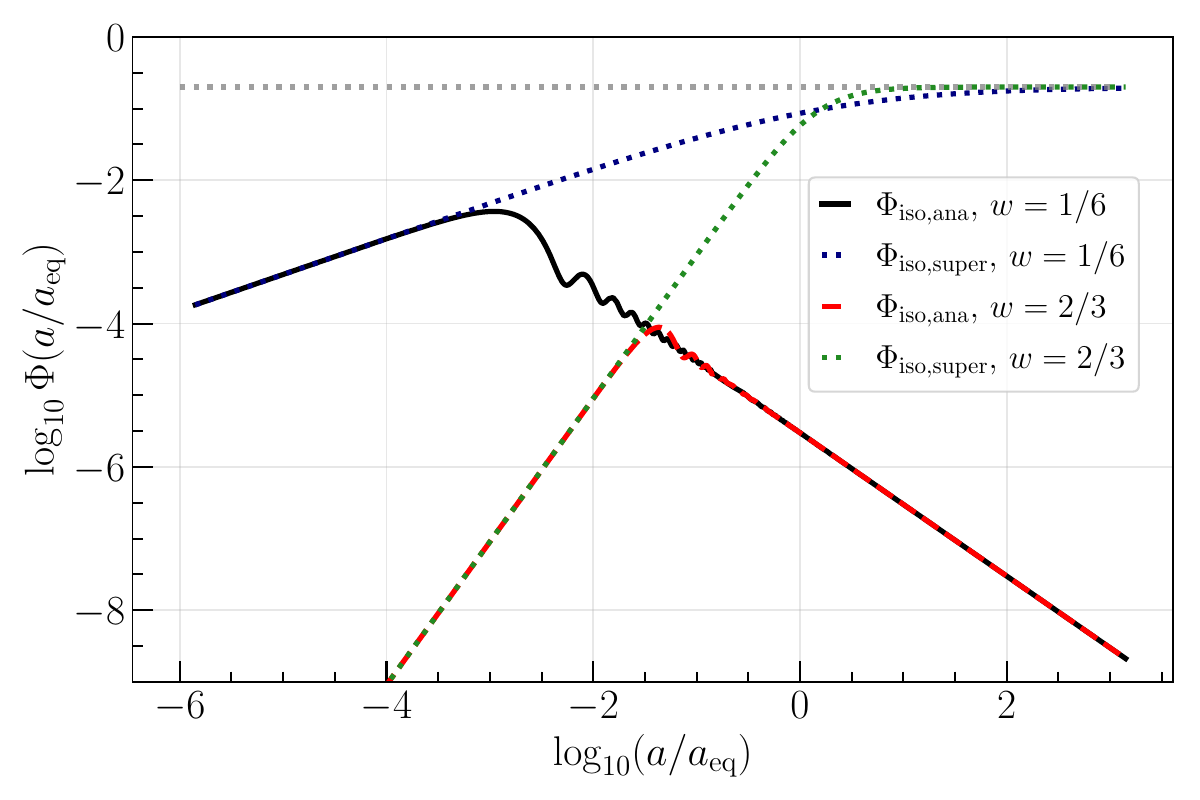}
  \caption{The isocurvature induced curvature perturbation $\Phi_{\rm iso}$ as a function of the scale factor $\chi=a/a_{\rm eq}$. We plot the analytical solution \labelcref{eq:Phi_Iso_wDE} and the superhorizon solution \labelcref{eq:Phi_Iso_Super} for two values of $w$ and $\kappa=5\times 10^2$. The plot shows the $w$-independent subhorizon ($x\gg 1$) behaviour found in \labelcref{eq:Phi_Iso_wDE_LargeX_Chi}. Note that the superhorizon solution is valid for the entire range, whereas the solution \labelcref{eq:Phi_Iso_wDE} is valid only for $a\ll a_{\rm eq}$ and does not capture the transition to the eMD. }
  \label{fig:PhiChiPlot}
\end{figure}
%

Using the Friedmann equation \labelcref{eq:FriedmannEq} and the expression for the total energy density \labelcref{eq:BackgroundRho}, we can express the perturbation equations \labelcref{eq:PhiEq,eq:SEq} in terms of the variable $y$ \labelcref{eq:yDefinition}, yielding
\begin{align} \label{eq:PhiEq_y}
\frac{d^2\Phi}{dy^2}+\frac{1}{6}
   \left(\frac{5+9w}{w y}-\frac{6}{w+y+1}+\frac{3}{y+1}\right)\frac{d\Phi}{dy}+\frac{ 3 w+2  (1+w)\kappa ^2
   y^{\frac{1}{3w}}}{9 w y (y+1) (w+y+1)}\Phi& \nonumber \\
   =\frac{1+w }{6 w y (1+y) (1+w+y)}S& \,,
\end{align}
and
\begin{align} \label{eq:SEq_y}
\frac{d^2S}{dy^2}+\frac{1}{6}
   \left(\frac{1+3w}{w y}-\frac{6}{1+w+y}+\frac{3}{1+y}\right)
   \frac{dS}{dy}+\frac{2 \kappa ^2 y^{\frac{1}{3w}}}{9 w (1+y)
   (1+w+y)}S& \nonumber \\
   =\frac{8 \kappa ^4 y^{\frac{2}{3w}} }{27 w (1+y) (1+w+y)}\Phi& \,.
\end{align}
Inserting the superhorizon solution \labelcref{eq:Phi_Iso_Super_SmallChi} for $\Phi$ in the integral \cref{eq:S_Iso_Implicit} for $S_{\rm iso}$ gives the behaviour of the isocurvature perturbation in the superhorizon ($x\ll 1$) regime
\begin{equation} \label{eq:S_Iso_Smallx}
    S_{\rm iso}(a\ll a_{\rm eq}; x\ll 1)\approx S_0-\frac{2^{\frac{b-3}{2}} (b-1) (b+1)^{b-1}  }{3 (b-3) (b+2)}S_0 \kappa ^{b-1} x^{3-b}\,.
\end{equation}

%
\begin{figure}
  \centering
  \includegraphics[width=\textwidth]{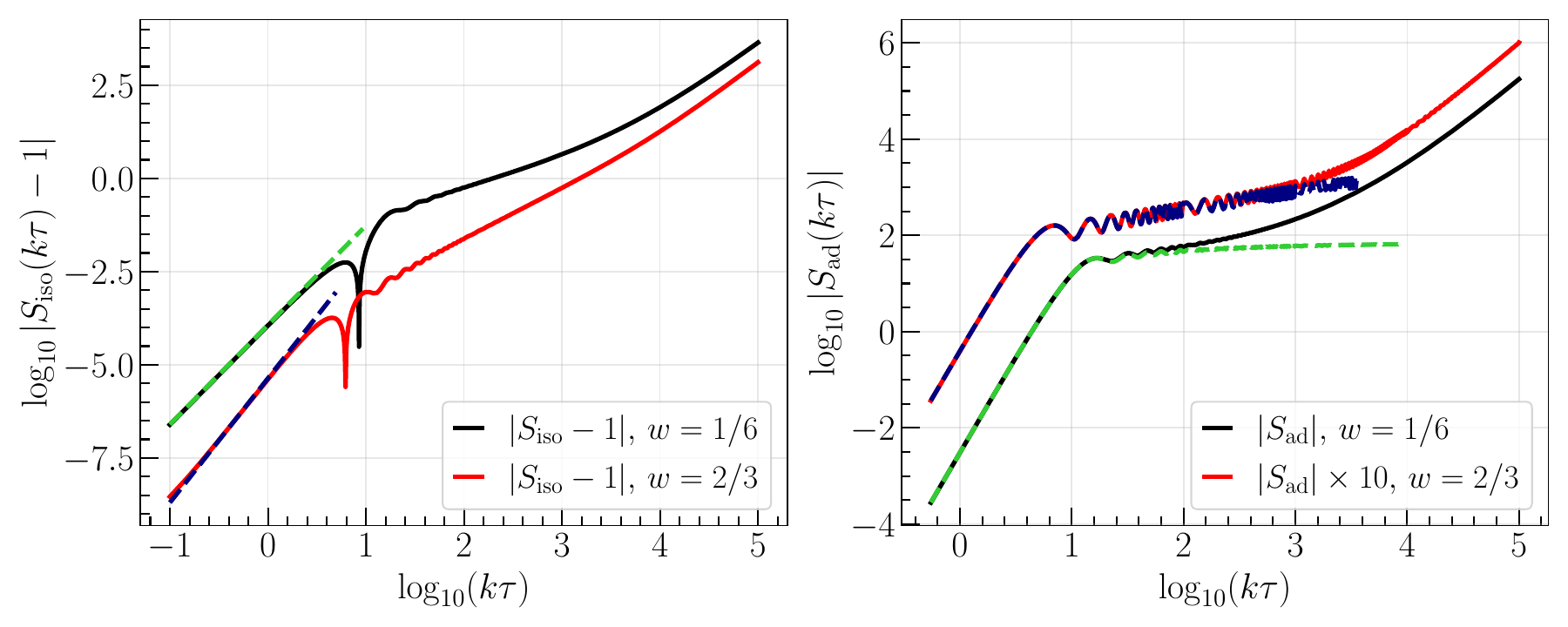}
  \caption{Here we show the isocurvature perturbation $S$ as a function of $x=k\tau$ for isocurvature (left) and adiabatic (right) initial conditions. We set $\kappa=10^3$ in both cases and $S_0=\Phi_0=1$. The solid lines show a numerical solution, the dashed lines mark the small-$x$ solution \labelcref{eq:S_Iso_Smallx} in the isocurvature case, and the analytical result \labelcref{eq:S_Adi_wDE} for the adiabatic one, respectively. Note that on the left we plot $|S-1|$ to show the growing first order correction $S_1$, and on the right we rescaled the curves for $w=2/3$ by a factor 10 for a clearer presentation.
  }
  \label{fig:Iso_plot}
\end{figure}
%

\section{Constructing the transfer function in the adiabatic case} \label{sec:TransferFct_Adi}
In this section we present in detail the derivation of the transfer function \labelcref{eq:Phi_Adi_eMD}.
To begin with, we consider early times where fluctuations in the $w$-fluid are dominant and we assume $\delta_w \gg \delta_{\rm PBH}$ for the respective density contrasts. This implies that the gravitational potential is determined by fluctuations in the dominant fluid and is described by \cref{eq:Phi_Adi_wDE}. We can derive a second order equation for $\delta_{\rm PBH}$ by differentiating the energy conservation equation \labelcref{eq:Energy_Cons_PBH} and using \cref{eq:Momentum_Cons_PBH} to replace $V_{\rm PBH}$. In this way we arrive at
\begin{equation} \label{eq:MatterPert_wDE}
    \delta_{\rm PBH}'' + \frac{2}{1+3w} \frac{\delta_{\rm PBH}'}{\tau}=k^2 \Phi - \frac{6}{1+3w}\frac{\Phi'}{\tau} - 3\Phi'' \,.
\end{equation}
Because of our assumption we can treat $\Phi$ as an external source. We first consider the homogeneous solutions to \cref{eq:MatterPert_wDE} by setting the right hand side to zero and find the two independent solutions
\begin{equation}
    \delta_{\rm h 1} = c_1 \quad \text{and} \quad \delta_{\rm h 2} = c_2 \frac{1-(k\tau)^{-b}}{b} \,,
\end{equation}
where we made explicit the factor $k^{-b}$ to leave the constant $c_2$ dimensionless and pulled out the constant $c_2/b$ necessary to recover the logarithmic solution in the radiation case ($b=0$) which appears due to the identity
\begin{equation}
    \displaystyle{\lim_{\alpha \to 0}} \, \frac{x^\alpha - 1}{\alpha} = \ln(x) \,.
\end{equation}
The Green's function can be obtained from the two homogeneous solutions by
\begin{equation}
    \mathcal{G}_\delta(\tau,\tilde{\tau}) = \frac{\delta_{\rm h 1}(\tau)\delta_{\rm h 2}(\tilde{\tau})-\delta_{\rm h 1}(\tilde{\tau})\delta_{\rm h 2}(\tau)}{\delta_{\rm h 1}'(\tilde{\tau})\delta_{\rm h 2}(\tilde{\tau})-\delta_{\rm h 1}(\tilde{\tau})\delta_{\rm h 2}'(\tilde{\tau})} = \tilde{\tau} \frac{1 - \tau^{-b}\tilde{\tau}^b}{b} \,,
\end{equation}
and the particular solution to \cref{eq:MatterPert_wDE} is obtained by a convolution of $\mathcal{G}_\delta(\tau,\tilde{\tau})$ with the source (i.e.~the right hand side of \cref{eq:MatterPert_wDE} evaluated at $\tilde{\tau}$). To solve the convolution integral we insert the solution \labelcref{eq:Phi_Adi_wDE} for $\Phi$ and its derivatives. Noting that the source term decays as $(k\tau)\rightarrow \infty$, we send the upper integration boundary to infinity, leaving the particular solution as the sum of two terms $\propto 1/b$ and $\propto \tau^{-b}/b$ with constant prefactors.
The initial conditions are fixed by noting that $\Phi(\tau\rightarrow 0)=\Phi_0$ is constant, and that \cref{eq:00EinsteinEq} in the superhorizon limit then implies $\delta = 2\Phi$ \cite{Baumann:2022mni}.
Using $\delta(\tau\rightarrow 0)\approx \delta_w=(1+w)\delta_{\rm PBH}$ we then require $c_1=2 \Phi_0/(1+w)$ and $c_2=0$. This results in the full solution for the PBH density contrast early during $w$-domination
\begin{align} \label{eq:DeltaPBH_wDE}
    \delta_{\rm PBH}= \frac{2}{1+w}\Phi_0 + \frac{B_1 - A_1 (k\tau)^{-b}}{b}  \Phi_0  \,,
\end{align}
with the constants $A_1$ and $B_1$ resulting from the convolution integrals given by
\begin{equation}
    A_1 = \frac{2^{b+1} 3^{\frac{b}{2}+1} b \left(\frac{b+1}{1-b}\right)^{b/2} \Gamma \left(\frac{b}{2}\right) \Gamma \left(b+\frac{5}{2}\right)}{(b-1) \Gamma \left(\frac{b+1}{2}\right)}  \quad \text{and} \quad B_1 = \frac{3 \left(b^2+6 b+3\right)}{(b-1)} \, .
\end{equation}
\Cref{eq:DeltaPBH_wDE} is valid from superhorizon scales through horizon crossing to subhorizon scales for $a\ll a_{\rm eq}$ as long as the fluctuations in the $w$-fluid are dominant.
Note that in the limit $b\rightarrow 0$, \cref{eq:DeltaPBH_wDE} reproduces the well-known logarithmic growth of matter density fluctuations during radiation domination \cite{Dodelson:2003ft}. \Cref{eq:DeltaPBH_wDE} shows that for a stiffer  EoS ($b<0$) matter density fluctuations grow faster on subhorizon scales than the logarithmic growth during radiation domination, leading to enhanced small-scale structure formation \cite{Allahverdi:2020bys}.
Interestingly, \cref{eq:DeltaPBH_wDE} also implies that during a kination period with $w=1$ density fluctuations grow linearly with the scale factor, the same as during matter domination, cf.~\cref{eq:DeltaPBH_Mesz_largey}.

Due to the non-zero pressure of the $w$-fluid, the potential $\Phi$ and perturbations $\delta_w$ decay after horizon reentry. Because the PBH density fluctuations grow according to \cref{eq:DeltaPBH_wDE}, or at least stay constant for $b>0$ (corresponding to $w<1/3$), they will at some point overtake the fluctuations of the $w$-fluid, even though the background evolution is still driven by $\rho_w>\rho_{\rm PBH}$. Therefore, we now consider the opposite limit $\delta_{\rm PBH}\gg \delta_w$ and focus on subhorizon scales $k\gg\cal{H}$.
The solutions derived in this way will be valid from $w$ domination through equality until deep inside the eMD, but only for subhorizon scales.

Combining again the energy-momentum conservation equations \labelcref{eq:Energy_Cons_PBH,eq:Momentum_Cons_PBH} and the Poisson equation derived from \cref{eq:00EinsteinEq}, we can derive a second order evolution equation for the density contrast in the regime when fluctuations $\delta_w$ have become negligible. Neglecting the terms with derivatives of $\Phi$ due to $k\gg\cal{H}$, and using $2k^2\Phi \approx a^2 \delta \rho_{\rm PBH}$ due to $\delta_{\rm PBH}\gg \delta_w$ we find \cite{Baumann:2022mni}
\begin{equation}
    \delta_{\rm PBH}'' + \mathcal{H} \delta_{\rm PBH}' - \frac{a^2}{2}\rho_{\rm PBH} \delta_{\rm PBH} = 0 \,.
\end{equation}
Transforming to the variable $y$ defined in \cref{eq:yDefinition} as time variable we obtain
\begin{equation}
   \frac{d^2 \delta_{\rm PBH}}{d y^2 } -\frac{(6 w  y+y+3 w +1) }{6 w  y(y+1)}\frac{d \delta_{\rm PBH}}{d y }-\frac{1}{6 y (y+1) w ^2}\delta_{\rm PBH} = 0 \,,
\end{equation}
which reduces to the well-known M\'esz\'aros equation for $w=1/3$ \cite{Meszaros:1974tb, Boehmer:2010hg}. The solution is given by a sum of hypergeometric functions as
\begin{equation}
    \delta_{\rm PBH}(y)=\tilde{c}_1 \, _2F_1\left(-\frac{1}{3 w },\frac{1}{2 w };\frac{1}{2} +\frac{1}{6w };-y\right)+\tilde{c}_2 \, y^{\frac{1}{2}-\frac{1}{6 w }} \, _2F_1\left(\frac{1}{2}+\frac{1}{3 w },\frac{w -1}{2 w };\frac{3}{2}-\frac{1}{6 w };-y\right) \,.
    \label{eq:DeltaPBH_Mesz}
\end{equation}
Expanding the hypergeometric functions for small argument $y\ll1$ and using the expression \labelcref{eq:ScaleFactor_wDE} for the scale factor to transform back to conformal time we find
\begin{equation} \label{eq:DeltaPBH_Mesz_smally}
    \delta_{\rm PBH}(a\ll a_{\rm eq})=\tilde{c}_1 + \tilde{c}_2 2^{b/2}(1+b)^b \kappa^b (k\tau)^{-b} \,.
\end{equation}
Comparing now \cref{eq:DeltaPBH_wDE,eq:DeltaPBH_Mesz_smally} we can directly read of the relation between the constants
\begin{equation}
    \tilde{c}_1 = \frac{2}{1+w} \Phi_0 + B_1/b \quad \text{and} \quad \tilde{c}_2=2^{-b/2}(1+b)^{-b} \kappa^{-b} A_1/ b \coloneq -\tilde{A} \kappa^{-b} / b\,.
\end{equation}
Remembering that \cref{eq:DeltaPBH_Mesz} is valid for subhorizon scales both before and after equality, we can also take the limit $y\gg 1$ to find
\begin{equation} \label{eq:DeltaPBH_Mesz_largey}
    \delta_{\rm PBH}(a\gg a_{\rm eq}) = \chi \left(\tilde{c}_1 c_3 + \tilde{c}_2 c_4  \right) \,,
\end{equation}
where we defined
\begin{equation} \label{eq:c34_Definition}
    c_3 = \frac{\Gamma \left(\frac{5}{6 w }\right) \Gamma \left(\frac{1}{2}+\frac{1}{6 w }\right)}{\Gamma \left(\frac{1}{2 w }\right) \Gamma \left(\frac{w +1}{2 w }\right)} \quad \text{and} \quad c_4 = \frac{\Gamma \left(\frac{5}{6 w }\right) \Gamma \left(\frac{3}{2}-\frac{1}{6 w }\right)}{\Gamma \left(\frac{1}{2}+\frac{1}{3 w }\right) \Gamma \left(1+\frac{1}{3 w }\right)} \,.
\end{equation}
\Cref{eq:DeltaPBH_Mesz_largey} reflects the well-known result that density fluctuations grow proportionally to the scale factor during matter domination \cite{Dodelson:2003ft, Baumann:2022mni}. Note that for an equation of state different from radiation, the amplitude of $\delta_{\rm PBH}$ carries a scale dependence $\kappa^{-b}$ through the constant $\tilde{c}_2$.
In the regime $a\gg a_{\rm eq}$ we can now use the Poisson equation
\begin{equation} \label{eq:PoissonEq_eMD}
    \Phi=\frac{3}{2}\left(\frac{\cal H}{k}\right)^2 \delta_{\rm PBH}
\end{equation}
to construct the plateau value of the potential $\Phi$ during the eMD from the PBH density contrast. We obtain
\begin{equation} \label{eq:Phi_Adi_eMD_Full}
     \Phi_{\rm ad,eMD}(a\gg a_{\rm eq}) = \frac{3}{2(1+w)}\Phi_0 c_3 \kappa^{-2} - \frac{3}{4} \Phi_0 \kappa^{-2-b}\left( \frac{c_4 \tilde{A} - c_3 B_1 \kappa^b}{b}\right) \,,
\end{equation}
which is valid on small scales $k\gg k_{\rm eq}$ because \labelcref{eq:DeltaPBH_Mesz_largey,eq:PoissonEq_eMD} were derived for subhorizon scales. The second term reproduces the well-known logarithmic $k$-dependence of the transfer function in the limit $b\rightarrow 0$ \cite{Dodelson:2003ft}.
For clarity of presentation in the main text we introduced the combinations
\begin{align}
    \alpha_1 &= -\frac{3}{4} c_3 B_1 =\frac{9 (b (b+6)+3) \Gamma \left(\frac{2-b}{1-b}\right) \Gamma \left(\frac{5 b+5}{2-2 b}\right)}{4 \Gamma \left(\frac{b+2}{1-b}\right) \Gamma \left(\frac{3 b+3}{2-2 b}\right)} \nonumber \\
    \alpha_2 &= \frac{c_4 \tilde{A}}{c_3 B_1}=\frac{2^{\frac{b}{2}+1} 3^{b/2} b \left(1-b^2\right)^{-\frac{b}{2}} \Gamma \left(\frac{2b-1}{b-1}\right) \Gamma \left(\frac{b}{2}\right) \Gamma \left(\frac{b+2}{1-b}\right) \Gamma \left(b+\frac{5}{2}\right) \Gamma \left(\frac{3 b+3}{2-2 b}\right)}{(b (b+6)+3) \Gamma \left(\frac{1}{1-b}\right) \Gamma \left(-\frac{2}{b-1}\right) \Gamma \left(\frac{b+1}{2}\right) \Gamma \left(\frac{b+3}{2-2 b}\right)} \,.  \label{eq:alphas}
\end{align}
Note that $\alpha_2(b=0)=1$, while for $b=1/2$ (corresponding to $w=1/9$) $c_4$ and therefore also $\alpha_2$ diverges.

The numerical fits for the amplitudes $A(w)$ and $A_\Phi(w)$ in \cref{eq:PhiAdLargeAFit,eq:PhiAd_Eff_PL} are given by
\begin{align} \label{eq:A(w)Fit}
    A(w)= 7.76 + 18.3 b + 12.5 b^2 \quad \text{and} \quad A_\Phi(1/5<w<2/3)= 15.6 + 39.2 b + 21.3 b^2 \,.
\end{align}

\section{Suppression due to finite duration of evaporation and non-linear scales} \label{sec:Suppresion_Seva}
Following \cite{Inomata:2019zqy, Inomata:2020lmk}, the additional suppression of $\Phi$ at the transition to radiation domination due to the finite duration can be understood by considering the Poisson equation
\begin{equation}
    k^2 \Phi \approx \frac{3}{2}\mathcal{H}^2\left(\frac{\rho_{\rm PBH}}{\rho_{\rm tot}}\delta_{\rm PBH} + \frac{\rho_{\rm rad}}{\rho_{\rm tot}}\delta_{\rm rad}\right) \,,
    \label{eq:PoissonEq}
\end{equation}
which is valid on subhorizon scales and close to $t_{\rm eva}$, when we can neglect $\rho_w$. Because perturbations in the radiation fluid are suppressed due to the non-zero pressure, $\Phi$ will be sourced by PBH density fluctuations until roughly $t_{\rm eva}$, even when radiation is already dominating. Thus, from \cref{eq:PoissonEq} we expect $\Phi(k\gg\Gamma)\propto \rho_{\rm PBH}\propto M(t)$. At later times, when $\Gamma\gg k$, $\Phi$ decouples from PBH fluctuations. While in the case of an instantaneous transition we essentially treat the time dependence of the PBH mass as a step function $M(t) \approx M_{\rm f} \, \Theta(t_{\rm eva}-t)$, the actual time dependence is as given in \cref{eq:MPBH(t)}, resulting in a correction factor
\begin{align}
    \frac{\Phi_{\rm RD}(t)}{\Phi_{\rm RD}^{\rm instant}}&\approx \left(1-\frac{t}{t_{\rm eva}}\right)^{1/3} \label{eq:SPhi_teva}\\
    &\simeq \exp \left(-\int_{t_{\rm f}} ^t dt' \, \Gamma(t') \right) \,.
\end{align}
In order for the subhorizon approximation used in the Poisson equation \labelcref{eq:PoissonEq} to hold, the following condition on the second time derivative of $\Phi$ (with respect to cosmic time $t$) must be satisfied
\begin{equation}
    \vert \ddot{\Phi} \vert \ll \frac{k^2}{3 a^2} \vert \Phi \vert \,,
\end{equation}
which can be used to set an upper bound on the time $t_{\rm dec}$, when the $\Phi$ modes decouple.
Computing the second time derivative of \labelcref{eq:SPhi_teva} and evaluating at $t_{\rm dec}$, we obtain for the decoupling time
\begin{equation}
    t_{\rm dec} \lesssim t_{\rm eva} - \sqrt{\frac{2}{3}}\frac{a_{\rm dec}}{k} \,.
\end{equation}
Evaluating now \cref{eq:SPhi_teva} at the bound for $t_{\rm dec}$ and using $a_{\rm dec}\approx a_{\rm eva}$, we obtain the suppression factor \labelcref{eq:SuppresionSeva} for $\Phi_{\rm RD}$ after evaporation due to the finite duration of the transition.

We can estimate the scale $k_{\rm NL}\coloneq k(\delta_{\rm PBH}=1)$\cite{Inomata:2020lmk}, above which non-linearities become relevant, using the Poisson equation during eMD \labelcref{eq:PoissonEq_eMD} and the transfer function \labelcref{eq:PhiIsoeMDInterpol} to obtain
\begin{align} \label{eq:kNL}
    k_{\rm NL}=&\sqrt{\frac{5 C(w)}{C(w)(k_{\rm eq} \tau) ^2 S_0 /6 -1}} \, k_{\rm eq}\\
    \approx& k_{\rm eq} \times
    \begin{cases} 
        3.3  / \sqrt{11.8 \left(\frac{M_{\rm f}}{10^4 \text{g}} \right) ^{4/3} \left(\frac{\beta}{10^{-5}}\right)^{7/3} \left(\frac{\gamma}{0.2}\right)^{2/3} \left(\frac{\tau}{\tau_{\rm eva}}\right)^{2} S_0(k_{\rm NL}) - 1} & \quad (w=1/6)\\
         2.4 / \sqrt{58.7 \left(\frac{M_{\rm f}}{10^4 \text{g}} \right) ^{4/3} \left(\frac{\beta}{10^{-8}}\right)^{4/3} \left(\frac{\gamma}{0.2}\right)^{2/3} \left(\frac{\tau}{\tau_{\rm eva}}\right)^{2} S_0(k_{\rm NL}) - 1} & \quad (w=1/3)\\
         1.9 / \sqrt{26.9 \left(\frac{M_{\rm f}}{10^4 \text{g}} \right) ^{4/3} \left(\frac{\beta}{10^{-13}}\right)^{5/6} \left(\frac{\gamma}{0.2}\right)^{2/3} \left(\frac{\tau}{\tau_{\rm eva}}\right)^{2} S_0(k_{\rm NL}) - 1}& \quad (w=2/3)
    \end{cases} \, ,
\end{align}
showing that for a stiffer EoS non-linearities become important at larger scales, or smaller frequencies conversely. A root of the denominator in the equations above implies that at that time $\tau$ no scales have become non-linear yet.
Note that the precise value of $k_{\rm NL}$ depends quite sensitively on the model parameters, and in particular also on the value of $S_0$ at that scale. One can roughly estimate the value of $S_0$ at a scale $k$ by taking $S_0\sim \mathcal{P}_{S_0}^{1/2}(k)$, but then the scale $k_{\rm NL}$ can only be determined numerically for fixed parameter values. For example, the values used in \cref{fig:OmegaGW_Forecast} yield $k_{\rm NL}\approx 2.5 \times 10^{-3} k_{\rm uv}$ ($w=1/6$) and $k_{\rm NL}\approx 2.8 \times 10^{-3} k_{\rm uv}$ ($w=2/3$).
Non-linear effects, such as binary PBH interactions, lead to a suppression of the power spectrum of PBH number density fluctuations at these small scales. As a consequence, the amplitude of the induced GWs will also be suppressed at frequencies larger than the corresponding $f_{\rm NL}$.

\section{Gravitational wave kernel} \label{sec:GW_kernel}
The Kernel $I(x,u,v)$ appearing in the tensor power spectrum \labelcref{eq:TensorPowerSpectrum} is defined as the convolution of the (retarded) tensor Green's function $\mathcal{G}_h(x,\tilde{x})$ and the source function $f(x,u,v)$, explicitly \cite{Domenech:2021ztg}
\begin{equation}
    I(x,u,v)=\int_{x_i}^x d\tilde{x} \ \mathcal{G}_h(x,\tilde{x}) f(\tilde{x},u,v) \,.
\end{equation}
The Green's function is found from the two homogeneous solutions $h_{1,2}(\tau)$ to the tensor equation of motion
\begin{equation}
    h_{k,\lambda}''(\tau) +  2\mathcal{H} h_{k,\lambda}'(\tau) + k^2 h_{k,\lambda}(\tau) = \mathcal{S}_{k,\lambda}(\tau) \,,
    \label{eq:TensorEoM}
\end{equation}
which holds for both polarisations $\lambda$ separately. In the isocurvature case, the source term $\mathcal{S}_{k,\lambda}(\tau)$, up to second order in perturbation theory and when a single fluid is dominant, is given by \cite{Domenech:2021ztg, Kohri:2018awv}
\begin{align}
    \mathcal{S}_{k,\lambda}(\tau)&= 4\int \frac{d^3q}{(2\pi)^3}e^{ij}_\lambda(k) q_i q_j S_{0,\textbf{q}}S_{0,\textbf{p}} f(\tau,q,p) \, , \\
    \text{where \ } f(\tau, q, p) &= T_\Phi(q\tau)T_\Phi(p\tau) + \frac{1+b}{2+b}\left(T_\Phi(q\tau)+\frac{T_\Phi'(q\tau)}{\mathcal{H}}\right) \left(T_\Phi(p\tau)+\frac{T_\Phi'(p\tau)}{\mathcal{H}}\right)
    \label{eq:SourceFctf}
\end{align}
with $p= |\textbf{k}-\textbf{q}|$ and the polarisation tensors $e^{ij}_\lambda$. The variables $u$ and $v$ appearing in \labelcref{eq:TensorPowerSpectrum} are defined in terms of the momenta $q$ and $p$ by $u=p/k$ and $v=q/k$. The transfer function $T_\Phi$ of the curvature perturbation is defined by $\Phi(\textbf{k},\tau)=S_{0,\textbf{k}} T_\Phi(k,\tau)$.
Let us note that the initial value of the isocurvature perturbation $S_{0,\textbf{k}}$ is drawn from a scale dependent distribution, denoted here by the subscript $\textbf{k}$.
The initial distribution is quantified by the (dimensionless) isocurvature power spectrum $\mathcal{P}_{S,\rm f}(k) \delta_D(k+k')=\frac{k^3}{2\pi^2}\langle S_{0,\textbf{k}} S_{0,\textbf{k}'} \rangle$, which is given in \labelcref{eq:InitialIsocurvaturePS}.
In the case of adiabatic initial conditions one has instead $\Phi(\textbf{k},\tau)=\Phi_{0,\textbf{k}} T_\Phi(k,\tau)$ with the initial curvature perturbation $\Phi_0$, which is distributed according to the power spectrum $\mathcal{P}_{\Phi_0}(k)$ given in \labelcref{eq:ScalarPS}.

The solution to the homogeneous part of \labelcref{eq:TensorEoM} yields a Green's function given by a sum of Bessel functions of the first and second kind as \cite{Domenech:2021ztg}
\begin{equation}
    \mathcal{G}_h(x,\tilde{x}) = \frac{\pi}{2} \frac{\tilde{x}^{b+3/2}}{x^{b+1/2}} \left(Y_{b+\frac{1}{2}}(x) J_{b+\frac{1}{2}}(\tilde{x})-J_{b+\frac{1}{2}}(x) Y_{b+\frac{1}{2}}(\tilde{x})\right) \Theta(x-\tilde{x}) \, ,
    \label{eq:TensorGreensFctGen}
\end{equation}
which is valid in a single-fluid Universe. The full solution for the tensor modes is then given by
\begin{equation}
    h_k(x) = \int_{x_i}^x d\tilde{x} \ \mathcal{G}_h(x,\tilde{x})\frac{\mathcal{S}_k(x)}{k^2} \, .
\end{equation}
During radiation domination, $b=0$, the Green's function \eqref{eq:TensorGreensFctGen} reduces to
\begin{equation}
    \mathcal{G}_h^{\rm RD}(x,\tilde{x})=\frac{\tilde x}{x} \sin(x-\tilde x) \ \Theta(x-\tilde{x})\,.
\end{equation}
Following the discovery of \cite{Inomata:2019ivs} that a sudden transition from matter domination to the radiation era leads to an enhanced production of GWs, we will consider only the contributions to \labelcref{eq:SourceFctf} after PBH evaporation.
Then, the dominant term is the one involving two time derivatives, i.e.~the one proportional to $\left(\mathcal{H}^{-1} T_\Phi'\right)^2$. This can be understood by considering the transfer function in the radiation period \labelcref{eq:PhiRD}, which entails $\Phi_{\rm RD}(\tau_{\rm eva})\propto \Phi_{\rm eMD}$ and $\mathcal{H}^{-1} \Phi_{\rm RD}'(\tau_{\rm eva})\propto (k \tau_{\rm eva})\Phi_{\rm eMD}$. As we are considering scales $k\gg k_{\rm eva}$, the second term is the larger one, originating from the enhanced oscillation of the curvature perturbation after the transition. Consequently, the dominant contribution to the Kernel $I(x,u,v)$ is given by \cite{Domenech:2020ssp}
\begin{align}
    I_{\rm RD}(\bar{x},u,v)\approx \frac{1}{2}uv\int_{x_{\rm eva}/2}^{\bar{x}} d\tilde{x} \ \tilde{x}^2 \mathcal{G}_h^{\rm RD}(\bar{x},\tilde{x})
    \frac{d T_\Phi^{\rm RD}(u\tilde{x})}{d (u\tilde{x})}
    \frac{d T_\Phi^{\rm RD}(v\tilde{x})}{d (v\tilde{x})} \,,
    \label{eq:KernelIRDGen}
\end{align}
where we introduced the shifted time coordinate $\bar{x}=k \bar \tau=x-x_{\rm eva}/2$ (resulting from the required continuity of the background) and used $q\tau=v x$ and $p\tau=u x$.
As we are interested in the late time limit of the GWs evolution, we will take the upper limit of the time integral $\bar{x}\rightarrow \infty$.
The transfer function for the radiation period $T_\Phi^{\rm RD}$ is defined from \labelcref{eq:PhiRD} and \labelcref{eq:SuppresionSeva} by splitting off the initial value $S_0$ or $\Phi_0$, respectively.
Inserting the transfer function into \cref{eq:KernelIRDGen}, changing the integration variable to $\tilde{x}\rightarrow \hat{x}=\tilde{x}-x_{\rm eva}/2$ and renaming again $\hat{x}\rightarrow \tilde{x}$ we obtain
\cite{Domenech:2020ssp}
\begin{align}
    I_{\rm RD}(\bar{x},u,v) \approx
    \frac{c_s^2 u v }{32 \bar{x}} x_{\rm eva}^4
   {\cal S}_{\Phi,\rm eva}(u k) {\cal S}_{\Phi,\rm eva}(v k)
    T_{\Phi}^{\rm eMD}(u k) T_{\Phi}^{\rm eMD}(v k) \, \mathcal{I}_{\rm osc}(\bar{x},u,v) \,,
    \label{eq:KernelRD}
\end{align}
at leading order in $x_{\rm eva}\gg1$ and we defined
\begin{align}
    \mathcal{I}_{\rm osc}(\bar{x},u,v) = \int_0^\infty \frac{d\tilde{x}}{\tilde{x}+x_{\rm eva}/2}
    \sin (\bar{\bar{x}} - \tilde{x})
    \sin \left(c_s u \tilde{x}\right)
    \sin \left(c_s v \tilde{x}\right) \,,
    \label{eq:OscIntegralRD}
\end{align}
where $\bar{\bar{x}} = \bar{x}-x_{\rm eva}/2$.
The time integral in \labelcref{eq:OscIntegralRD} can be done analytically and results in a sum of 8 terms involving sine and cosine integrals, given explicitly in eq.~(C4) of \cite{Domenech:2020ssp}.
Taking the oscillation average of $I_{\rm RD}^2$ and inserting the resulting Kernel $\overline{I_{\rm RD}^2}$ in the expression \labelcref{eq:TensorPowerSpectrum} leaves us with the expression \labelcref{eq:TensorPowerSpectrumFull} for $\overline{\mathcal{P}_{h,\rm RD}}$ quoted in the main text.

\section{Gravitational wave integrals} \label{sec:GW_integrals}
In both the isocurvature and adiabatic cases, the momentum integrals in \cref{eq:TensorPowerSpectrumFull} can be computed numerically for fixed $M_{\rm f}$ and $\beta$ by defining a grid of values $k_{\rm num}=k/k_{\rm uv}$, and using the relations \labelcref{eq:Relationsk_kUV} to express the scales $k_{\rm eq}$ and $k_{\rm eva}$ in terms of $k_{\rm uv}$. For each grid point, we then define a range of $v$-values (respecting the UV-cutoff ${v_{\rm uv}=u_{\rm uv}=k_{\rm uv}/k}$), resulting in a purely numerical expression and allowing us to perform the integral over $u$. We then compute the integral for every value of $v$, interpolate the resulting data to obtain the integrand as a function of $v$, and finally perform the remaining integral over $v$. In this way we obtain $\overline{\mathcal{P}_{h,\rm RD}}$ as a function of $k_{\rm num}$.
Using that during radiation domination $\mathcal{H}=1/\bar{\tau}$, we have $k^2/\mathcal{H}^2 = \bar{x}^2$, and the GW energy density is easily obtained from the tensor power spectrum by \labelcref{eq:OmegaGWDefinition}.

Selecting only the dominant part of the kernel \labelcref{eq:OscIntegralRD} near the resonance - the term involving a divergent cosine integral \cite{Inomata:2019ivs, Domenech:2020ssp} - and considering the $k\gg k_{\rm eq}$ part of the transfer function yields
\begin{align}
    \mathcal{I}_{\rm osc, res}(\bar{x};u+v\sim c_s^{-1}) = \frac{1}{4}\text{Ci}\left(\left(1-c_s(u+v)\right) x_{\rm eva}/2\right)\sin (\bar{\bar{x}}+\left(1-c_s(u+v)\right) x_{\rm eva}/2) \,.
\end{align}
Taking the oscillation average, i.e.~integrating over half a period and dividing by $\pi$, we obtain
\begin{align}
    \overline{\mathcal{I}^2_{\rm osc, res}}(u,v) = \frac{1}{32} \text{Ci}\left(|(1-c_s(u+v))| x_{\rm eva}/2\right)^2 \,.
    \label{eq:OscIntegralRD2_res}
\end{align}
Si$(x)$ and Ci$(x)$ denote the sine and cosine integrals, respectively, and are defined by
\begin{equation}
    \text{Si}(x)=\int_0^x \frac{\sin(t)}{t}dt \quad \text{and} \quad \text{Ci}(x) = -\int_x^\infty\frac{\cos(t)}{t} dt \,,
\end{equation}
with limiting values Si$(x\rightarrow\infty)=\pi/2$, Ci$(x\rightarrow\infty)=0$, Si$(x\rightarrow 0)=0$ and Ci$(x\rightarrow0)=-\infty$.

At low frequencies $k\ll k_{\rm uv}$ the momentum integrals are dominated by the regime $u\approx v \gg1$.
Setting $u=v$, we select the terms \cite{Domenech:2020ssp}
\begin{align}
    \mathcal{I}_{\rm osc, LV}(x;u=v\gg 1) = -
    \frac{1}{2} \left(\text{Ci}(x_{\rm eva}/2) \sin (x)+\left(\pi/2 - \text{Si}(x_{\rm eva}/2)\right) \cos (x)\right) \,,
\end{align}
which are most relevant at $u \sim v \gg 1$. After taking the oscillation average we have
\begin{align}
    \overline{\mathcal{I}^2_{\rm osc, LV}}(u,v) = \frac{1}{8} \left(\text{Ci}(x_{\rm eva}/2)^2+\left(\pi/2 -\text{Si}(x_{\rm eva}/2)\right)^2\right) \,.
    \label{eq:OscIntegralRD2_LV}
\end{align}

\subsection{Isocurvature induced GWs}
Inserting the dominant part of the Kernel near the resonant peak, \cref{eq:OscIntegralRD2_res}, into the expression \cref{eq:TensorPowerSpectrumFull} we obtain the resonant contribution to the tensor power spectrum
\begin{align}
    \overline{\mathcal{P}_{h,\rm RD, res}}&(k,\tau,x\gg1) \approx
   C(w)^4 \frac{c_s^4}{2^{20/3}3^{4/3}\pi ^2  \bar{x}^2}
    \left(\frac{k_{\rm eq}}{k_{\rm uv}}\right)^8
    \left(\frac{k_{\rm uv}}{k_{\rm eva}}\right)^{2}
    \left(\frac{k}{k_{\rm eva}}\right)^{14/3} \nonumber \\
    & \times \int_{0}^{v_{\rm uv}} dv \int_{|1-v|}^{\text{min}(1+v,v_{\rm uv})} du
    \frac{\left(\left(1+v^2-u^2\right)^2-4 v^2\right)^2}{(u v)^{5/3}} \text{Ci}^2\left(| (1-(u+v)c_s)| \frac{k}{k_{\rm eva}}\right) \, ,
    \label{eq:TensorPowerSpectrumRes}
\end{align}
where $v_{\rm uv}=k_{\rm uv}/k$.
The double integral in \labelcref{eq:TensorPowerSpectrumRes} can be approximately solved analytically by transforming to the new variables $y=((u+v)c_s-1)k/k_{\rm eva}$ and $s=u-v$. We then effectively treat the divergent cosine integral like a Dirac delta function by evaluating the integrand at $y=0$, except for the Ci-term itself, which we integrate from $-\infty$ to $\infty$, yielding a factor of $\pi$. This approximation is found to agree excellently with a numerical solution of the integral.
Applying this approximation, the second line in \labelcref{eq:TensorPowerSpectrumRes} results in
\begin{align}
    \Theta_{\rm uv}(k) \coloneq & \int_{-s_0(k)}^{s_0(k)} ds
    \frac{\left(s^2-1\right)^2}{\left(1-c_s^2 s^2\right)^{5/3}} \label{eq:ThetaUV} \\
    =&\frac{3 s_0 \left(5 c_s^4-2 c_s^2 \left(2 s_0^2+5\right)+9\right)-\left(5 c_s^2 \left(c_s^2+6\right)-27\right) s_0 \left(c_s^2 s_0^2-1\right) \, _2F_1\left(\frac{5}{6},1;\frac{3}{2};c_s^2 s_0^2\right)}{10 c_s^4 \left(1-c_s^2 s_0^2\right)^{2/3}}\nonumber
\end{align}
times some prefactor, and we introduced
\begin{align}
 s_0(k)\coloneq
 \begin{cases}
        1  \quad & \frac{k_{\rm uv}}{k}\geq \frac{1+c_s^{-1}}{2}\\
        2\frac{k_{\rm uv}}{k}-c_s^{-1} \quad & \frac{1+c_s^{-1}}{2}\geq \frac{k_{\rm uv}}{k}\geq\frac{c_s^{-1}}{2}\\
        0 \quad & \frac{c_s^{-1}}{2}\geq\frac{k_{\rm uv}}{k}
    \end{cases} \,,
\end{align}
resulting from the integration boundaries in \labelcref{eq:TensorPowerSpectrumRes} due to momentum conservation, combined with the UV-cutoff of the isocurvature power spectrum \labelcref{eq:InitialIsocurvaturePS} at $k_{\rm uv}$. $s_0(k)$ acts as a step function that is equal to 1 for small $k/k_{\rm uv}$ and then smoothly drops until it vanishes at and above $k/k_{\rm uv}=2 c_s$.

In the IR regime $k\ll k_{\rm uv}$ the contribution from $u\approx v \gg1$ dominates the momentum integrals. We split the integrals into two parts $v\in(0,v_{\rm eq}=k_{\rm eq}/k)$ and $v\in(v_{\rm eq},v_{\rm uv}=k_{\rm uv}/k)$, where the respective branches of the transfer function \labelcref{eq:PhiIsoeMDCases} apply. Noting that the power of the integrand in \cref{eq:TensorPowerSpectrumFull} is positive, the integrals are dominated by the upper boundaries.
After setting $u=v$ we can solve the remaining momentum integral to find for the two regimes
\begin{align}
    \label{eq:OmegaGWLVLk}
\overline{\mathcal{P}_{h,\rm RD, LV}}&(v > v_{\rm eq}) \approx C(w)^4 \frac{c_s^4 }{5\ 2^{2/3} 3^{1/3} \pi ^2 \bar{x}^2} 
    \left(\frac{k_{\rm eq}}{k_{\rm eva}}\right)^8
    \left(\frac{k_{\rm eva}}{k_{\rm uv}}\right)^{10/3}
    \left(\frac{k}{k_{\rm uv}}\right) \\
\overline{\mathcal{P}_{h,\rm RD, LV}}&(v < v_{\rm eq}) \approx \frac{c_s^4 }{217500\ 2^{2/3} 3^{1/3} \pi ^2 }
\left(\frac{k_{\rm eq}}{k_{\rm eva}}\right)^{14/3}
\left(\frac{k_{\rm eq}}{k_{\rm uv}}\right)^{5}
\left(\frac{k}{k_{\rm uv}}\right)\,,
\end{align}
where we also expanded the Ci and Si terms for large argument (i.e.~$x_{\rm eva} \gg 1$). The first contribution \labelcref{eq:OmegaGWLVLk} is significantly larger and determines the IR tail $\Omega_{\rm GW, IR}$.

\subsection{Adiabatic induced GWs}
In the adiabatic case, the cutoff function $\tilde{\Theta}_{\rm uv}(k)$ for the resonant part \labelcref{eq:OmegaGWresAdi} is defined by
\begin{align}
    \tilde{\Theta}_{\rm uv}&(k) \coloneq  
    \int_{-s_0(k)}^{s_0(k)} ds\left(s^2-1\right)^2\left(1-c_s^2 s^2\right)^{n_{\rm eff}} \label{eq:ThetaUV_ad} \\
    =&\frac{2}{5} s_0^5 \, _2F_1\left(\frac{5}{2},-n_{\rm eff};\frac{7}{2};c_s^2 s_0^2\right)-\frac{4}{3} s_0^3 \, _2F_1\left(\frac{3}{2},-n_{\rm eff};\frac{5}{2};c_s^2 s_0^2\right)+2 s_0 \, _2F_1\left(\frac{1}{2},-n_{\rm eff};\frac{3}{2};c_s^2 s_0^2\right) \,. \nonumber
\end{align}

The factors $\xi_{1,2}$ appearing in \labelcref{eq:OmegaGWmidAdi,eq:OmegaGWIRAdi} are introduced to account for the fact that we treat the transfer function in a piecewise manner when splitting the $v$-integral in \cref{eq:TensorPS_Adi_Gen} into large and small $v$ pieces. However, the main contribution in these cases stems from the boundary resulting from the split and is thus sensitive to this approximation.
For \labelcref{eq:OmegaGWmidAdi} the relevant contribution results from the $v>v_{\rm eq}$ piece, where the transfer function \labelcref{eq:PhieMDAdi} is scale-dependent, and we find that $\xi_1$ varies with $\beta$ and $w$. In this case it has to fixed by comparison to numerical results on a case-by-case basis.

In the IR tail \labelcref{eq:OmegaGWIRAdi} the $v<v_{\rm eq}$ part dominates, where the transfer function \labelcref{eq:PhieMDAdi} is constant.
In this case we determine $\xi_2$ by equating the large and small $k/k_{\rm eq}$ branches of \labelcref{eq:PhieMDAdi} and solving for the corresponding scale, where the two branches cross. We find that evaluating the momentum integrals at this scale agrees well with the numerical solution, where the interpolated transfer function \labelcref{eq:PhiAdieMDInterpol} is fed into the integrals. The resulting $\xi_2$ is given by
\begin{equation} \label{eq:xi_matching}
    \xi_2(w>1/3) = \left(\frac{A(w)}{\Phi_{\rm ad}^{\rm super}}\right)^{\frac{1+3w}{3+3w}} \quad \text{and} \quad
    \xi_2(w<1/3) = \left(\frac{A(w)}{\Phi_{\rm ad}^{\rm super}}\right)^{\frac{2+6w}{3(1+5w)}} \,,
\end{equation}
where we defined $\Phi_{\rm ad}^{\rm super} = (3 w +5)/(5w +5)$ from the superhorizon solution \cref{eq:PhiAdSuperMD}.

The lower integral boundary of the $v<v_{\rm eq}$ part of the momentum integrals in the large-$v$ approximation would give an additional contribution determined by the IR cutoff of the power spectrum. We neglect this part, as for the scales we are considering this contribution is deep inside the $v\ll 1$ regime where our approximation is invalid, and we find that \cref{eq:OmegaGWIRAdi} alone yields a good description of the IR tail.

\bibliography{references}

\end{document}